\documentclass[usenatbib,useAMS]{mn2e}
\newcommand{\bib}{\bibitem[\protect\citeauthoryear}

\newcommand{\ba}{beam$^{-1}$}
\usepackage{epsf}
\usepackage[dvips]{graphicx}
\usepackage[T1]{fontenc}
\usepackage{ae,aecompl}
\usepackage{placeins}

\begin{document}
\title[Decelerating relativistic jets]{Systematic properties of decelerating relativistic jets in 
  low-luminosity radio galaxies}
\author[R.A. Laing, A.H. Bridle]
   {R.A. Laing \thanks{E-mail: rlaing@eso.org}$^{1}$, A.H. Bridle $^2$\\
    $^1$ European Southern Observatory, Karl-Schwarzschild-Stra\ss e 2, D-85748 
    Garching-bei-M\"unchen, Germany \\ 
    $^2$ National Radio Astronomy Observatory, Edgemont Road, Charlottesville,
    VA 22903-2475, U.S.A. \\
}

\date{Received }
\maketitle

\begin{abstract}
We model the kiloparsec-scale synchrotron emission from jets in 10
Fanaroff-Riley Class I radio galaxies for which we have sensitive,
high-resolution imaging and polarimetry from the Very Large Array.  We
assume that the jets are intrinsically symmetrical, axisymmetric,
decelerating, relativistic outflows and we infer their inclination
angles and the spatial variations of their flow velocities, magnetic
field structures and emissivities using a common set of fitting
functions.  The inferred inclinations agree well with independent 
indicators.  The spreading rates increase rapidly, then decrease, 
in a flaring region. The jets then recollimate to form conical outer 
regions at distance $r_0$ from the active galactic nucleus (AGN).  
The flaring regions are homologous when scaled by $r_0$. 
At $\approx$0.1$r_0$, the jets brighten abruptly
at the onset of a high-emissivity region and we find an outflow speed
of $\approx$0.8$c$, with a uniform transverse profile. Jet
deceleration first becomes detectable at $\approx$0.2$r_0$ and the
outflow often becomes slower at its edges than it is
on-axis. Deceleration continues until $\approx$0.6$r_0$, after which
the outflow speed is usually constant. The dominant magnetic-field
component is longitudinal close to the AGN and toroidal after
recollimation, but the field evolution is initially much slower than
predicted by flux-freezing.  In the flaring region, acceleration of
ultrarelativistic particles is required to counterbalance the effects 
of adiabatic losses and account for observed X-ray synchrotron emission, 
but the brightness evolution of the outer jets is consistent with 
adiabatic losses alone.
We interpret our results as effects of the interaction between the
jets and their surroundings.  The initial increase in brightness
occurs in a rapidly falling external pressure gradient in a hot,
dense, kpc-scale corona around the AGN.  We interpret the
high-emissivity region as the base of a transonic `spine' and suggest
that a subsonic shear layer starts to penetrate the flow there.  Most
of the resulting entrainment must occur before the jets start to
recollimate.
\end{abstract}


\section{Introduction}
\label{Introduction}

Jets from active galactic nuclei (AGNs) are important in many areas of
astrophysics: they extract energy from supermassive black holes,
produce the most energetic photons (and perhaps cosmic rays) we
observe, act as conveyors of ultrarelativistic particles and magnetic
fields from the parsec-scale environments of AGN to the
multi-kiloparsec scales of extended radio galaxies and quasars, and
supply copious amounts of energy to their surroundings, thereby
preventing cooling and profoundly affecting the evolution of massive
galaxies and clusters.

AGN jets are relativistic where they are first formed \citep[and
references therein]{book}.  In this paper, we are concerned
with jets in low-luminosity radio galaxies, whose flows are initially
relativistic (e.g. \citealt{Biretta95,Giov01,Hard03}), but rapidly
decelerate on kiloparsec scales (e.g.\ \citealt{LPdRF}).

We have made deep Very Large Array (VLA) observations of twin radio
jets in nearby, low-luminosity radio galaxies with \citet{FR74} Class
I (FR\,I) morphology, in which we can image both jets with high
angular resolution transverse to their axes.  Our goal is to
understand the kinematics and dynamics of these jets. There are, as yet,
no predictive theoretical models for FR\,I jets on kiloparsec
scales. The problem of
simulating the propagation of a very light,
relativistic, magnetized jet in three dimensions is computationally
prohibitive, with poorly known initial conditions: no simulation can
yet hope to follow a jet all the way from its formation on scales
comparable with the gravitational radius of the central black hole
to the kiloparsec scales for which the most detailed observations
are available.
We have therefore adopted an empirical approach 
to jet modelling
in which we
attempt to infer basic flow parameters without introducing too many
preconceptions about the underlying physics.

The jets we have observed exhibit {\sl systematic} side-to-side
asymmetries in total intensity and linear polarization that can be
recognized as large-scale manifestations of Special Relativistic
aberration.  Our key assumption is that {\em apparent asymmetries due
  to aberration are much larger than any intrinsic asymmetries over
  the faster parts of the jets}.  Specifically, we assume that the
jets can be approximated as intrinsically symmetrical, axisymmetric,
relativistic, stationary flows in which the magnetic fields are
disordered but anisotropic.  We adopt simple, parametrized functional
forms for the geometries, velocity fields, intrinsic emissivity
variations and three-dimensional magnetic field configurations of the
outflows, calculate model brightness distributions and optimize the
parameters by fitting to our observed $I$, $Q$ and $U$ images.  Our
kinematic models are described in a series of papers
\citep{LB02a,CL,CLBC,LCBH06,backflow}, where we present evidence for
deceleration from relativistic to sub-relativistic speeds on
kiloparsec scales. Starting from these kinematic models, we have also
addressed the dynamics of jet deceleration using a conservation-law
approach \citep{LB02b,Wang09}.

We now wish to look for systematic similarities and differences
between the flow properties deduced by these methods for FR\,I radio
galaxies with different luminosities and in different environments.
During the course of our project, it became clear that some of the
functional forms we had used in earlier papers were insufficiently
general while others were unnecessarily complicated.  Changes to the
fitting functions that we had made as we refined our approach made it
difficult to compare our results systematically across all of the
sources we had observed.  In this paper, we use the same set of
fitting functions for all of the sources.

In Section~\ref{obs-red}, we give the essential information about the
sources and our observational material. We outline the model-fitting
technique in Section~\ref{fit-method} and show comparisons between
data and models in Section~\ref{compare} in a way that emphasizes the
systematic variation of the appearance of these jets with their
inferred orientation to the line of sight. The results of the model
fits are presented and described in Section~\ref{results}.
Consistency tests are outlined in Section~\ref{consistency} and the
effects of intrinsic asymmetries on our results are explored in
Section~\ref{intrinsic}.  We discuss the implications for jet physics
in Section~\ref{discuss} and summarize our conclusions in
Section~\ref{Conclusions}.  The fitting functions are listed for
reference in Appendix~\ref{function-details}.  The $\chi^2$ values for
the fits are tabulated and plotted in Appendix~\ref{chisq} and
Appendix~\ref{notes} gives notes on individual model fits, emphasizing
any differences from our published work. Vector images illustrating
the polarization fits are shown in Appendix~\ref{vectors}. The model
parameters and their errors are tabulated in
Appendix~\ref{parameter-tables}.

\section{Observations and Image Processing}
\label{obs-red}

\subsection{Source selection}
\label{sources}

We seek to model the jets in the subset of FR\,I radio galaxies whose
large scale structure is currently fed by jets that: (a) have
propagated far from their AGN, (b) are detectable and resolvable by
polarimetry with the VLA on both sides of the nucleus and (c) are
reasonably straight. These are the {\em twin-jet} sources as 
classified by \citet{L93} and \citet{Leahy93}.  Our selection
eliminates some classes of radio source entirely (e.g.\ relic emission
without jets, relaxed doubles whose jets disrupt very close to the
nucleus, wide-angle tails and other objects whose jets remain
well collimated and asymmetric until they disrupt, narrow-angle tails
and any sources that are confined to the nuclear regions of their
galaxies).  For the 3CRR and B2 samples, roughly 35\% and 46\%,
respectively, of the FR\,I sources are of the twin-jet type
\citep*{LRL,Parma87,3CRRAtlas}.

\begin{center}
\begin{table}
\caption{Basic data for the sources in this paper. (1) Name as used in
  this paper; (2) alternative names; (3) redshift (bracketed if the
  distance is derived from a redshift-independent indicator); (4)
  metric distance; (5) linear scale, in kpc\,arcsec$^{-1}$; (6)
  references for redshift and distance.\label{tab:sources}}
\begin{minipage}{80mm}
\begin{tabular}{llllll}
\hline
Name     & Alt. &  $z$     &$D$& Scale         & Ref \\
         & name        &          &Mpc  & kpc           &\\
         &             &          &     & arcsec$^{-1}$ &\\
\hline                                    
1553+24  &             & 0.04263  &180.8& 0.841 & 7 \\
0755+37  & NGC\,2484   & 0.04284  &181.7& 0.845 & 3 \\
0206+35  & UGC\,1651   & 0.03773  &160.2& 0.748 & 2 \\
&4C35.03&&&\\                             
NGC\,315 &             & 0.01649  & 70.4& 0.335 & 5 \\
3C\,31   & NGC\,383    & 0.01701  & 72.6& 0.346 & 6 \\
NGC\,193 & UGC\,408    & 0.01472  & 62.8& 0.300 & 4 \\
M\,84    & 3C\,272.1   &(0.0035)  & 18.5& 0.089 & 1,6 \\
&NGC\,4374&&&\\                           
0326+39  &             & 0.02430  &103.5& 0.490 & 3 \\
3C\,296  & NGC\,5532   & 0.02470  &105.2& 0.498 & 3 \\
3C\,270  & NGC\,4261   &(0.007465)& 30.8& 0.148 & 1,6 \\
3C\,449  &             & 0.017085 & 72.9& 0.347 & 7 \\
\hline 
\end{tabular}

References:
(1)  \citet{ATLAS3D};
(2)  \citet{Falco99}; 
(3)  \citet{Miller02}; 
(4)  \citet{Ogando}; 
(5)  \citet{Smith2000}; 
(6)  \citet{Trager}; 
(7)  \citet{3RC}
\end{minipage}
\end{table}
\end{center}

We selected 10 twin-jet sources by the following criteria.
\begin{enumerate}
\item The jets are either straight and antiparallel or bend by sufficiently
  small angles that the images can be `straightened' by a simple
  transformation (Section~\ref{bends}; Laing \& Bridle, in preparation).
\item Any surrounding lobe emission is sufficiently weak to be subtracted using
  a simple linear interpolation across the jets \citep[Section~\ref{lobesub};][]{backflow}.
\item There are significant 
brightness and polarization
asymmetries in the jet bases for our models to fit.  
\item The polarized emission from both jets can be imaged with adequate
  signal-to-noise ratio. This is essential in order to break the degeneracy
  between velocity and angle to the line of sight (Section~\ref{principles}).
\end{enumerate}
 
We also show observations for one source we cannot model,
3C\,449. This has highly symmetrical jets and is likely to be a
side-on counterpart of the other sources, so we include it 
in studies of trends with inclination.

The sources considered here are
listed in Table~\ref{tab:sources} in order of increasing fitted angle
to the line of sight (Section~\ref{results}), together with their
redshifts and linear scales.  For the two nearest galaxies, we
adopted redshift-independent distances \citep[and references
  therein]{ATLAS3D}. In the remaining cases, we derived the
distances directly from the quoted redshifts without further
correction, assuming a concordance cosmology with $H_0$ =
70\,$\rm{km\,s^{-1}\,Mpc^{-1}}$, $\Omega_\Lambda = 0.7$ and $\Omega_{\rm M} =
0.3$.

\subsection{Images}
\label{Images}

High-quality VLA images at 4.9 or 8.5\,GHz are available for all of
the sources.  We started from images of Stokes $I$, $Q$ and $U$ at one
or two resolutions.  We then corrected the observed ${\bmath E}$-vector
position angles for Faraday rotation using multifrequency,
high-resolution rotation-measure (RM) images (except for 0326+39 and
1553+24, for which the corrections are close to zero) and checked that
residual depolarization is negligible.  From the corrected position
angles, $\chi_0$, we derived zero-wavelength $Q$ and $U$ images: $Q_0
= P\cos 2\chi_0$ and $U_0 = P\sin 2\chi_0$, where $P =
(Q^2+U^2)^{1/2}$ is the polarized intensity. We fit to $I$, $Q_0$ and
$U_0$ (from now on we drop the suffices).  We define the direction of
the {\em apparent magnetic field} to be $\chi_0 + \upi/2$ (this is the
same as the projected position angle on the sky for a uniform magnetic
field, but not in the general case of an integration along the line of
sight) and the scalar degree of polarization to be $p = P/I$.

In Table~\ref{tab:images}, we
list the model fields and the sizes, resolutions and intensity ranges
of the images displayed in this paper. We also give references to
descriptions of the observations, data reduction, Faraday rotation
correction and modelling.

\begin{center}
\begin{table*}
\caption{Radio data, image parameters and associated references. (1)
  Name; (2) label used in the figures; (3) $\log$ of the luminosity in extended
  emission at an emitted frequency of 1.4\,GHz, in W\,Hz$^{-1}$; (4)
  observing frequency for the modelled images, in GHz; (5) model
  field, in arcsec;  (6) -- (8) refer to the lower-resolution images
  used for modelling and specifically to the plots in
  Figs~\ref{fig:icomposite} -- \ref{fig:qicomposite} and
  \ref{fig:vectors}; (6) beamwidth (FWHM, arcsec); (7) field size
  along the jet axis as plotted in the figures; (8) total-intensity
  maximum for Fig.~\ref{fig:icomposite}; (9) -- (11): as (6) -- (8),
  but for the higher-resolution images;  (12) references to radio
  observations and models.\label{tab:images}}
\begin{minipage}{160mm}
\begin{tabular}{llllrrrrrrrl}
\hline
&&&&&&&&&&&\\
Name & & $\log (P_{\rm ext}$  &$\nu$&Model  &\multicolumn{3}{c}{Resolution 1}&\multicolumn{3}{c}{Resolution 2}& Refer-\\ 
     & &W\,Hz$^{-1})$&GHz &field  & FWHM & Field & $I_{\rm max}$ &FWHM & Field & $I_{\rm max}$   & ences \\
     & &            &    &arcsec &arcsec& arcsec & mJy         &arcsec& arcsec & mJy & \\ 
     & &            &    &       &      &        &\ba          &      &        & \ba & \\
&&&&&&&&&&&\\
\hline
&&&&&&&&&&&\\     
1553+24 &a &23.79 &8.5 &60.0 &0.75& 57.5       & 1.0 & 0.25  & 9.6 & 2.00            & 1 \\
0755+37 &b &25.02 &4.9 &66.0 &1.30& 63.0       & 3.0 & 0.40  &10.0 & 5.0             & 8,10 \\
0206+35 &c &24.82 &4.9 &22.0 &0.35& 21.0       & 3.0 &       &     &                 & 8,10 \\
NGC\,315&d &24.58 &4.9 &200.0&2.35& 190.0      &10.0 & 0.40  &38.0 &1.25             & 2, 7,11 \\
3C\,31  &e &24.54 &8.4 &56.0 &0.75& 54.0       & 3.0 & 0.25  &20.0 & 1.00            & 5, 8  \\
NGC\,193&f &23.96 &4.9 &120.0&1.35& 117.0      & 3.0 & 0.45  &17.0 & 1.50            & 9,11 \\
M\,84   &g &23.42 &4.9 &40.0 &0.40& 25.0       & 3.0 &       &     &                 & 9,11 \\
0326+39 &h &24.27 &8.5 &43.5 &0.50& 42.0       & 0.5 & 0.25  &15.0 & 0.2             & 1, 9 \\
3C\,296 &i &24.76 &8.5 &83.4 &0.75& 81.9       & 1.5 & 0.25  &15.0 & 0.5             & 6, 9 \\
3C\,270 &j &24.32 &4.9 &82.0 &0.60& 80.0       & 0.75&       &     &                 & 12  \\
3C\,449 &k &24.38 &8.5 &     &1.25& 90.0       & 1.5 & 0.80  &52.5 & 1.0             & 3, 4\\
&&&&&&&&&&&\\     
\hline
\end{tabular}

References for Table~\ref{tab:images}: 
1  \citet{CL}; 
2  \citet{CLBC}; 
3  \citet{Feretti99}; 
4  \citet{Guidetti10}; 
5  \citet{LB02a}; 
6  \citet{LCBH06}; 
7  \citet{ngc315ls}; 
8  \citet{3c31ls}; 
9  \citet{lobes}; 
10 \citet{backflow}; 
11  Laing \& Bridle (in preparation); 
12  Laing, Guidetti \& Bridle (in preparation); 
\end{minipage}
\end{table*}
\end{center}

\subsection{Complications: the symmetry assumption and bent jets, lobe subtraction, small-scale structure and backflow}

\subsubsection{The symmetry assumption and bent jets}
\label{bends}

Even if jets are exactly symmetrical where they are
first formed, interactions between them and their environment 
introduce asymmetries and often become the dominant shapers of the
large-scale radio structure.  We aim to identify and model only those
parts of the jets near the nucleus where relativistic effects dominate
the observed asymmetries and to quantify the errors introduced by
residual environmental effects.

For that reason, we initially restricted our modelling to straight,
antiparallel jets. In some FR\,I sources, the jets bend in
projection on the sky by small angles while maintaining their
collimation, and
these bends occur at discrete locations rather than forming
continuously-curved ridge-lines.
  In such cases it is possible
to `straighten' the jets by a simple image transformation (Laing \&
Bridle, in preparation) in which the brightness distribution maintains its
initial position angle up to some distance from the AGN, after which
it is sheared in a constant direction in such a
way that the ridge-line is rotated by a constant angle $\Delta$.  This
type of distortion is a position-dependent translation
and therefore preserves surface brightness.  We implement the transformation by 
polynomial interpolation of the brightness distributions in all three 
Stokes parameters, together with a rotation of the ${\bmath E}$-vector position 
angle by $-\Delta$ which in turn modifies $Q$ and $U$.  After this
transformation, $I$, $Q$ and $U$ all have the expected symmetries with
respect to the projection of the jet axis on the sky.

This approach has allowed us to extend our earlier modelling of
NGC\,315 \citep{CLBC} to larger distances, to improve our model for
3C\,296 \citep{LCBH06} and to derive new models for NGC\,193 and
M\,84.  The images for these four sources shown below have all been
corrected in this way.  Additional uncertainties are obviously
introduced by the use of a bending correction: we cannot determine 
the extent to which jets bend perpendicular to the plane of the sky 
(and hence any change in Doppler factor) and there may be other deviations 
from intrinsic symmetry.  
In NGC\,315, 3C\,296 and NGC\,193, we expect these
uncertainties to be small: the bends are $\leq$5$^\circ$ in projection
and/or restricted to the outermost parts of the jets which have low
weight in the modelling. For M\,84, the bends are both larger in
amplitude ($5^\circ$ and $14^\circ$ for the main and counter-jets,
respectively) and much closer to the nucleus, so larger errors are
likely. Full details will be given by Laing \& Bridle (in
preparation).

\subsubsection{Lobe subtraction}
\label{lobesub}

The jets in the majority of sources discussed here are surrounded, at
least in projection, by extended lobes. In order to model the jets,
any superposed lobe emission must be removed in all Stokes parameters.
One possible approach is to exploit the spectral differences between
jet and lobe emission; a second is to assume that the lobe emission
varies slowly across the jet, and to use spatial interpolation to
perform the subtraction (see \citealt{backflow} for a comparison of
these methods). We found that the latter approach gave superior
results in all cases.  We have applied it to all of the images in
which there is significant lobe emission at the lower (or only)
resolution used for modelling. The sources affected are: 0206+35,
0755+37, NGC\,193, M\,84, 3C\,296 and 3C\,270. We show only images
after lobe subtraction. Full details are given by \citet[and in
  preparation]{backflow} and Laing et al. (in preparation).

\subsubsection{Small-scale structure}
\label{smallscale}

Conceptually, we assume stationary flow. In reality, all jets develop
small-scale, stochastic structure.  Our aim is not to describe the
fluctuations, but rather to average over a complex and presumably
time-variable flow pattern in such a way as to recover global
structure in the brightness and polarization distributions.  Our
estimates of the flow parameters will be inaccurate if the brightness
distributions are dominated by small numbers of compact features,
especially if, as we would expect, they are not symmetrically located
with respect to the AGN.  We seek to mitigate this problem by
identifying and fitting common features in the brightness and
polarization distributions of multiple sources.  We do not expect stochastic variations to bias the mean
flow parameters for a sample of sources, since they should be uncorrelated with inclination.

\subsubsection{Backflow}
\label{backflow-intro}

Many FR\,I radio galaxies have outer structures
resembling the lobes of FR\,II sources, but without the compact
hot-spots that are thought to mark the terminations of
high-Mach-number jets.  Our observations of two sources of this type,
0206+35 and 0755+37 \citep{lobes,backflow}, revealed that the jets in
both sources have two-component structures transverse to their
axes. Close to the axis, the main jets are centre-brightened whereas
the counter-jets are centre-darkened. Both are surrounded by broader
collimated emission that is brighter on the counter-jet side. We
modelled these jets as decelerating, relativistic outflows surrounded
by slower (but still mildly relativistic) backflows \citep{backflow}.
In this paper, we are concerned primarily with the outflows and
their relation to similar structures in other sources.  When comparing
models with observations, we perforce include the backflow component
(otherwise we could not find an acceptable fit), but with the primary
purpose of isolating and fitting the outflow. We do not duplicate our
earlier, detailed discussion of backflow properties \citep{backflow}.

\section{Model fits}
\label{fit-method}

\subsection{Assumptions}
\label{assumptions}

We make the following assumptions when calculating model brightness distributions.
\begin{enumerate}
\item The jets are intrinsically symmetrical, axisymmetric, stationary
  flows. After cosmological corrections are applied, the jet boundary
  is at rest in the observer's frame.
\item The flow is laminar. If (as seems likely) the real flow has significant
  random motions, then we will determine average parameters weighted by
  the observed-frame emissivity.
\item The emission is optically thin synchrotron radiation.  We do not
  include optically thick emission from the core in the model.
\item The radiating particles have a power-law energy distribution
  \begin{eqnarray}
  n(E){\rm d}E &=& n_0 E^{-q}{\rm d}E \label{eq:energy}\\ \nonumber
  \end{eqnarray}
  with $q = 2\alpha + 1$, corresponding to a constant
  spectral index $\alpha$ with $I(\nu) \propto \nu^{-\alpha}$. In practice,
  we use the mean spectral index for the jets in a given source over
  the modelled region in our frequency range. This is a good
  approximation for all of our sources, for which $\alpha \approx 0.6$
  \citep{spectra}.
\item The pitch-angle distribution of the radiating electrons is
  random, in which case the maximum observed degree of polarization is
  $p_0 = (3\alpha+3)/(3\alpha+5) \approx 0.7$.
\item The magnetic field is disordered on small scales, with many
  reversals, but anisotropic. We quantify the anisotropy using the two
  independent ratios of the rms field components along three
  orthogonal directions defined with respect to the flow streamlines.
  Both vector-ordered and disordered fields can produce high degrees
  of polarization.  As we will show, the dominant field components are
  toroidal and longitudinal. Large-scale helical fields (with
  significant longitudinal as well as toroidal components) are
  inconsistent with observations because the distributions of angles
  between the field and the line of sight are different on opposite
  sides of the jet ridge-line and the transverse profiles of total
  intensity and linear polarization consequently show systematic
  asymmetries \citep*{L81,LCB}.  They are also unlikely on such large
  scales because the longitudinal magnetic flux close to the AGN would
  then be unreasonably large \citep*{BBR}. The combination of ordered
  toroidal and disordered longitudinal components would produce the
  same emission as a purely disordered configuration, however.  We
  cannot distinguish between these two cases (and others with
  symmetrical field-angle distributions), but our estimates of field
  component ratio are essentially independent of the details of the
  configuration.
\item   We define the scalar {\em emissivity function} 
  \begin{eqnarray}
  \epsilon &=& n_0 B^{1+\alpha},\label{eq:em-function}\\\nonumber
  \end{eqnarray}
  where $B$ is the rms total magnetic field (all quantities are
  defined in the rest frame).  It is multiplied by a constant and by
  functions depending on the field structure to give the true
  emissivities $e$ in $I$, $Q$ or $U$ (Section~\ref{principles}).
\item Variations of velocity and field-ordering with position are
  smooth. We allow limited discontinuities in the emissivity function.
\end{enumerate}

\subsection{Principles}
\label{principles}

The key to our method is to determine the velocity and inclination angle 
independently by comparing emission from the main and
counter-jets in {\em both total intensity and linear polarization}. For an
emitting element moving at an angle $\theta$ to the line of sight ($0
\leq \theta \leq \upi$, with $\theta = 0$ towards the observer) and
observed at frequency $\nu$, the emissivity $e$ in the
observer's frame is given by
\begin{eqnarray}
e(\theta,\nu) &=& D^2(\theta) e^\prime(\theta^\prime,\nu^\prime), \label{eq:beam} \\ \nonumber
\end{eqnarray}
where $e^\prime$, $\theta^\prime$ and $\nu^\prime$ are measured
in the rest frame of the emitting material \citep[equation~C6]{BBR}.
$D$ is the Doppler factor, which also relates the frequencies in the
observed and rest frames:
\begin{eqnarray}
D(\theta) &=& [\Gamma(1-\beta\cos\theta)]^{-1} \label{eq:doppler} \\ 
 \nu &=& D(\theta)\nu^\prime.\\  \nonumber
\end{eqnarray}
Here, $\beta c$ is the bulk flow speed and $\Gamma =
(1- \beta^2)^{-1/2}$ is the bulk Lorentz factor.
The angles to the line of sight in the two frames are related by
\begin{eqnarray}
\sin\theta^\prime &=& D(\theta)\sin\theta .\label{eq:aberration} \\ \nonumber
\end{eqnarray}
For an optically thin source at distance $d$ with a power-law spectrum of spectral
index $\alpha$ as defined in Section~\ref{assumptions}, the observed
flux density $S$ is 
\begin{eqnarray}
S(\theta,\nu) &=& D^{2+\alpha} \int e^\prime(\theta,\nu) {\rm d}V / d^2 \\ \nonumber
\end{eqnarray}
The integration is performed in the observer's frame, for which $dV$
is the volume element \citep[equation~C7]{BBR}.

From now on, we consider antiparallel jets and take $\theta$ ($0 \leq
\theta \leq \upi/2$) to be the angle to the line of sight for the
approaching one. The Doppler factors for the approaching and receding
jets, $D_{\rm j}$ and $D_{\rm cj}$ are
\begin{eqnarray}
D_{\rm j} &=& [\Gamma(1-\beta\cos\theta)]^{-1} \label{eq:dj} \\
D_{\rm cj} &=& [\Gamma(1+\beta\cos\theta)]^{-1} \label{eq:dcj} \\ \nonumber
\end{eqnarray} 
For isotropic emission in the rest frame, the jet/counter-jet ratio is then given by
the well-known formula
\begin{eqnarray}
\frac{I_{\rm j}}{I_{\rm cj}} & = & \left( \frac
{1+\beta\cos\theta}{1-\beta\cos\theta}\right)^{2+\alpha} \label{eq:iratio}\\ \nonumber 
\end{eqnarray}
We cannot determine $\beta$ and $\theta$ independently from this ratio alone. In
general, however, emission is {\em anisotropic in the rest frame} and the
angular dependences are different for the three Stokes parameters.  This allows us to separate the two quantities.

In order to illustrate this point, we analyse two simple field
configurations for which there are analytical expressions for the
total and polarized intensity and which are good initial approximations
to those we find in FR\,I jets (Section ~\ref{field-results}).  We consider the idealized case of
cylindrical, antiparallel jets with constant velocity. The
field configurations are assumed not to vary along the jets and the
particle densities and rms field strengths are independent of
position. We also take $\alpha = 1$ in order to simplify the formulae.
We choose the zero-point of ${\bmath E}$-vector position angle to be the
projection of the jet axis on the plane of the sky, so that $U
= 0$. The two configurations are as follows:
\begin{enumerate}
\item a field which has no longitudinal component, but which is
  orientated randomly in planes perpendicular to the jet axis
  \citep[Section IIIa]{L81} and
\item a field which is orientated randomly within shells of given
  radius, but with no component perpendicular to the jet axis \citep[Section IIIb, model
    B]{L81}.
\end{enumerate}
In the first configuration, the apparent magnetic field is always perpendicular
to the jet axis ($Q > 0$):
\begin{eqnarray}
Q(\theta^\prime) &=& Kp_0 \sin^2\theta^\prime \label{eq:P2D}\\
I(\theta^\prime) &=& K(2 - \sin^2\theta^\prime)\label{eq:I2D}\\ \nonumber
\end{eqnarray}
($K$ varies across the jet, but is the same for both Stokes parameters).  In the
observed frame, the ratios of polarized and total intensity are
\begin{eqnarray}
\frac{Q_{\rm j}}{Q_{\rm cj}} &=& \left(\frac{D_{\rm j}}{D_{\rm cj}}\right)^5 \label{eq:P2D2} \nonumber\\ &=&
\left(\frac{1+\beta\cos\theta}{1-\beta\cos\theta}\right)^5 \\ 
\frac{I_{\rm j}}{I_{\rm cj}} &=& \left(\frac{D_{\rm j}}{D_{\rm cj}}\right)^3 \frac{2 -
  D_{\rm j}^2}{2 - D_{\rm cj}^2}\nonumber\\ &=&
\left(\frac{1+\beta\cos\theta}{1-\beta\cos\theta}\right)^3
\frac{2 - (1-\beta^2)(1+\beta\cos\theta)^{-2}}{2 - (1-\beta^2)(1-\beta\cos\theta)^{-2}}\label{eq:I2D2} \\\nonumber
\end{eqnarray}
The degree of polarization in the approaching jet always exceeds that in the receding jet.

In the second configuration, the degree of polarization varies perpendicular to
the jet. On-axis, the apparent magnetic field is again always
transverse, and we have the same relations as in equations~(\ref{eq:P2D}) and
(\ref{eq:I2D}), but with $\theta^\prime$ offset by $\upi/2$:
\begin{eqnarray}
Q(\theta^\prime) &=& Kp_0 (1-\sin^2\theta^\prime)\label{eq:PmodB}\\
I(\theta^\prime) &=& K(1 + \sin^2\theta^\prime)\label{eq:ImodB}\\ \nonumber
\end{eqnarray}
Consequently, the emission ratios in the observed frame are
\begin{eqnarray}
\frac{Q_{\rm j}}{Q_{\rm cj}}&=& \left(\frac{D_{\rm j}}{D_{\rm cj}}\right)^3 \frac{1-D_{\rm j}^2\sin^2\theta}{1-D_{\rm cj}^2\sin^2\theta}\nonumber\\
&=&\left(\frac{1+\beta\cos\theta}{1-\beta\cos\theta}\right)^3\frac{1-(1-\beta^2)(1-\beta\cos\theta)^{-2}}{1 - (1-\beta^2)(1+\beta\cos\theta)^{-2}} \label{eq:PmodB2}\\
\frac{I_{\rm j}}{I_{\rm cj}}&=& \left(\frac{D_{\rm j}}{D_{\rm cj}}\right)^3 \frac{1+D_{\rm j}^2\sin^2\theta}{1+D_{\rm cj}^2\sin^2\theta}\nonumber\\
&=&\left(\frac{1+\beta\cos\theta}{1-\beta\cos\theta}\right)^3\frac{1+(1-\beta^2)(1-\beta\cos\theta)^{-2}}{1 + (1-\beta^2)(1+\beta\cos\theta)^{-2}}\label{eq:ImodB2}\\ \nonumber
\end{eqnarray}
If $\beta = \cos\theta$, the degree of polarization in the approaching jet $p_{\rm j} = |Q_{\rm j}|/I_{\rm j} = 0$  (equation~\ref{eq:PmodB2}), whereas the receding jet
has $p_{\rm cj} = 2p_0\cos^2\theta/(1+\cos^4\theta)$ with a
transverse apparent field (equations~\ref{eq:PmodB} and \ref{eq:ImodB}).  At the edges of both jets, $Q/I = -p_0$ (longitudinal apparent field) and
\begin{eqnarray}
\frac{I_{\rm j}}{I_{\rm cj}}&=&\frac{Q_{\rm j}}{Q_{\rm cj}}\nonumber\\
&=& \left(\frac{D_{\rm j}}{D_{\rm cj}}\right)^3\nonumber\\
&=&\left(\frac{1+\beta\cos\theta}{1-\beta\cos\theta}\right)^3\label{eq:PmodBedge}\\ \nonumber
\end{eqnarray}
as in the isotropic case.

If we knew the field configuration a priori, we could use pairs of
equations such as (\ref{eq:P2D2}) and (\ref{eq:I2D2}) or (\ref{eq:PmodB2})
and (\ref{eq:ImodB2}) to determine $\beta$ and $\theta$ independently.
In practice, of course, the field configuration is unknown.  The
additional constraints we use to determine it are the transverse
variations of total intensity and linear polarization across the two
jets: profiles across both jets are necessary in order to provide
integrations through the field distributions at different angles to
the line of sight in the rest frame.

\subsection{Resolving degeneracies using transverse profiles}

It is clearly important to check whether the same observed brightness
distributions (in $I$, $Q$ and $U$) could be produced by alternative 
combinations of velocity, magnetic field and emissivity
function.  First, we note that differences between the approaching and 
receding jets are barely affected by the form adopted for the emissivity 
function, which divides out from the jet/counter-jet ratio in any Stokes 
parameter in the limit of a uniform flow.  There is a potential
degeneracy between velocity and field structure, however.  We infer
the presence of a velocity gradient from the centrally-peaked
transverse profile of intensity ratio $I_{\rm j}/I_{\rm cj}$ but there
is a way to mimic this profile purely from the anisotropy of the
rest-frame emission for one special field configuration, which we 
now discuss.

Even if the jets have uniform
velocities, aberration can cause the angle between a well-ordered
field and the line of sight in the rest frame, $\psi^\prime$, to be
much smaller in the approaching jet.  The emissivity ($\propto
\sin^{1+\alpha}\psi^\prime$) is then much lower.  If this happens at
the edges of the jets but not on-axis, then the jet/counter-jet 
sidedness-ratio profile will be centrally peaked.  Of the field 
configurations which
are qualitatively consistent with the observed linear polarization,
the only one that can produce this effect has a dominant toroidal
component, with the field loops seen close to edge-on in the rest
frame in the approaching jet ($\psi^\prime \approx 0$, so
$\theta^\prime \approx \upi/2$ and $\beta \approx \cos\theta$).
Profiles of sidedness ratio and $Q/I$ for this case are shown in
\citet[their fig.~A1]{backflow}.  If we mistakenly assumed isotropic
emission in the rest frame and did not look at the polarization, we
might indeed conclude (incorrectly) that there is a transverse
velocity gradient. However, the polarization of the approaching jet
produced by this field configuration is not consistent with the
observed one. A pure toroidal field always gives transverse apparent
field with $p = p_0$ on-axis. If the field loops are viewed edge-on in
the rest frame, then this extends over the entire width of the jet. In
less perfectly aligned cases, transverse field and high polarization
are still seen over much of the width of the jet
\citep[Fig.~A1]{backflow}.  This is rarely observed.

In practice, solutions of this type do not fit our data: even starting
with a purely toroidal field and the appropriate velocity to generate
the observed transverse sidedness ratio gradients, our fitting
algorithm (Section~\ref{opt}) converges to solutions with a transverse
velocity gradient and a mixture of longitudinal and toroidal field, as
only these can reproduce the observed polarization.  We have also
verified that the velocity gradient is still required even if the
field component ratios are allowed to vary across the jets.

We are therefore confident that there are no significant degeneracies
between transverse velocity profile and field structure for the jets we have
observed and modelled, because the possibilities can be distinguished by their 
different polarization profiles. 

\subsection{Terminology}
\label{Jet-terms}

The term {\it flaring} is used in two contexts when describing the properties
of kiloparsec-scale radio jets: to describe changes in jet geometry and of jet 
brightness.

{\it Geometrical} flaring of a jet refers to significant
increases in its apparent spreading rate (opening angle) with
increasing distance from the AGN.  These changes (inferred from 
observing the outer isophotes of the jets) appear to be a continuous
process on the scales we can resolve (i.e.\ the opening angle gradually 
increases, then decreases).  Geometrical flaring can thus be ascribed 
to an extended region of the propagating jet, and the only observed 
discontinuity appears to be where the jet opening angle becomes 
constant.

{\it Brightness} flaring of kiloparsec-scale jets refers to significant increases
in their apparent brightness with increasing distance from the AGN, often
following an initial `gap', or extended region in which the radio emission is weak or 
undetectable.  Unlike geometrical flaring, brightness flaring usually has a 
well-defined onset (especially considering the effects of projection), so we can 
often define a single brightness flaring point with some precision.

The two phenomena are evidently connected in that the brightness flaring point generally 
occurs in a part of the flow where the jet opening angle is increasing with distance, 
i.e.\ within the geometrically flaring region.  The term `flaring' has been 
used elsewhere (e.g.\ \citealt{Bridle82, Roberts86, Loken1995, LPdRF, Jetha2006, Krause2012, lobes, 
backflow}), to describe both phenomena. 
 We continue this practice, but we will explicitly 
distinguish geometrical flaring and brightness flaring in what follows, to clarify 
the relationship(s) between them.  
 
\subsection{Fitting functions}
\label{functions}

The functions used to fit the jet outflows have been chosen
empirically to have simple algebraic forms which together allow good
fits to the brightness and polarization distributions and
straightforward estimation of key physical parameters.  The
characterization of variations along the jets reflects the observation
that there are distinct regions within which the quantities that we
model (geometry, velocity, emissivity function and field structure)
must vary in different ways. The regions are identifiable by changes
in:
\begin{description} 
\item [{\em geometry}] (the shapes of the outer isophotes);
\item [{\em velocity}] (the gradient of the sidedness ratio);
\item [{\em emissivity function}] (the logarithmic slope of the surface brightness) and
\item [{\em field structure}] (the gradient of $Q/I$ on-axis).
\end{description} 
In the latter two cases, the changes must be common to both jets.
The observations thus lead to the concepts of {\em fiducial locations} (the boundaries
between regions) and {\em fiducial values} defined at these locations,
both at the centre and edge of the jet.  The functional forms are
chosen to interpolate smoothly between the fiducial locations and
between the centre and edge. The precise form used for the interpolation
functions is not critical (within reason) provided that the values at
the fiducial locations are correct. For example, we have used
different functions to fit the longitudinal and transverse velocity
variations in 3C\,31, but the inferred velocity field is very
similar in all cases \citep[][their fig. 17, and this paper]{LB02a}.

The fiducial distances and values, together with the functional forms,
are defined in Sections~\ref{coords} -- \ref{mag-functions}, below.
For completeness, we tabulate the complete coordinate definitions and
functional forms in Appendix~\ref{function-details}
(Table~\ref{tab:functions}).  Distances, angles and velocities are
defined in the observer's frame and intrinsic parameters for field and
emissivity function refer to the rest frame of the emitting plasma.

\subsubsection{Geometry and coordinate systems}
\label{coords}

\begin{figure}
  \epsfxsize=8.5cm
  \epsffile{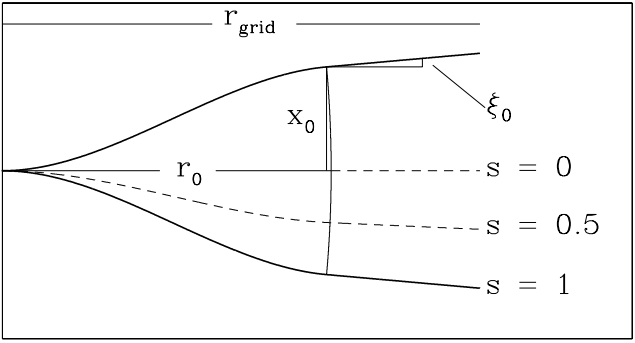}
  \caption{The geometry of the model jet outflow, showing the flaring
    and outer regions and the three quantities $r_0$, $x_0$ and $\xi_0$
    which define the shape of its outer surface. Three example
    streamlines are shown: $s = 0$ (on-axis), $s = 0.5$ and $s = 1$ (the
    outer boundary).
    \label{fig:geomsketch}}
\end{figure}

The jet axis is inclined by an angle $\theta$ to the line of sight; 
$z$ and $x$ are coordinates along and transverse to the jet axis,
respectively.  We model on a grid whose size, set by the observed
image, is fixed in projection on the sky. The corresponding physical
size measured along the jet axis, $r_{\rm grid}$, then depends on
$\theta$.  Motivated by the discussion in Section~\ref{Jet-terms}, we
divide a jet into {\em geometrically flaring} and {\em outer} regions,
as shown in Fig.~\ref{fig:geomsketch}.  The geometry is
completely defined by the transition distance between the two regions,
$r_0$, the radius of the jet at the transition between the regions,
$x_0$, and the opening angle of expansion in the outer region,
$\xi_0$.

In order to parametrize the spatial variations of velocity,
emissivity function and field ordering, we use a coordinate system $(r,s)$
where the index $s$ is constant for a given streamline,
running from 0 on-axis to 1 at the jet boundary, and $r$ increases
monotonically with distance along it. The distance of a streamline
from the jet axis is:
\begin{eqnarray}
  x(z,s) & = & a_2(s) z^2 + a_3(s) z^3 \makebox{~~~~(flaring region)} \label{eq:flare}\\
  x(z,s) & = & (z+A)\tan (\xi_0 s) \makebox{~~~~~~(outer region)}\\
  \nonumber
\end{eqnarray}
where $A = x_0/ \sin\xi_0 - r_0$.  In the outer region, $s =
\xi/\xi_0$, where $\xi$ is the angle between the flow vector and the
jet axis. $a_2(s)$ and $a_3(s)$ are constant along a given streamline
and are defined by the conditions that $x(z,s)$ and its derivative
with respect to $z$, $x^\prime(z,s)$, are continuous at the transition
between the two regions.  The vertex of the flow in the outer region is displaced from the
nucleus by a distance $A$ and the boundary surface between geometrically flaring and
outer regions is a sphere of radius $r_0+A$ centred on the 
vertex. This geometry has the natural feature that the streamlines are
orthogonal to the boundary surface where they cross it.

The coordinate along a streamline:
\begin{eqnarray}
  r & = & \frac{zr_0}{(r_0 + A) \cos (\xi_0s) - A} \hspace{1.0cm}
  r \leq r_{0} \\ r & = & \frac{z + A}{\cos (\xi_0s)} - A
  \hspace{2.25cm} r \geq r_{0} \\ \nonumber
\end{eqnarray}
increases monotonically from 0 at the nucleus. The boundary between the flaring and outer
regions is at $r = r_0$ regardless of the value of $s$.  On the jet
axis, $r$ is just the distance from the nucleus, $z$.

The geometry and coordinate system are exactly as used in earlier papers in this
series (with the special case $A=0$ for 3C\,31; \citealt{LB02a}).

We chose these functional forms as the simplest which match the observed outer
isophotes of the jets on scales which we can resolve: an
extrapolation of the flaring region geometry to smaller scales would not be
consistent with higher-resolution observations, however.

\subsubsection{Velocity}
\label{vel-functions}

\begin{figure}
  \epsfxsize=8.5cm
  \epsffile{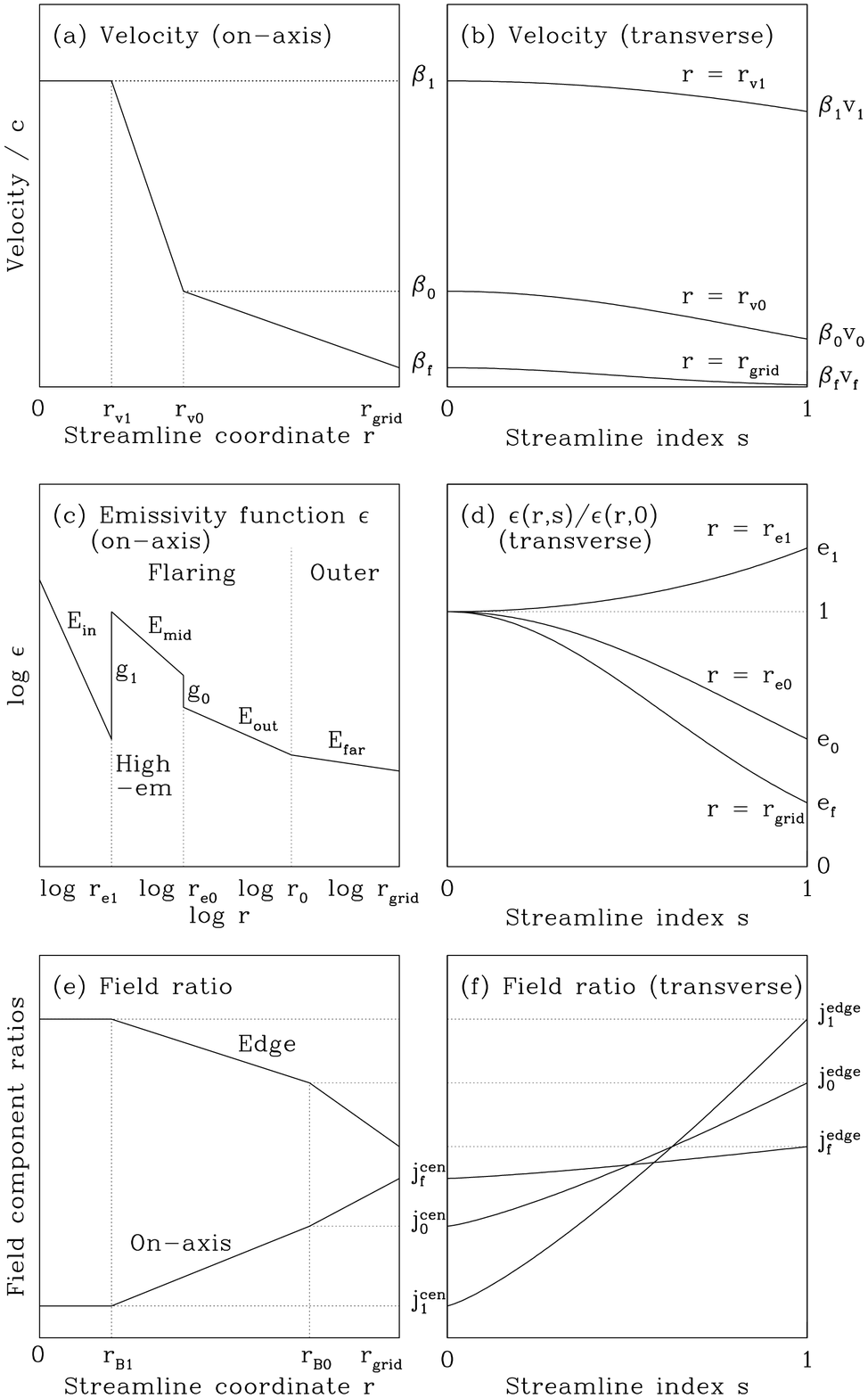}
  \caption{Examples of the functional forms used to fit the jet
    velocity, emissivity function and field-component ratios. (a)
    On-axis velocity profile. (b) Transverse velocity profiles at the
    three fiducial distances, with the same on-axis velocities as in
    panel (a).  (c) On-axis profile of the emissivity function
    $\epsilon = n_0B^{1+\alpha}$ (note the logarithmic scales).  (d)
    Normalized transverse profiles of the emissivity function at three
    fiducial locations. (e) Profiles of the field ratio $j = \langle
    B^2_r \rangle^{1/2}/\langle B^2_t \rangle^{1/2}$ on-axis and at
    the edge of the jet. (f) Corresponding transverse profiles for the
    field ratios at the three fiducial distances. The functional form
    for the field ratio $k = \langle B^2_l \rangle^{1/2}/\langle B^2_t
    \rangle^{1/2}$ is identical.
    \label{fig:sketches}}
\end{figure}

Our model velocity field is simplified significantly from those used
in earlier papers, where we adopted rather complicated functional
forms purely to enforce continuity in acceleration as well as
velocity. In fact, a good fit does not require the velocity to vary
smoothly, but merely to be continuous.  We assume that the velocity is
a separable function of distance coordinate and streamline index,
$\beta(r,s) = \beta_r(r)\beta_s(s)$, with $\beta_s(0) = 1$. The
on-axis velocity $\beta_r(r)$ (Fig.~\ref{fig:sketches}a) is taken to
have a constant value $\beta_1$ out to a distance $r_{v1}$ and to
decrease linearly to $\beta_0$ at $r_{v0}$.  Thereafter, either
uniform acceleration or deceleration to velocity $\beta_f$ at $r =
r_{\rm grid}$ is allowed.  The transverse velocity variation
$\beta_s(s)$ (Fig.~\ref{fig:sketches}b) has a truncated Gaussian form
$\beta_s(s) = \exp[s^2\ln v(r)]$, specified by the fractional edge
velocity, $v(r) \leq 1$ [we found that allowing $v(r) > 1$ led to
  problems with the optimization as $\beta$ approached 1]. The precise
form assumed for $\beta_s(s)$ does not make much difference either to
the quality of the fit or to the derived values of $v$, provided that
it is reasonably flat on-axis and decreases smoothly towards the edge:
two alternatives were compared by \citet{LB02a}. The values of $v(r)$
at the three fiducial distances $r_{v1}$, $r_{v0}$ and $r_{\rm grid}$
are $v_1$, $v_0$ and $v_f$, respectively. $v(r) = v_1$ for $r \leq
r_{v1}$; intermediate values are determined by linear interpolation in
$r$.  The complete form for the velocity function is given in
Table~\ref{tab:functions}. It is defined by the fiducial distances
$r_{v1}$ and $r_{v0}$, the on-axis velocities $\beta_1$, $\beta_0$ and
$\beta_f$ and the fractional edge velocities $v_1$, $v_0$ and $v_f$.

\subsubsection{Emissivity function}
\label{em-functions}

As in the case of the velocity, we
take the emissivity function $\epsilon$ to be separable: $\epsilon(r,s) =
\epsilon_r(r)\epsilon_s(s)$.  $\epsilon_r(r)$ has a piecewise
power-law dependence on $r$ (Fig.~\ref{fig:sketches}c).  Close to the
nucleus ($r \leq r_{e1}$), the index is $-E_{\rm in}$. At $r =
r_{e1}$, the {\em brightness flaring point}, a discontinuity of a
factor of $g_1$ is allowed. The flaring point marks the beginning of
the {\em high-emissivity region}, with index $-E_{\rm mid}$, which
again may end in a discontinuity (a factor of $g_0$).  Thereafter,
$\epsilon_r(r)$ is continuous, with indices $-E_{\rm out}$ from $r =
r_{e0}$ up to recollimation ($r = r_0$) and $-E_{\rm far}$ from $r_0$
until the end of the grid.

The transverse variation of emissivity function again has a truncated Gaussian
form, $\epsilon_s(s) = \exp[s^2 \ln e(r)]$ (Fig.~\ref{fig:sketches}d),
but the value of $e(r)$ is allowed to be $\leq$1 (centre-brightening)
or $>$1) (limb-brightening).  $e(r) = e_1$ for $r \leq r_{e1}$ and
takes the values $e_0$ and $e_f$ at $r_{e0}$ and $r_{\rm grid}$,
respectively. Intermediate values for $r > r_{e1}$ are determined by
linear interpolation.  The complete form  of the emissivity function
is given in Table~\ref{tab:functions}. The defining parameters are the
fiducial distances $r_{e1}$, $r_{e0}$; the on-axis slopes $E_{\rm
  in}$, $E_{\rm mid}$, $E_{\rm out}$, $E_{\rm far}$; the fractional
edge emissivities $e_1$, $e_0$, $e_f$ and the discontinuities $g_1$,
$g_0$.

The model emissivity is set to zero within a fixed projected
distance from the nucleus to prevent confusion with the unresolved
radio core emission, which we do not attempt to model.  This
corresponds to a linear distance of $r_{\rm min}$ along the jet axis.

\subsubsection{Magnetic-field structure}
\label{mag-functions}

The three rms magnetic-field components in the rest frame are: $\langle
B_l^2\rangle^{1/2}$ (longitudinal, parallel to a streamline), $\langle
B_r^2\rangle^{1/2}$ (radial, orthogonal to the streamline and outwards
from the jet axis) and $\langle B_t^2\rangle^{1/2}$ (toroidal,
orthogonal to the streamline in an azimuthal direction). The rms
total field strength is $B = \langle B_l^2 + B_r^2 +
B_t^2\rangle^{1/2}$.  The magnetic-field structure is parametrized by
the ratio of rms radial/toroidal field, $j = \langle
B_r^2\rangle^{1/2}/\langle B_t^2\rangle^{1/2}$ and the
longitudinal/toroidal ratio $k =\langle B_l^2\rangle^{1/2}/\langle
B_t^2\rangle^{1/2}$.

We found that the truncated Gaussian form used for velocity and
emissivity function did not provide a good description of the transverse
variation of the field ratios.  A field component ratio is therefore
described in terms of its values at the centre and edge as functions
of $r$, with a power-law interpolation between them. For the
radial/toroidal ratio, $j(r,s) = j^{\rm cen}(r) + [j^{\rm edge}(r) -
  j^{\rm cen}(r)]s^{w_j}$. $w_j$ may be positive or negative. The
longitudinal variation is defined by values at three fiducial
locations. $j^{\rm edge}(r) = j_1^{\rm edge}$ for $r \leq
r_{B1}$ and then varies linearly to $j_0^{\rm edge}$ at $r = r_{B0}$
and $j_f^{\rm edge}$ at $r = r_{\rm grid}$
(Fig.~\ref{fig:sketches}e). $j^{\rm cen}(r)$ is identical in form, and
examples of the resulting transverse variations are plotted in
Fig.~\ref{fig:sketches}(f).  The full functional form for $j(r,s)$ is
again given in Table~\ref{tab:functions}.  The longitudinal/toroidal
field ratio $k(r,s)$ is described in an identical way.

The free parameters describing the field ordering are the fiducial
distances $r_{B1}$, $r_{B0}$; indices $w_j$, $w_k$ and six values 
per ratio (three each for the centre and edge).

\subsubsection{Fits close to the nucleus}
\label{narrow}

Close to the nucleus (in practice upstream of the brightness flaring
point, $r < r_{e1}$), the jets are often faint (at least on one side
of the AGN) and poorly resolved.  This violates the conditions needed
for us to estimate inclination, emissivity function, velocity and field structure
independently. The inclination is well determined from fits at larger
distances, but we have chosen to {\em assume} that the velocity
remains constant for $r < r_{v1}$ and that the emissivity function may have a
discontinuity. This is not a unique choice, although it allows
reasonable fits close to the AGN.  For this reason, the parameters
$E_{\rm in}$ (the emissivity function slope upstream of the flaring point) and
$g_1$ (the emissivity function jump there) should not be taken too seriously.
The faintness of the jets in this region means that this region has
low weight in the modelling, so the remaining parameters are
essentially determined by the brightness and polarization
distributions at larger distances (where they are well constrained),
and assumed to remain constant close to the AGN.

\subsubsection{Minimal models}
\label{minimal}

Although we need to retain the complete parameter set described above
in order to compare all of the sources, the full complexity is not
always required. Fits of essentially the same quality can be obtained
using a limited subset of parameters, which may then be better
constrained. One important example is the form of the velocity
variation for $r > r_{v0}$.  Deceleration is required by the data in
one case (3C\,31), and we therefore allow the velocity to increase or
decrease linearly with distance until the end of the model grid.  For
the majority of the sources, the quality of the fit assuming a
constant velocity at $r > r_{v0}$ is only slightly worse.  Similarly,
the data for some of the sources are fully consistent with an absence
of transverse variation in the field-ordering parameters.  We have
therefore derived a set of {\em minimal models} for all except the two
sources that require the full parameter set (3C\,31 and M\,84), as follows.
\begin{enumerate}
\item The on-axis velocity and its transverse profile remain constant
  for $r > r_{v0}$ ($\beta_f = \beta_0$ and $v_f = v_0$).
\item The transverse variation of emissivity function remains constant for $r >
  r_{e0}$ ($e_f = e_0$).
\item There is no further change in the field ordering parameters with
  distance for $r > r_{B0}$; their transverse profiles also remain
  constant.
\end{enumerate}
This means that all of the parameters defined at the edge of the model
grid ($r = r_{\rm grid}$; subscript $f$) become redundant. In a subset
of cases, we make additional simplifications, as follows.
\begin{enumerate}
\item The power-law slope of the emissivity function variation with distance
  remains the same for $r > r_{e0}$, i.e.\ $E_{\rm far} = E_{\rm
    out}$.
\item There is no transverse variation of the field-ordering
  parameters, so the $j^{\rm edge}$, $k^{\rm edge}$, $w_j$ and $w_k$ parameters are not
  needed.
\end{enumerate}
We use the minimal models explicitly in the discussion of
flux-freezing and adiabatic models (Sections~\ref{Bevolution} and
\ref{Adiabatic}).

\subsubsection{Backflow fits}
\label{backflow-fits}

As outlined in Section~\ref{backflow-intro}, our models for 0206+35
and 0755+37 include backflowing components. The functional forms used
to fit backflow are exactly as described by \citet{backflow}, but are
also listed for completeness in Appendix~\ref{function-details}
(Table~\ref{tab:backflow_functions}).

\subsection{Optimization}
\label{opt}

Having chosen a set of functional forms, we optimize the parameters by
minimizing $\chi^2$ between the model and observations. The `noise' on the observed images is dominated by small-scale
brightness fluctuations (e.g.\ knots and filaments), and we estimate
its value, 
$\Sigma$, by measuring the deviation from reflection symmetry.  Our
prescription for $\Sigma$ is $1/\sqrt 2$ times the rms difference
between the image and a copy of itself reflected across the jet axis
for $I$ and $Q$ and $1/\sqrt 2$ times their sum for $U$ ($I$ and $Q$
are symmetric under reflection and $U$ is antisymmetric for an
axisymmetric model flow).  These estimates of $\Sigma$ are dominated by real small-scale structure, but also include 
contributions from receiver noise and deconvolution artefacts: they are
usually much larger than the off-source noise levels.  Some small-scale
features are mirror-symmetric, and we will underestimate their
contributions to $\Sigma$.

We fit to images at one or two resolutions.  The higher (or only)
resolution is always the maximum possible. If the brightness
sensitivity is too low to allow accurate imaging of the fainter parts
of the jets, then we also use a second, lower resolution.  We fit to
the higher-resolution images over the central bright regions and the
lower-resolution images elsewhere. We average the values of $\Sigma$
over the regions used in the fits at each resolution (this is a fairly
crude approximation for the inner jet regions, where the surface
brightness varies rapidly with position).

The algorithm works as follows:
\begin{enumerate}
\item At each pixel, determine the boundaries of the emission and
  integrate $I$, $Q$ and $U$ along the line of sight in the observed
  frame. At each evaluation of the integrand:
  \begin{enumerate}
  \item account for relativistic aberration given the model velocity field.
  \item determine the geometry,
  field-ordering and emissivity function from the formulae given earlier;
  \item calculate the proper emissivity from the emissivity function and field 
  ordering using a look-up table for the appropriate spectral index \citep{L02}.
  \end{enumerate}
\item Normalize to the observed total intensity at the lower (or only)
  resolution, excluding the core.
\item Convolve the resulting $I$, $Q$ and $U$ images with the observing beam(s).
\item Evaluate $\chi^2$ and sum over resolutions and Stokes parameters.
\item Iterate using the downhill simplex algorithm \citep{NM} as implemented by \citet{NR}
  to optimize the parameters. 
\end{enumerate}
Finally, we add the convolution of a point source with the observing beam at
the position of the core (this is purely cosmetic).

Aside from the effect of projection, the fits to the geometry
parameters $r_0$, $x_0$ and $\xi_0$ are essentially determined by the
shapes of the observed outer isophotes. Fits to the transition
distances for velocity, $r_{v1}$ and $r_{v0}$, are mostly affected by
variations in the jet/counter-jet sidedness and $Q/I$ ratios with
distance from the nucleus and those for emissivity function transitions ($r_{e1}$ and $r_{e0}$) 
by sharp changes in brightness gradient.  We
actually optimize all of the distances from the nucleus in projection on the sky, only converting
afterwards to the jet frame. Equation~(\ref{eq:iratio}) with $\beta = 1$
gives an approximate upper limit to $\theta$. Finally, reproducing the observed
asymmetry in linear polarization requires $\langle B^2_l\rangle
\approx\langle B^2_t\rangle \gg \langle B^2_r\rangle$ near the AGN and dominant toroidal field at larger distances, so a good
starting approximation for the field-ratio parameters is $j = 0$
everywhere, with $k = 1$ close to the AGN and $k = 0$ at large
distances.  Finding an approximate starting point for the optimization
is therefore reasonably straightforward.

The downhill simplex algorithm is a remarkably robust method for
minimizing multidimensional functions whose derivatives are not known,
but has the disadvantage that it is not guaranteed to converge to a
global minimum. A particular issue for our problem is the coupling
between $\theta$ and other parameters via the Doppler factor.  We
adopted a four-stage process to locate a global minimum. First, we
made a coarse, but systematic exploration of possible starting
conditions subject to the simple physical constraints identified above
and allowing the parameters defining the outer boundary of the
emission to vary, with $\chi^2$ measured over fixed areas including
all of the emission.  This always led to an acceptable model, but
additional stages were required to refine it. The second step was to 
fix the outer boundary
in projection and only to evaluate $\chi^2$ within it. We also found
empirically that the downhill simplex algorithm, once close to the
correct values of $\theta$, tended to `get stuck', in the sense that
it left the input $\theta$ unchanged and optimized all of the other
parameters. The third stage was therefore to run a set of optimizations
with fixed values of $\theta$ (and various starting simplexes), to
plot $\chi^2$ against $\theta$ and to find the lower bound of the
distribution.  This always showed a clear minimum. Depending on the
starting simplex, the algorithm often converged to values of $\chi^2$
slightly above the bound; occasionally, it found noticeably worse 
solutions.  Once the global minimum was accurately located, the fourth and final
stage was to verify its stability by optimizing all of the parameters,
including $\theta$.

The full outflow models have up to 40 free parameters; the minimal
models between 26 and 32 (Appendix~\ref{function-details};
Tables~\ref{tab:functions} and \ref{tab:backflow_functions}).  In
addition, we use nine parameters to fit the backflow components in
0755+37 and 0206+35. Our images have 1200 -- 2700 independent points
with adequate signal-to-noise in each of $I$, $Q$ and $U$, or 3600 --
8000 measurements in total, so the solutions are well constrained.  A
table of minimum $\chi^2$ values and numbers of independent points is
given in Appendix~\ref{chisq}
(Table~\ref{tab:chisq}). Fig.~\ref{fig:chisq} shows plots of $\chi^2$
against $\theta$ from the third stage of optimization.

\subsection{Error estimation}
\label{errors}

In multi-dimensional optimization problems of the type described here,
estimates of some of the parameters are strongly correlated.  We have
also imposed additional constraints by our choice of fitting
functions. Finally, we do not know the statistics (or even the rms
level) of the `noise' a priori.  The use of the $\chi^2$ statistic
allows effective optimization, but assessing confidence limits on
parameters is extremely difficult. A full Bayesian Markov chain Monte Carlo analysis is
becoming feasible on relatively modest clusters (each model evaluation
takes between 6 and 15\,s on a single Intel i5 core) and we plan to
carry this out in the future. In the mean time, we adopted a simple ad
hoc procedure whereby we scale the noise to make $\chi^2$ equal to the
number of degrees of freedom, set a $\chi^2$ threshold corresponding
to the formal 99\% confidence limit for independent Gaussian errors
and that number of degrees of freedom and rescale the threshold for
the original noise level.  We then vary single parameters in turn
until $\chi^2$ reaches that threshold. The error estimates are
qualitatively reasonable, in the sense that varying a parameter by its
assigned error leads to a visibly unacceptable fit, and we believe
that they give a good general impression of the range of allowed
models. They should not be taken as referring to a specific confidence
level.

Given the special role of the inclination, $\theta$, in optimization
(Section~\ref{opt}), we also evaluated the range of $\theta$ over
which we could find any solution with $\chi^2$ below the threshold,
allowing all other parameters to vary (a crude marginalization over
these parameters).  The inclination range $\Delta\theta$ from this
analysis is typically 5\degr\ -- 15\degr, compared with the 2\degr\ --
5\degr\ range from our single-variable analysis, but acceptable fits
can be found for 1553+24 over a 30\degr\ range of inclination
(Fig.~\ref{fig:chisq}a; Appendix~\ref{notes}).  The remaining
parameters vary very little from their best-fitting values over this
range.

\section{Model-data comparisons}
\label{compare}

\begin{figure*}
  \epsfxsize=15cm
  \epsffile{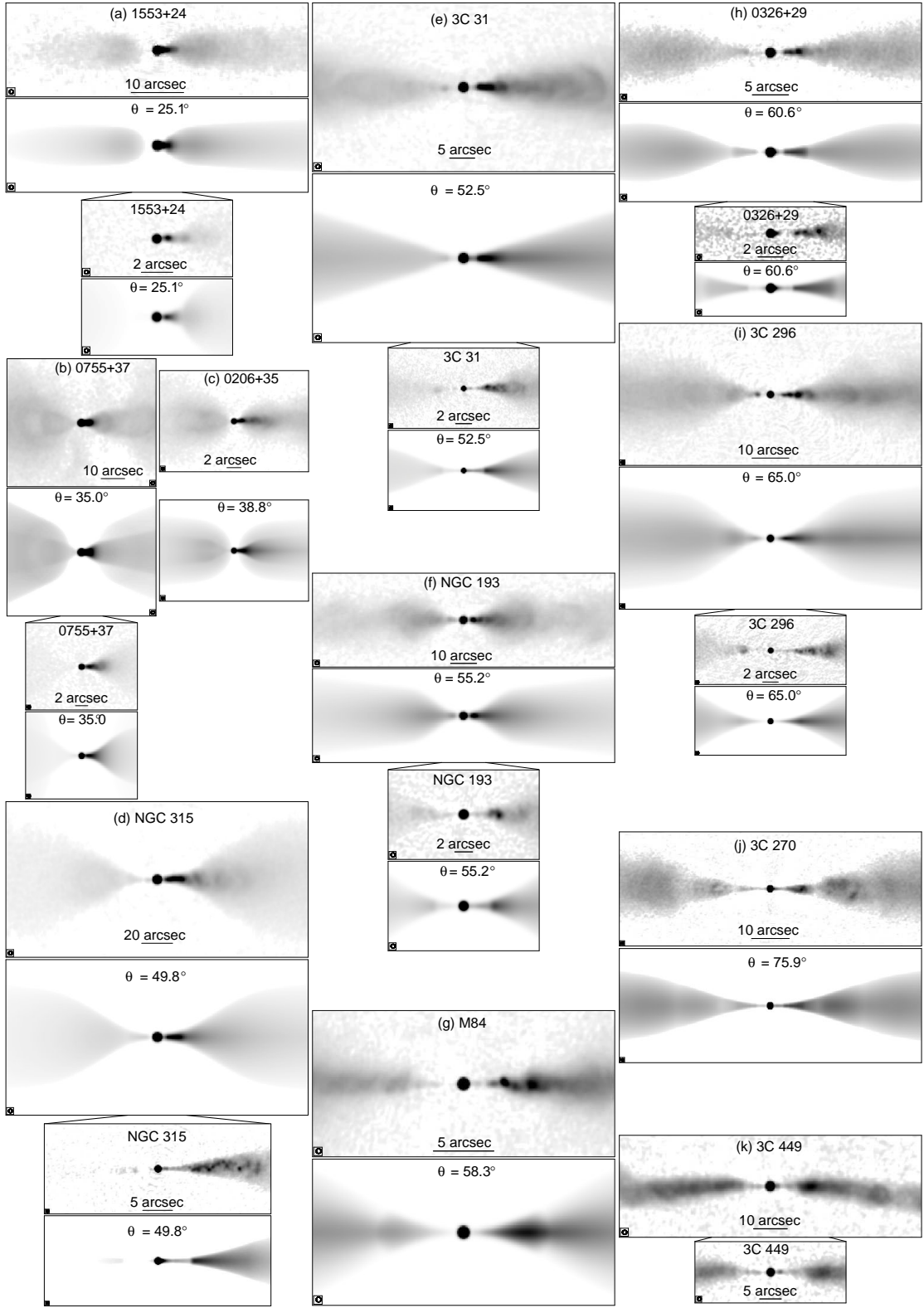}
  \caption{Comparison between observed and model total-intensity
    images. The plots are arranged in pairs. The upper and lower
    panels of each pair show the observed image (labelled with the
    source name) and the model image (with the fitted value of
    $\theta$), respectively. The angular scale is indicated by a
    labelled bar on the upper panel and the FWHM of the beam by a
    circle in one of the lower corners.  If two image resolutions were
    used, then the comparison at high resolution is shown below that
    at low resolution with the relative areas indicated.  The panels
    are in order of increasing angle to the line of sight,
    $\theta$. No model is shown for 3C\,449 (panel k).  Field sizes,
    grey-scale ranges and resolutions are all given in
    Table~\ref{tab:images}.
    \label{fig:icomposite}}
\end{figure*}

\begin{figure*}
  \epsfxsize=15cm
  \epsffile{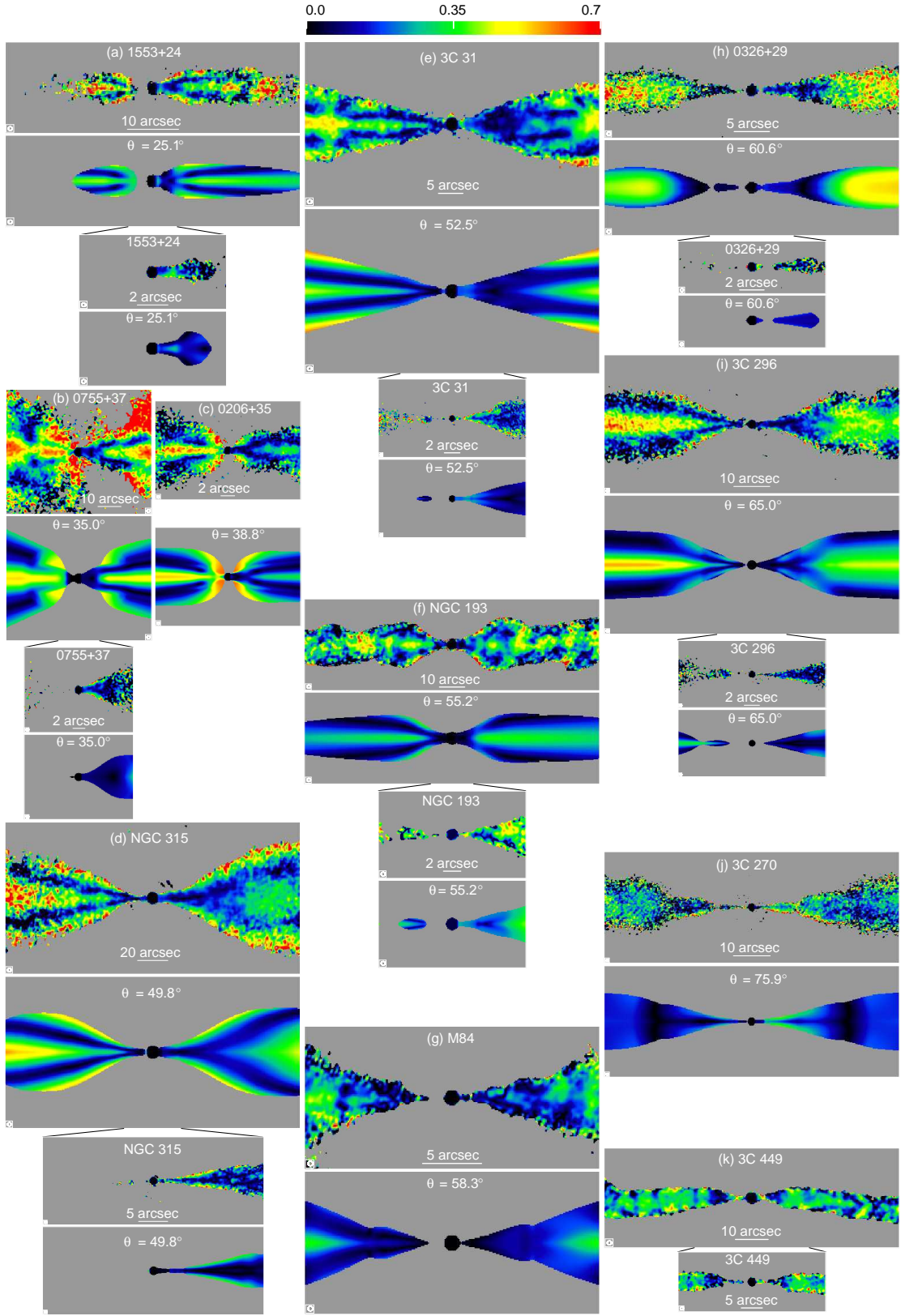}
  \caption{Comparison between observed and model images of the degree
    of polarization, $p = P/I$, in the range 0 -- 0.7 as indicated by
    the labelled wedge. The layout is identical to that in
    Fig.~\ref{fig:icomposite}. The observed values of $P$ have been
    corrected for Ricean bias \citep{WK}. The observed and model
    images are both blanked (grey) wherever $I < 5\sigma_I$
    ($\sigma_I$ is the off-source noise level).  There are
    systematic errors in $p$ around the edges of the structure of
    0755+37 (panel b), where the signal-to-noise ratio is low and
    there are large uncertainties in lobe subtraction.
    \label{fig:pcomposite}}
\end{figure*}

\begin{figure*}
  \epsfxsize=15cm
  \epsffile{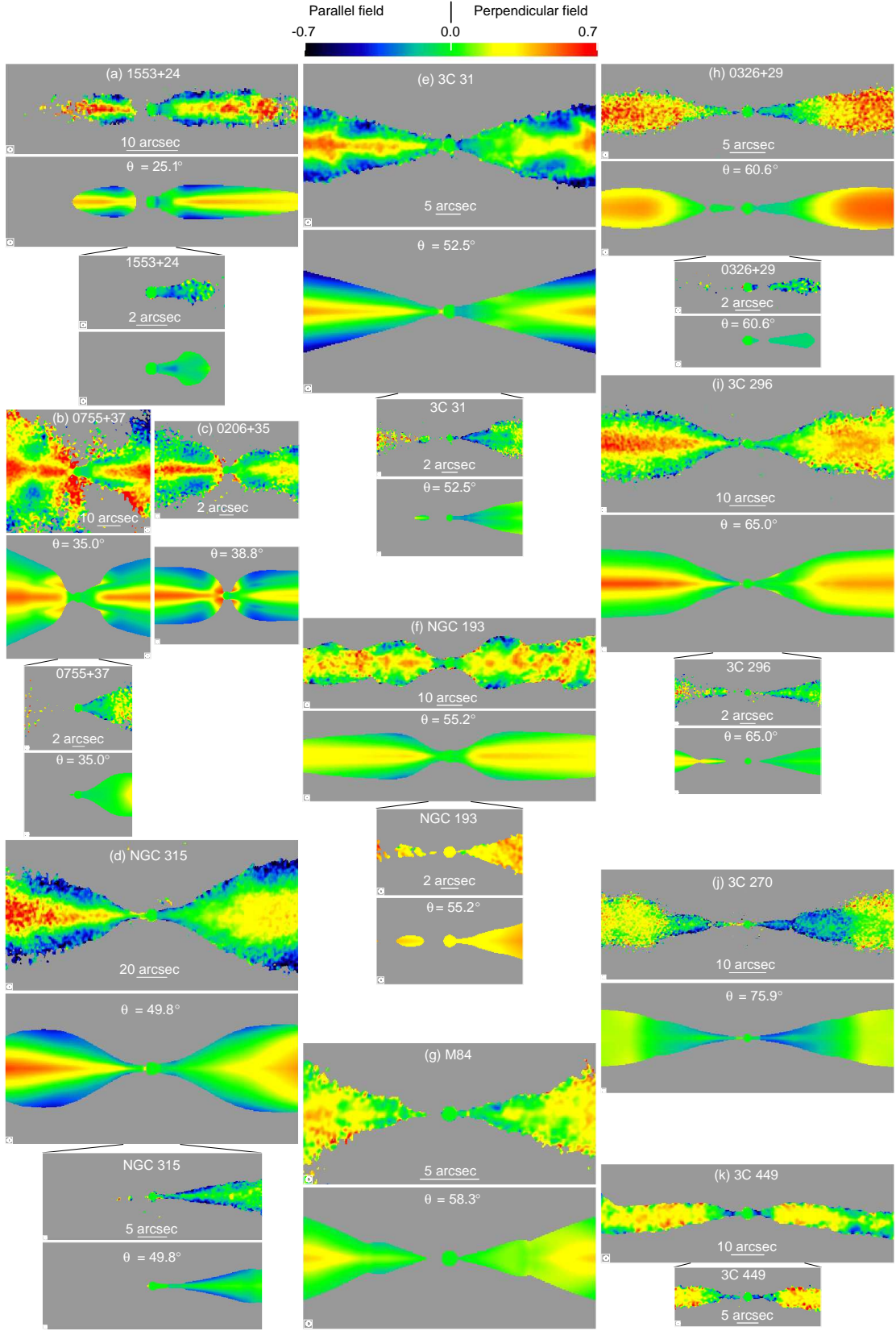}
  \caption{Comparison between observed and model images of the ratio
    $Q/I$ in the range $-0.7$ to 0.7 as indicated by the labelled
    wedge. The layout is identical to that in
    Fig.~\ref{fig:icomposite}. The observed and model images are both
    blanked (grey) wherever $I < 5\sigma_I$ ($\sigma_I$ is the
    off-source noise level).
    \label{fig:qicomposite}}
\end{figure*}

Detailed comparisons between data and model fits (including profiles
along and transverse to the jet axes) are presented elsewhere
\citep[Laing \& Bridle, in preparation; Laing et al., in preparation]{LB02a,CL,CLBC,LCBH06,backflow}. Here, we show images which
summarize the results in such a way as to emphasize general features
and trends with inclination (Figs~\ref{fig:icomposite} --
\ref{fig:qicomposite} and \ref{fig:vectors}).  In all of the plots, the radio core component is
at the centre and the brighter (approaching) jet is to the
right. Panels (a) -- (j) show model and observed images and are
arranged in order of increasing fitted angle to the line of sight,
$\theta$, which is indicated on the model panels.  The final panel (k)
shows the observations only for 3C\,449\footnote{The images of 3C\,449
  have not been `straightened'.}.  Leaving aside the small-scale
structure which we cannot model, the overall quality of the fits is extremely good and a clear pattern of
inclination-dependent features has emerged.

Fig.~\ref{fig:icomposite} shows the observed and model total-intensity
images over identical brightness ranges (the peak intensities are
listed in Table~\ref{tab:images}). All of the sources show initial
geometrical flaring followed by recollimation to a uniformly expanding
flow.  The location of the brightness flaring point is clear at high
resolution in all of the main jets. The jet/counter-jet ratio
decreases monotonically with distance from the brightness flaring
point, often reaching $I_{\rm j}/I_{\rm cj} \approx 1$ at the edges of
the plots, as expected for flows decelerating to subrelativistic
velocities.  Our model fits require similar velocities at the
brightness flaring point for all of the sources
(Section~\ref{velocity-results}), so the jet/counter-jet ratio there
is anticorrelated with angle, as is evident from the sequence of
plots.  This sequence is completed by 3C\,449, whose jet structure is
highly symmetrical, and which we believe to have $\theta \approx
90^\circ$.  The transverse intensity profiles also differ
systematically, in the sense they tend to be centrally peaked in the
main jets but flatter or even centre-darkened in the counter-jets.
The outer isophotes on both sides of the nucleus are quite
symmetrical, even if the on-axis brightness distributions are
not. These phenomena are naturally interpreted as the effects of
transverse velocity gradients: the flow is faster on-axis than at the
jet boundaries.

In Fig.~\ref{fig:pcomposite}, we present images of the degree of
polarization, $p = P/I$, in the range $0 \leq p \leq 0.7$, with
identical blanking for the observed and model images. In the
coordinate system of Section~\ref{principles}, where the zero-point of
${\bmath E}$-vector position angle is the jet axis, the linear
polarization is dominated by the $Q$ Stokes parameter: if the jets are
approximately cylindrical, then the polarization ${\bmath E}$-vectors are
either parallel or perpendicular to the axis, and $U \approx 0$.  A
clearer picture of the polarization asymmetries is therefore provided
by images of $Q/I$, which we show in the range $-0.7 \leq Q/I \leq
0.7$ in Fig.~\ref{fig:qicomposite}. Parallel and perpendicular
apparent magnetic fields have $Q/I < 0$ and $Q/I > 0$, respectively.
A full description of the linear polarization state requires all three
Stokes parameters, and this is particularly important where $U$ and
$Q$ are both significant, for example at the edges of the flaring
regions: we display vectors with lengths proportional to $p$ and
directions along the apparent magnetic field in
Fig.~\ref{fig:vectors}.  The vector plots have a similar
format to Figs~\ref{fig:icomposite} -- \ref{fig:qicomposite}, but are
on larger scales.

All of the modelled sources 
show {\em a common pattern of asymmetry in} $p$ {\em which correlates with that
seen in total intensity} (Fig.~\ref{fig:pcomposite}). In the main
(approaching) jet bases, $p$ is low close to the AGN on the jet axis,
drops to $p \approx 0$ and then rises gradually with distance.  It is
larger at the same distance from the nucleus in the counter-jet,
increasing monotonically with distance. $p$ is high on the jet axis
(particularly in the counter-jet) and at the edges of both jets,
dropping to low values at intermediate radii.  In $Q/I$
(Fig.~\ref{fig:qicomposite}), this characteristic pattern becomes
clearer. On-axis in the main jet, $Q/I$ is negative close to the
nucleus, goes through 0 and becomes positive farther out. This is the
well-known transition from longitudinal to transverse apparent field
in the approaching jet bases of FR\,I sources \citep{Bridle84}. The
counter-jets behave differently: $Q/I$ is generally $>$0 everywhere on-axis
(predominantly perpendicular apparent field), with a magnitude that
increases with distance.     $Q/I$ tends to be negative
(longitudinal apparent field) at the edges of both jets, but
particularly on the counter-jet side. These patterns are also clear
in the vector plots (Fig.~\ref{fig:vectors}), where  high
degrees of polarization and close alignment of the field vectors with
the outer boundary at the edges of the jets (particularly in the
flaring region) are often evident.   

The asymmetries in linear polarization at the bases of the jets are
perfectly correlated with those in total intensity and well fitted by
our models, consistent with the hypothesis that both are caused by
relativistic aberration. 3C\,449 is symmetrical in polarization structure, just as
it is in total intensity, consistent with expectations for a source
close to the plane of the sky.
   
At larger distances from the AGN, the pattern of transverse apparent
field on-axis and longitudinal field at the edges persists in most of
the modelled sources, but (like the total intensity) becomes more
symmetrical as the jets decelerate. 0326+39 shows a different
polarization distribution, with less transverse variation in $Q/I$ and
no evidence for a parallel-field edge, indicating a qualitatively
different intrinsic field configuration (Figs~\ref{fig:qicomposite}h
and \ref{fig:vectors}h).  

The polarization images for 3C\,270 show large deviations from
axisymmetry, and the fits are therefore poor (Appendix~\ref{notes}).

\section{Model results}
\label{results}

The values of the fitted parameters, their estimated errors and the
angle range $\Delta\theta$ are tabulated in
Appendix~\ref{parameter-tables}.

In order to compare the sources, we show plots of outer isophotes,
velocity, emissivity function and fractional field components over
fixed multiples of the recollimation distance, $r_0$, in
Figs~\ref{fig:geomresults} -- \ref{fig:Brad}, below.

\subsection{Geometry}

\begin{figure}
  \epsfxsize=8.5cm
  \epsffile{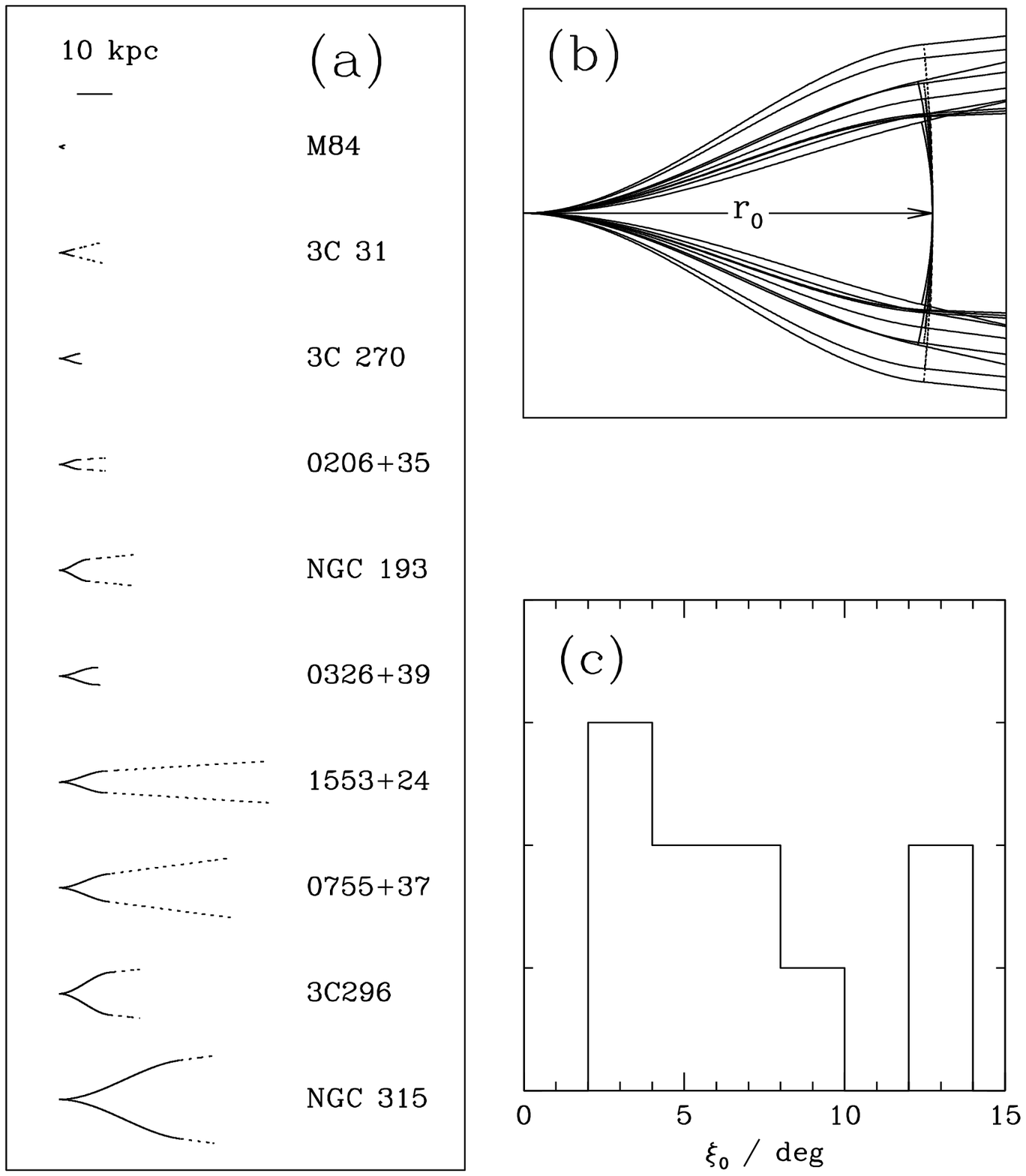}
  \caption{Plots of model jet geometry. (a) The outer boundaries of the
    model jet outflows in the plane containing the jet axis, drawn to
    the same linear scale and ordered by recollimation distance,
    $r_0$. The flaring and outer regions are plotted as full and dotted
    lines, respectively. (b) The outer boundaries (full lines), scaled
    to the same value of the recollimation distance, $r_0$.  The dashed
    curves represent the boundaries between outer and flaring
    regions. (c) Histogram of the half-opening angle of the outer
    region, $\xi_0$.
    \label{fig:geomresults}}
\end{figure}

Fig.~\ref{fig:geomresults}(a) shows the profiles of the model jet
boundaries to the same linear scale, emphasizing that the majority of
sources have recollimation distances, $r_0$, between 5 and
15\,kpc. The conspicuous outliers are M\,84 ($r_0 = 1.8$\,kpc; the
closest and least luminous of the sample members) and NGC\,315 ($r_0 =
35$\,kpc).  The shapes of the geometrically flaring regions\footnote{This geometrical form is clearly more complex than that
  of the self-similar flows of opening angle 23--24$^\circ$ described
  by \citet{DeYoung10}.} are
remarkably similar: Fig~\ref{fig:geomresults}(b) shows the outer
boundaries of the jet outflows scaled to the same value of
$r_0$. The ratio of width to length of the flaring
region, $x_0/r_0$, has a mean value of 0.29 with an rms of 0.06.  The
majority of the outer jets have half-opening angles, $\xi_0$ in the
range $2^\circ$ -- $10^\circ$, the two exceptions with $\theta >
10^\circ$ being 3C\,31 and M\,84 (Fig.~\ref{fig:geomresults}c).

\subsection{Velocity}
\label{velocity-results}

\begin{figure*}
  \epsfxsize=15cm
  \epsffile{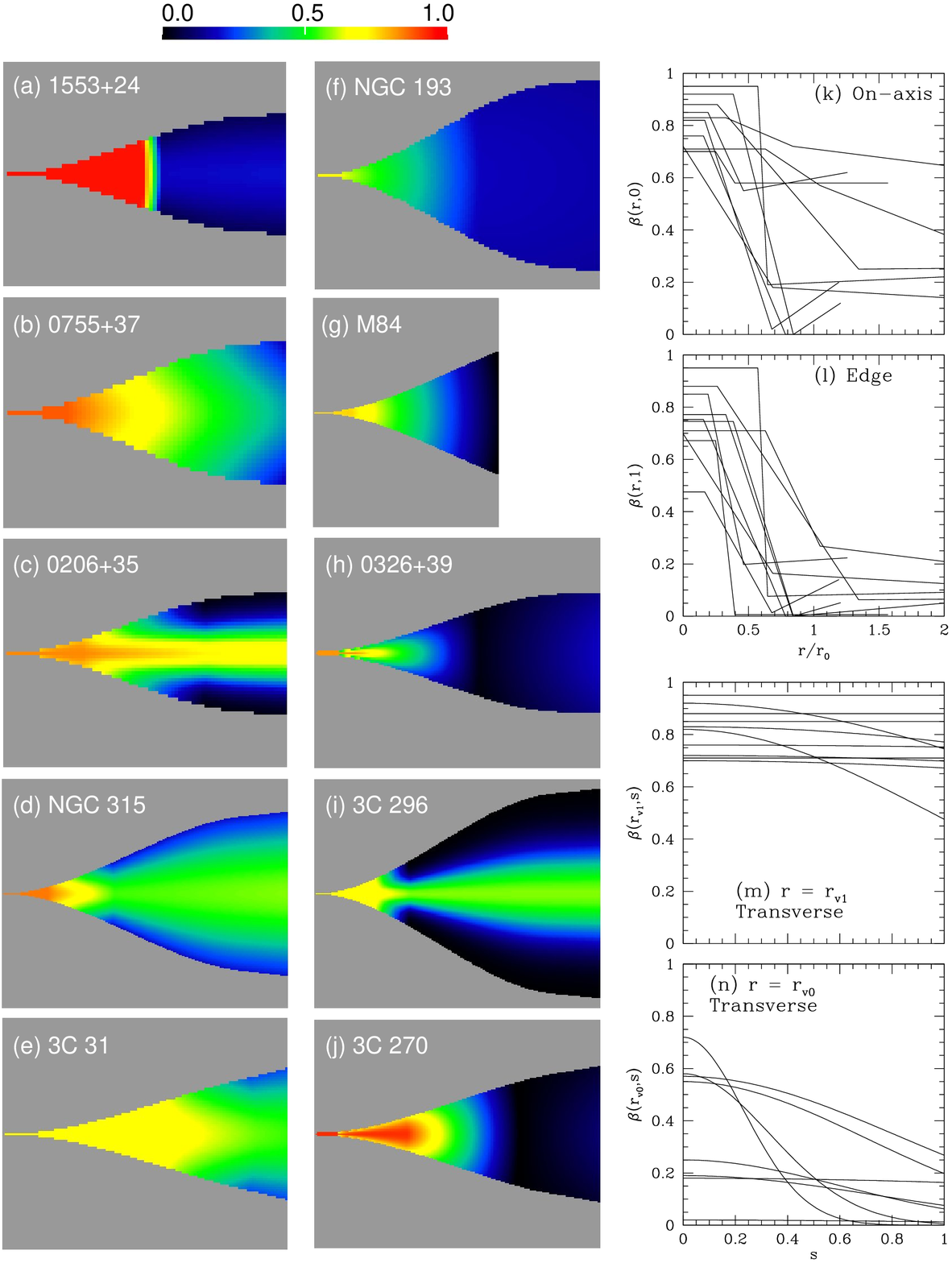}
  \caption{(a) -- (j): false-colour plots of the model velocity
    fields, in the range $0 \leq \beta \leq 1$. The maximum plot width
    is scaled to a distance of $r_0/0.85$ from the nucleus (a smaller
    region is plotted for M\,84). The backflow components of
    the models for 0755+37 and 0206+35 are not shown. (k) and (l):
    longitudinal velocity profiles for all of the modelled
    sources. The distance coordinate, $r$, is normalized by the
    recollimation distance, $r_0$, and the maximum range is $2r_0$.
    (k) on-axis; (l) edge.  (m) and (n): transverse velocity
    profiles. Velocity is plotted against streamline index $s$ at
    constant distance $r$. (m) profile at the start of deceleration,
    $\beta(r_{v1},s)$; (n) profile at the end of rapid deceleration,
    $\beta(r_{v0},s)$.  $\beta_0 = 0$ for M\,84 and 3C\,270
    and 0.02 for 0326+39. Even though the values of $v_0$ are
    essentially unconstrained for these sources
    (Table~\ref{tab:fullparams}), the errors in the transverse
    profiles are small.
    \label{fig:beta3d}}
\end{figure*}

The model velocity fields for the jet outflows are plotted in
Fig.~\ref{fig:beta3d}(a) -- (j) and longitudinal profiles on-axis and
at the edges of the jets are shown in Fig.~\ref{fig:beta3d}(k) and
(l), respectively. The on-axis velocity first becomes well determined
just downstream of the brightness flaring point, where it has a mean
value $\langle \beta_1 \rangle = 0.81$, with a rms of 0.08 (compared
with 0.06 expected from the estimated errors alone). All sources show
unambiguous evidence for deceleration.  There is usually a short
region beyond the flaring point over which the velocity field shows no
detectable variation with distance, although deceleration begins
almost immediately in NGC\,193.  Rapid deceleration occurs over a
limited range of distance: in most cases, the evidence for further
deceleration or acceleration at $r > r_{v0}$ is weak, and the velocity
is consistent with a constant value. In particular, the apparent
accelerations in 0326+39 and 1553+24 are marginally significant
\citep{CL}: minimal models with $\beta_f = \beta_0$ provide almost as
good a fit (Table~\ref{tab:chisq}) and there are indications from the
emissivity function evolution that they are physically more plausible
(Section~\ref{Adiabatic}).  In 3C\,270, the velocity is consistent
with 0 for $r > r_{v0}$ and in M\,84 it is undetermined there. Only 3C\,31
decelerates significantly after recollimation.

The sources can be divided into two groups by on-axis speed after
deceleration, $\beta_0$.  Four (3C\,31, NGC\,315, 0206+35 and 3C\,296)
have $\beta_0 > 0.5$. The remaining sources have $\beta_0 < 0.3$.

\begin{figure*}
  \epsfxsize=15cm
  \epsffile{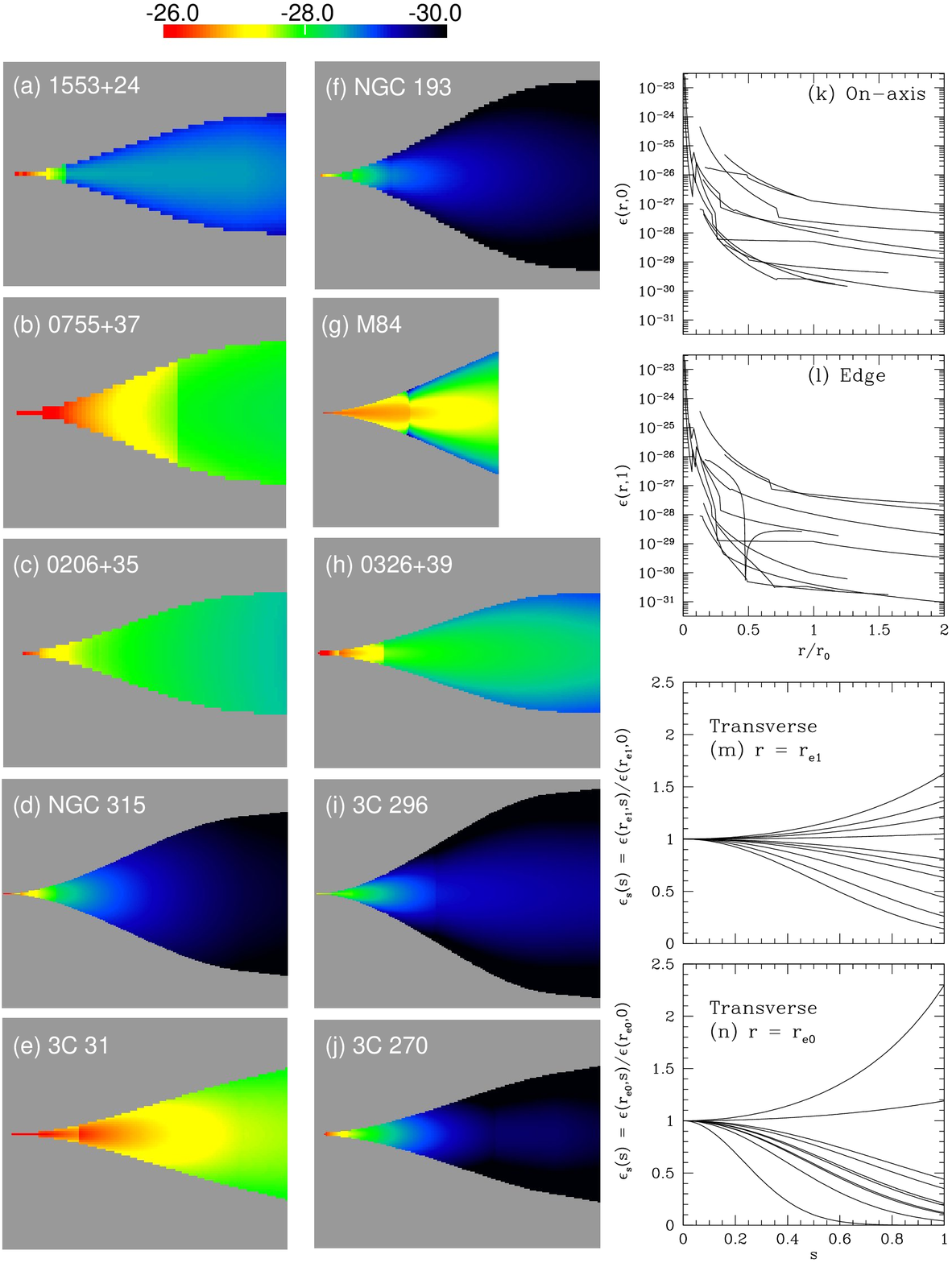}
  \caption{(a) -- (j): false-colour plots of the emissivity function
    $\log \epsilon(r,s) = \log(n_0 B^{1+\alpha})$ (with $n_0$ and $B$ in SI units).  The plotted areas are
    the same as in Fig.~\ref{fig:beta3d}(a) -- (j). (k) and (l):
    longitudinal profiles of $\epsilon$ for all of the modelled
    sources. The distance coordinate, $r$, is normalized by the
    recollimation distance, $r_0$.  (k) on-axis, $\epsilon(r,0)$; (l) edge $\epsilon(r,1)$.  The
    profiles for emission upstream of the brightness flaring point, $r
    < r_{e1}$ are only plotted if the exponent $E_{\rm in}$ is
    well determined. (m) and (n): normalized transverse emissivity function
    profiles, plotted against streamline index $s$ at
    constant distance $r$. (m) profile at the brightness flaring point,
    $\epsilon_s(s) = \epsilon(r_{e1},s)/\epsilon(r_{e1},0)$; (n) profile at the end of the high-emissivity
    region, $\epsilon_s(s) = \epsilon(r_{e0},s)/\epsilon(r_{e0},0)$.
    \label{fig:emiss}}
\end{figure*}

At or slightly before the start of rapid deceleration ($r \la
r_{v1}$), the transverse velocity variations become
well determined. Transverse profiles at $r = r_{v1}$ are plotted in
Fig.~\ref{fig:beta3d}(m).  Flat (`top-hat') profiles are consistent
with the fits for all sources except 0326+39. Profiles in
which the velocity increases slightly towards the edges of the jet are
not allowed by the fitting software (Section~\ref{vel-functions}), but
would also be consistent with the data in some cases.  We see no
evidence for any sharp velocity gradient at the jet edge, subject to
the limits set by transverse resolution.

Transverse profiles at the end of rapid deceleration, $r = r_{v0}$,
are plotted in Fig.~\ref{fig:beta3d}(n).  The normalized transverse
velocity profiles clearly evolve with distance from the nucleus in
0206+35, NGC\,315, 3C\,31 and 3C\,296 (where $\beta_0 > 0.5$).  There
is a hint of a relation between edge velocity and environment for
these four sources: the jets in 0206+35 and 3C\,296 propagate within
lobes and their edge velocities drop rapidly to values consistent with
zero whereas those in NGC\,315 and 3C\,31 ($v_0 = 0.36$ and 0.47,
respectively) appear to be in direct contact with the surrounding hot
gas.  The transverse velocity profiles for 0206+35, NGC\,315 and
3C\,296 remain well determined beyond $r = r_{v0}$ and do not evolve
significantly.

If the on-axis velocity is low, transverse variations in Doppler
factor are slight, and the velocity difference between centre and edge
is harder to measure, particularly if $\theta$ is large.  Three other
sources show evidence for transverse velocity gradients, but with
larger errors: 1553+24, 0755+37 and 0326+39.  The first two have small
on-axis velocities $\beta_0 \approx 0.2$, but low inclinations, so
evolution of the profile is still detectable. As mentioned above,
0326+39 is unusual in showing a transverse gradient at $r =
r_{v1}$. This persists over the first half of the deceleration region
(consistent with the initial value of $v_1 \approx 0.6$), after which
the velocity becomes too low to measure a gradient and $v_0$ is
unconstrained.  The velocity profile of NGC\,193 is consistent with a
constant value, but with large errors.

Finally, $v_0$ is undetermined for M\,84 and 3C\,270, which decelerate rapidly to 
speeds at which relativistic aberration is negligible.

To summarize: evolution of the transverse velocity profiles is
measured accurately in four cases, and is required in a further two.
Relative transverse velocity variations of the same form are not
excluded in any of the remaining four sources.  The unweighted mean
fractional edge velocity after deceleration is $\langle v_0\rangle =
0.35$ with the three undetermined values excluded, compared with
$\langle v_1\rangle = 0.92$ at its start.

The velocity fields are not well determined between the nucleus and
the brightness flaring point (Sections~\ref{narrow} and ~\ref{inner}).

\begin{figure*}
  \epsfxsize=15cm
  \epsffile{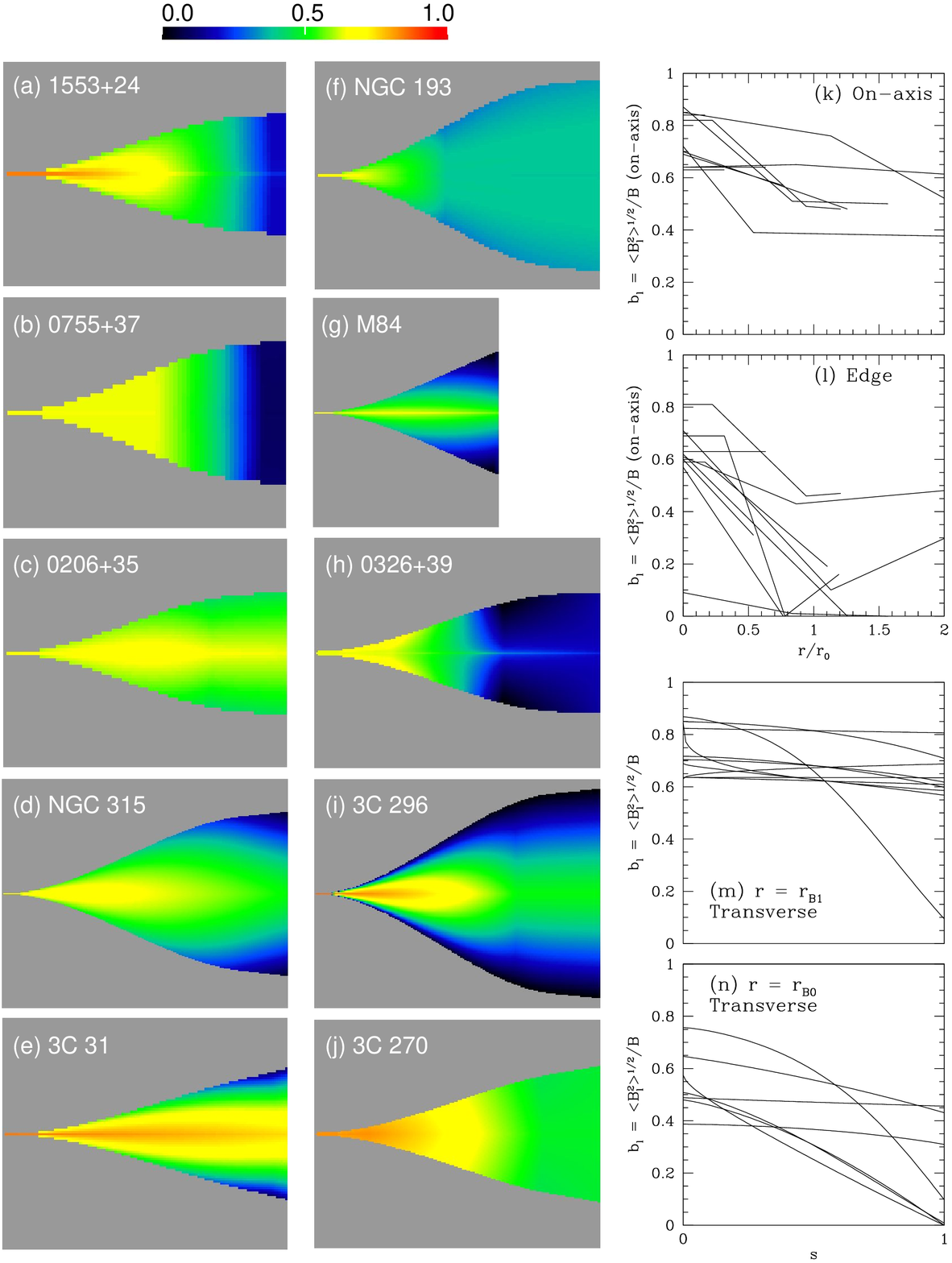}
  \caption{(a) -- (j): false-colour plots of the fractional
    longitudinal component of the magnetic field, $b_l = \langle B^2_{\rm
      l}\rangle^{1/2}/B$, in the range 0 -- 1. The plotted areas are the
    same as in Fig.~\ref{fig:beta3d}.  (k) and (l): longitudinal
    profiles of $b_l = \langle B^2_{\rm l}\rangle^{1/2}/B$. The distance
    coordinate, $r$, is normalized by the recollimation distance,
    $r_0$.  (k) on-axis; (l) edge. (m)
    and (n): transverse profiles of $b_l = \langle B^2_{\rm l}\rangle^{1/2}/B$,
    plotted against streamline index $s$ at constant distance $r$. (m) profile 
    at the inner fiducial distance, $r_{B1}$; (n) profile at the outer
    fiducial distance, $r_{B0}$.   Profiles are only plotted where
    the field component fraction has a range of $<$0.5 as deduced from
    the errors on the field ratios in Table~\ref{tab:fullparams}. 
    \label{fig:Blong}}
\end{figure*}

\begin{figure*}
  \epsfxsize=15cm
  \epsffile{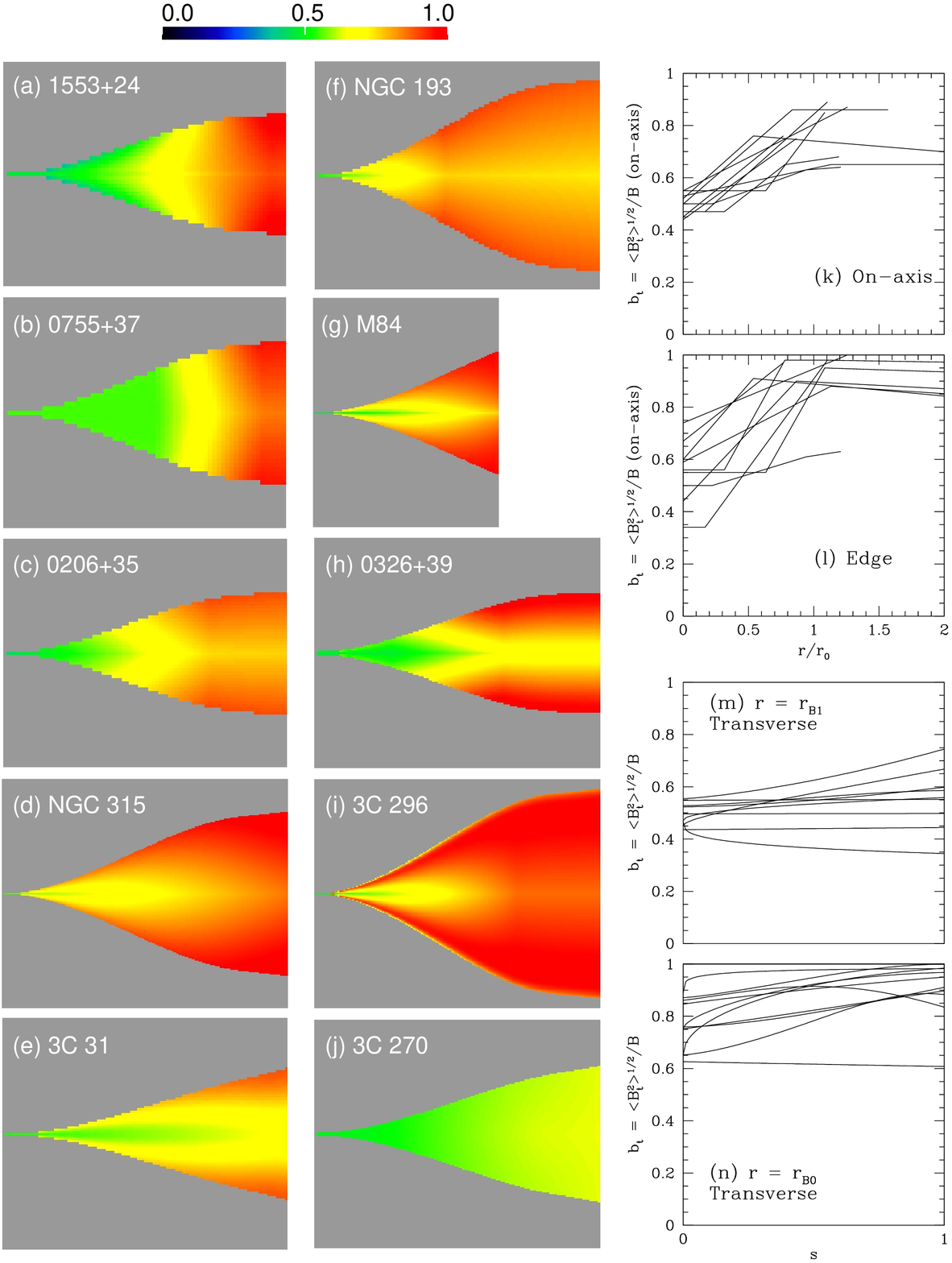}
  \caption{False-colour plots and profiles of the fractional toroidal
    component of the magnetic field, $b_t = \langle B^2_{\rm t}\rangle^{1/2}/B$. The layout is identical to that 
    in Fig.~\ref{fig:Blong}.
    \label{fig:Btor}}
\end{figure*}

\begin{figure*}
  \epsfxsize=15cm
  \epsffile{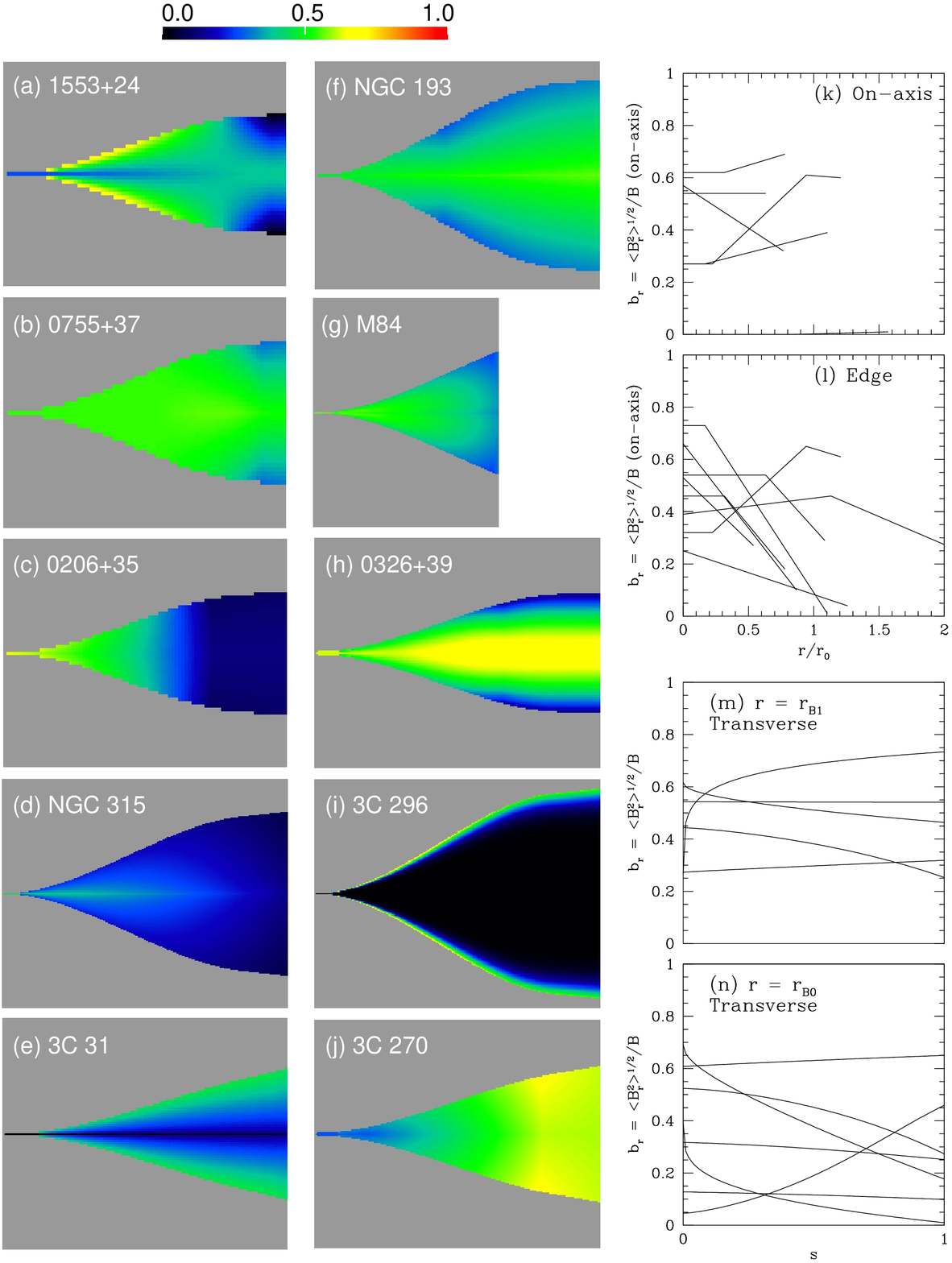}
  \caption{False-colour plots and profiles of the fractional radial component of the
    magnetic field, $b_r = \langle B^2_{\rm r}\rangle^{1/2}/B$.  The layout is identical to that 
    in Fig.~\ref{fig:Blong}. 
    \label{fig:Brad}}
\end{figure*}

\subsection{Emissivity function}
\label{emissivity-results}

Model distributions for the emissivity function $\epsilon(r,s) = n_0B^{1+\alpha}$ are plotted in Fig.~\ref{fig:emiss}(a)
-- (j) and longitudinal profiles on-axis and at the edges of the jets
are shown in Fig.~\ref{fig:emiss}(k) and (l), respectively.

The emissivity structure up to the brightness flaring point is not well
constrained (Section~\ref{narrow}). Subject to our assumption of
constant velocity at $r \leq r_{v1}$, an increase of emissivity function from
upstream to downstream of the flaring point is required by the data
for 1553+24, NGC\,315, 3C\,31, NGC\,193 and 0326+39.  In the remaining
cases, the emissivity function is consistent with being continuous across the
flaring point, but with a change of slope: there will be a marked 
increase in brightness purely as a result of the rapid
spreading of the jet in this vicinity provided that the emissivity function fall-off is not too
steep.

The end of the high-emissivity region is usually marked by one or both of a
discontinuous drop in emissivity function ($g_0 < 1$; 1553+24, 0755+37, NGC\,315, 0326+39,
3C\,296) or a significant flattening in the slope of the longitudinal emissivity function
profile ($E_{\rm out} < E_{\rm mid}$; 1553+24, 0755+37, 0206+35, 3C\,31, NGC\,193, 0326+39).

There is a general tendency for the power-law slope of the emissivity function
variation to flatten with distance from the nucleus
(Figs~\ref{fig:emiss}k and l). In three cases ($E_{\rm out}$ for
1553+24 and 3C\,270; $E_{\rm mid}$ for M\,84), this progression is
interrupted by short regions of roughly constant emissivity function.  Values
of the power-law slope after recollimation are between 0.9 and 2.2.

Figs~\ref{fig:emiss}(m) and (n) illustrate the tendency for the
transverse emissivity function profile to evolve from uniform (or perhaps even
slightly limb-brightened in some cases) to centrally peaked (0206+35
and 0755+37 remain uniform, with even a hint of a thin layer of
enhanced emission at the boundary between outflow and backflow).  The
(unweighted) mean values of the fractional edge emissivity function are $\langle
e_1 \rangle = 0.82$ at the brightness flaring point and $\langle e_0
\rangle = 0.50$ at the end of the high-emissivity region.

\subsection{Magnetic field structure}
\label{field-results}

False-colour plots of the fractional longitudinal, toroidal and radial
field components, $b_{\rm l} = \langle B^2_{\rm l}\rangle^{1/2}/B$,
$b_{\rm t} = \langle B^2_{\rm t}\rangle^{1/2}/B$ and $b_{\rm r} =
\langle B^2_{\rm r}\rangle^{1/2}/B$, are plotted in panels (a) -- (j)
of Figs.~\ref{fig:Blong} -- \ref{fig:Brad}, respectively.  Longitudinal and
transverse profiles are shown in panels (k) -- (n) of the same Figures.
The errors in the field-component ratios (particularly the
radial/toroidal ratio $j$) can be large, and there are real
differences between sources. Nevertheless, some clear trends
emerge. We quantify these using values of the fractional field
components $\langle b_{\rm l} \rangle$, $\langle b_{\rm t} \rangle$
and $\langle b_{\rm r} \rangle$ computed from the error-weighted mean
field ratios at the fiducial distances.
\begin{enumerate}
\item The largest single field component close to the AGN is longitudinal; the
  toroidal component dominates at large distances.
\item The radial component does not show any obvious systematic trends and is
  usually the weakest of the three.
\item Consequently, the images of toroidal and longitudinal field fraction are
  strikingly anticorrelated, except in a few locations where the radial field is
  significant (Figs~\ref{fig:Blong} and \ref{fig:Btor}).
\item Close to the nucleus ($r \approx r_{B1}$):
  \begin{enumerate}
  \item the longitudinal component tends to be slightly stronger than the
    toroidal component on-axis and the radial component is small:
    $\langle b_{\rm l} \rangle = 0.78$, $\langle b_{\rm t} \rangle =
    0.55$ and $\langle b_{\rm r} \rangle = 0.29$; 
  \item the field approaches isotropy at the edges: $\langle b_{\rm l}
    \rangle = 0.62$, $\langle b_{\rm t} \rangle = 0.61$ and $\langle
    b_{\rm r} \rangle = 0.50$.
 \end{enumerate}
\item At larger distances $r \approx r_{B0}$, the field configuration becomes
mostly toroidal.
  \begin{enumerate}
  \item The toroidal component is always dominant at the edge of the jet:
 $\langle b_{\rm l}
    \rangle = 0.05$, $\langle b_{\rm t} \rangle = 0.97$ and $\langle
    b_{\rm r} \rangle = 0.23$.
  \item It is also usually the largest single component on-axis,
    although the longitudinal component remains significant: $\langle
    b_{\rm l} \rangle = 0.55$, $\langle b_{\rm t} \rangle = 0.80$ and
    $\langle b_{\rm r} \rangle = 0.23$.
  \item 3C\,296 and NGC\,315 are particularly striking, in that the
  field is almost purely toroidal over most of their outer jets, with only 
  small longitudinal components on-axis.
  \end{enumerate}
\item There is little evidence for further evolution in the field
  components at larger distances $r > r_{B0}$.
\end{enumerate}

The approximate equality of longitudinal and toroidal field on-axis in
the middle of the flaring region is the key to understanding the clear
asymmetry in polarization between the main and counter-jets seen in
Figs~\ref{fig:pcomposite} and ~\ref{fig:qicomposite}.  If the radial
component is negligible, $\langle b_l \rangle
\approx \langle b_t \rangle \approx 2^{-1/2}$, so the field
forms a two-dimensional sheet with equal components in the two
directions.  This is the case described by equations~(\ref{eq:PmodB}) --
(\ref{eq:PmodBedge}).  The zero polarization point on
the axis of the main jet occurs where $\beta = \cos\theta$ in this
approximation.  For example, we would expect $p = 0$ where $\beta =
0.57$ for $\theta = 50^\circ$, typically in the deceleration region.
At the corresponding distance from the AGN in the counter-jet, the
degree of polarization would be $p \approx 2p_0\cos^2\theta/(1+\cos^4\theta) \approx
0.5$ with a transverse apparent field (equations~\ref{eq:PmodB} and
\ref{eq:ImodB}).  We also expect longitudinal apparent field with $p$
approaching $p_0$ at the edges of both jets in this model, again
consistent with the observations.

There is one special case in which the radial and toroidal components are
similar in magnitude over a significant volume: on-axis in 0326+39 at large distances
(Figs~\ref{fig:Btor}h and \ref{fig:Brad}h). This part of the jet 
resembles a two-dimensional field sheet with $\langle B^2_{t}
\rangle^{1/2} \approx \langle B^2_{r} \rangle^{1/2} \gg \langle B^2_{l}
\rangle^{1/2}$, as described by equations~(\ref{eq:P2D}) -- (\ref{eq:I2D2}).

\section{Consistency tests}
\label{consistency}

\subsection{General}

There are obvious selection effects in our choice of source: it is
hard for us to model jets which are highly projected (in which case
slight bends appear amplified) or close to the plane of the sky (so
that intrinsic or environmental asymmetries exceed relativistic
effects).  Our sources are selected from parent samples with random
distributions of inclination, but the
distribution of orientations we derive is biased in the sense that
values of $\theta$ between $\approx$30$^\circ$ and $\approx$65$^\circ$
are over-represented.  Our objectives in this section are to test
whether the distributions of orientation indicators for our sources are consistent with those of their parent samples 
 -- i.e.\ that the sample members we have {\em not}
observed are predominantly at higher and lower inclinations -- and to look
for correlations between the values of $\theta$ we derive and
independent measures.

Eight of the 10 modelled sources are drawn from two complete
samples, as follows.
\begin{description}
\item [{\em B2}] \citet{LPdRF} selected a complete sample of 38 nearby
  FR\,I radio sources with jets from the B2 catalogue.  Of these, we
  modelled four (0206+35, 0326+39, 0755+37 and 1553+24). 
\item [{\em 3CRR}] In order to define a similar sample starting
  from the 3CRR catalogue \citep{LRL}, we selected FR\,I sources with
  kpc-scale jets on at least one side of the nucleus and $z < 0.05$,
  adding NGC\,315, which meets the selection criteria on the basis of
  later flux-density measurements \citep{Mack97}. We observed and
  modelled 4 of these (3C\,31, NGC\,315, M\,84 and 3C\,296) from a
  total of 15.  3C\,449 is also a member of this sample (3C\,270
  satisfies the flux-density criterion but is outside the Declination
  range).
\end{description}
The jet inclinations for sources in these two parent samples are
expected to be isotropically distributed to a good approximation,
since the emission at the selection frequencies (178\,MHz for 3CRR and
408\,MHz for B2) should come primarily from slowly moving, extended
components such as outer jets, lobes or tails. Deviations from
isotropy caused by dependences of the total flux density and angular
size on orientation are likely to be slight (post hoc estimates based
on our jet models are given by \citealt{LPdRF} and \citealt{CL}).

We use three orientation indicators: jet sidedness (i.e.\ jet/counter-jet intensity ratio; 
 Section~\ref{sided}) fractional core flux density
(Sections~\ref{coredist} and \ref{corefractioncorrelation}), and the
ratio of Faraday rotation or depolarization (Sections~\ref{DPdist} and
\ref{RMrmsratio}).  The jet/counter-jet ratio is expected to be the
most accurate of the three orientation indicators, but is used
implicitly in our modelling and thus does not provide an independent test.
The core fraction is known to vary with time, but is not used in the
model and has a predictable dependence on angle.  The Faraday ratio is
also independent of the model, but its variations with $\theta$ are
determined by the host galaxy environment, in which there is a wide
range.  We can usefully check the distributions of all three
indicators for our modelled sources against those for the parent
samples and the correlations of core fraction and Faraday ratio with
$\theta$ for the modelled sources alone.

\subsection{Jet sidedness distribution}
\label{sided}

\begin{figure}
  \epsfxsize=5cm
  \epsffile{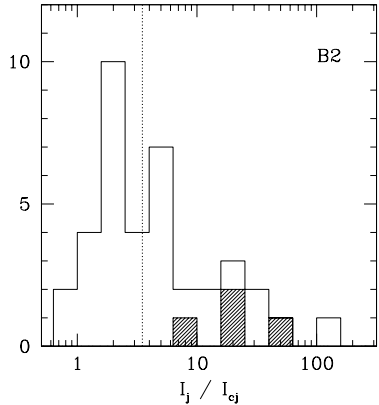}
  \caption{A histogram of the jet/counter-jet ratios at the brightness flaring
    point, $I_{\rm j}/I_{\rm cj}$, for the 38 sources in the B2 jet sample
    defined by \citet{LPdRF}. The four modelled sources from this sample
    (for which the inclination range is $25^\circ$ to $61^\circ$) are
    hatched and the median value of $I_{\rm j}/I_{\rm cj}$ is indicated
    by a dotted line.
    \label{fig:irb2}}
\end{figure}

We deliberately chose to model sources with significant brightness
asymmetries (at least 5:1 and more usually $\ga$10:1) in their
jet bases.  For a single-velocity flow with $\beta = 0.81$ (the mean
initial velocity we estimate) and $\alpha = 0.6$ emitting
isotropically in the rest frame, $I_{\rm j}/I_{\rm cj} \geq 5$
corresponds to $\theta \leq 68^\circ$ (equation~\ref{eq:iratio}), in
adequate agreement with our inferred inclination range of $25^\circ
\leq \theta \leq 76^\circ$.

Next, we ask whether the ratios for the modelled sources are consistent with
their membership of an isotropic parent sample.  A homogeneous
set of measurements of the jet/counter-jet ratio at the brightness flaring point
is available for the B2 jet sample \citep{LPdRF} and their distribution is
shown in Fig.~\ref{fig:irb2}.  All of the modelled sources in this
sample have ratios above the median, consistent with their derived
inclination range of $25^\circ - 61^\circ$.

\subsection{Core fraction distribution}
\label{coredist}

A second, widely-used, orientation indicator is the ratio $f$ of radio core to extended 
flux density (or luminosity) at fixed emitted frequency.  The core emission is
partially optically thick and comes from the bases of the  jets
\citep{BK79}. A simple model in which there is a constant intrinsic
ratio of core to extended flux density (or luminosity) and the
parsec-scale emission comes from a pair of antiparallel
jets\footnote{It may be that the receding jet also suffers free-free
  absorption, in which case the second terms in both the numerator and
  denominator will be reduced. We do not analyse
  this case here.} with velocity $\beta_{\rm c}$ and spectral index
$\alpha_{\rm c}$ predicts
\begin{eqnarray}
  f &=& f_0\frac{(1-\beta_{\rm c}\cos\theta)^{-(2+\alpha_{\rm c})} + (1+\beta_{\rm c}\cos\theta)^{-(2+\alpha_{\rm c})}}
  {(1-\beta_{\rm c}/2)^{-(2+\alpha_{\rm c})} + (1+\beta_{\rm c}/2)^{-(2+\alpha_{\rm c})}} \label{eq:corefraction} \\ \nonumber
\end{eqnarray}
again assuming isotropic emission in the rest frame. $f_0$ is the core
fraction at $\theta = 60\degr$ (the median value for an isotropic
sample).

One potential complication is that the relation between core and
extended luminosity is non-linear \citep{Giov88,DeR90}.  For this
reason, \citet{LPdRF} defined an alternative orientation indicator,
the normalized core power, $P_{\rm cn}$. This is the ratio of $f$ to
its median value at given extended luminosity.  Given that the sample
used to establish the slope of the median relation has a much larger
luminosity range than we consider here, is dominated by types of
source other than twin jets and includes powerful FR\,II sources, it
is not clear whether this normalization is valid for our sample. We
therefore prefer to use $f$ rather than $P_{\rm cn}$. The range of
extended luminosity for the sources in this paper is a factor
of $\approx$40, with the majority having $\log (P_{\rm ext}/{\rm
  WHz}^{-1})$ close to the median value of 24.3
(Table~\ref{tab:images}), so the normalization will not, in any case,
affect our results significantly.

\begin{figure}
  \epsfxsize=6.5cm
  \epsffile{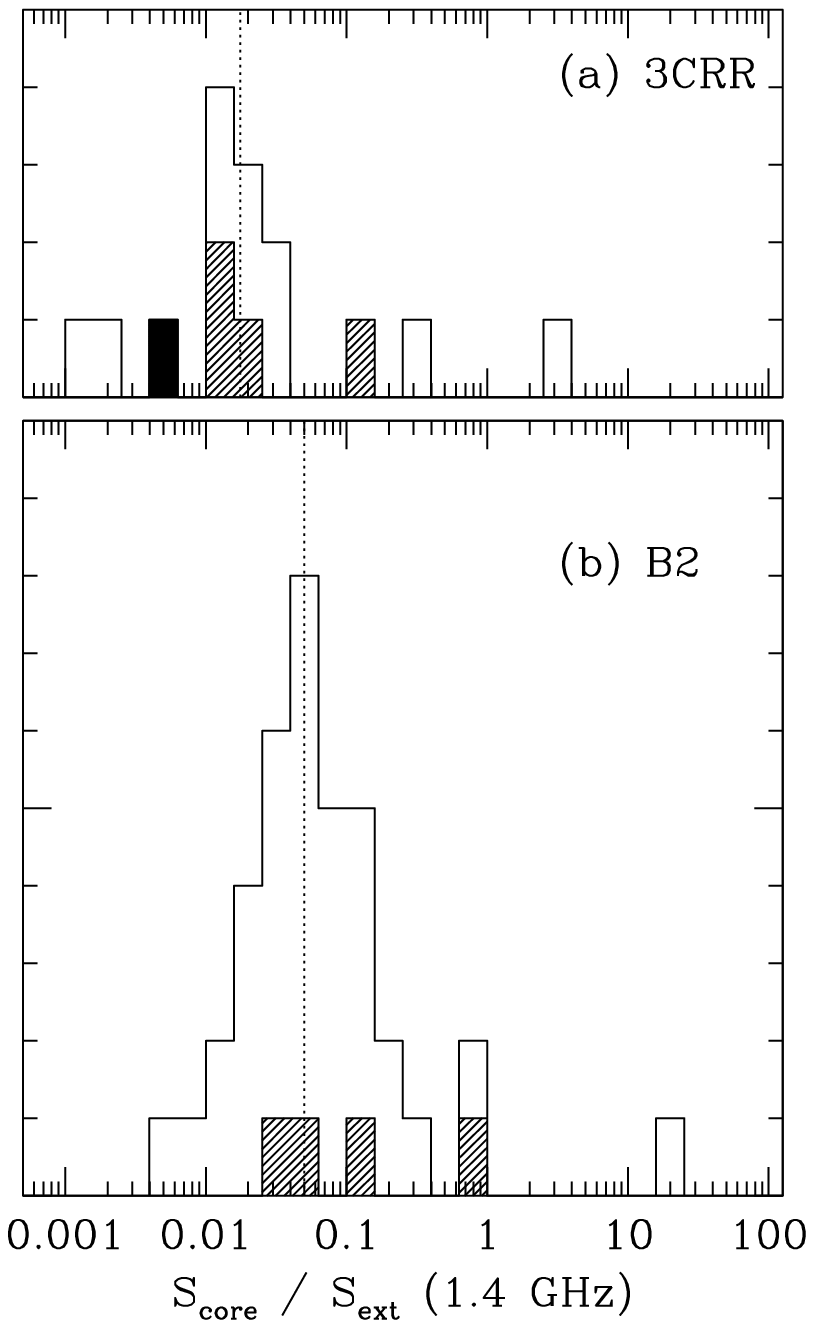}
  \caption{(a) and (b): histograms of core/extended flux-density ratio,
    $f$, for the B2 and 3CRR samples. The modelled sources are hatched
    and the dotted line indicates the median ratio for the sample. (a)
    3CRR sample \citep{LRL}. The four modelled sources have estimated
    inclinations between $50^\circ$ and $65^\circ$. 3C\,449, for which
    we estimate $\theta \approx 90^\circ$, is shaded black. (b) B2
    jet sample \citep{LPdRF}. The inclination range for the modelled sources
    is $25^\circ$ to $61^\circ$.
    \label{fig:fc}}
\end{figure}

In Fig.~\ref{fig:fc}, we show the distributions of the core fraction
$f$ at 1.4\,GHz emitted 
frequency\footnote{This frequency was chosen to minimize the effects
  of core variability.} for the 3CRR and B2 jet samples, with the modelled
sources and 3C\,449 indicated.  For
the modelled sources, the inclination ranges are $50^\circ < \theta <
65^\circ$ (3CRR) and $25^\circ < \theta < 61^\circ$ (B2); we expect
$\theta \approx 90^\circ$ for 3C\,449. We therefore predict core
fractions from just below to significantly above the median for the
modelled sources and close to the lower end of the distribution for
3C\,449.  The observed and predicted distributions are reasonably
consistent, especially considering the possibility of dispersion in the
intrinsic core fraction.

\begin{figure}
  \epsfxsize=6.5cm
  \epsffile{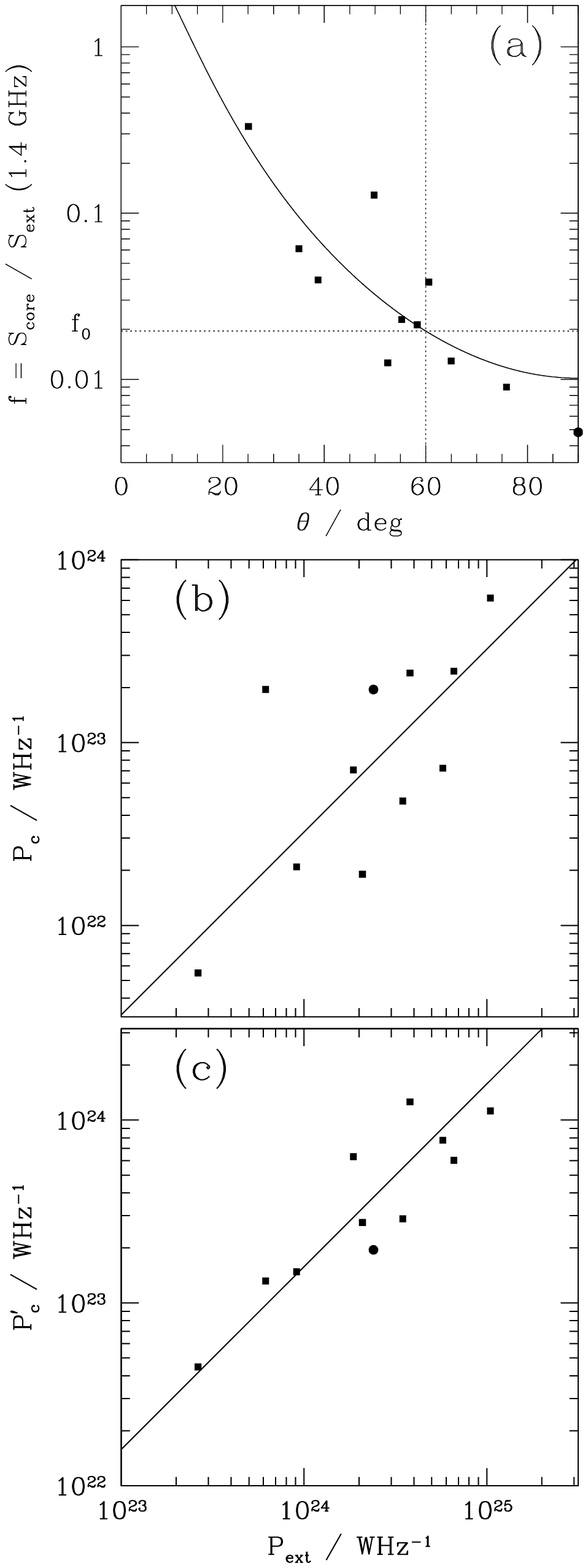}
  \caption{(a) A plot of the ratio, $f$, of core to extended flux density at
    1.4\,GHz rest frequency against $\theta$ for the sources in
    our sample.  The full curve is the expected relation if the core
    emission comes from a pair of antiparallel jets with a spectral
    index of $-0.2$ (the mean for our sample), the best-fitting
    bulk speed, $\beta_{\rm c} = 0.98$, and $f_0 = 0.0195$
    (equation~\ref{eq:corefraction}). The dotted lines represent the median
    values of $\theta$ ($= 60^\circ$) and core flux density ratio,
    $f_0$, for an isotropic sample. (b) A plot of core luminosity,
    $P_{\rm c}$ against extended luminosity $P_{\rm ext}$, both at an
    emitted frequency of 1.4\,GHz. The best-fitting linear relation is
    shown. (c) As (b), but with the core luminosity $P^\prime_{\rm c}$
    corrected for beaming using equation~(\ref{eq:debeam}). In all three
    panels, data for the modelled sources are plotted as filled
    squares.  The filled circles represent 3C\,449. This source, which
    we take to have $\theta = 90^\circ$, was not included in any of
    the fits.
    \label{fig:tfc}}
\end{figure}

\subsection{Correlation of core fraction with inclination}
\label{corefractioncorrelation}

We plot the relation between inclination and core fraction at an
emitted frequency of 1.4\,GHz in Fig.~\ref{fig:tfc}(a). There is a
clear anticorrelation (significant at the 99.8\% level according to
the Spearman rank test).

The simple model of equation~(\ref{eq:corefraction}) with $\alpha_{\rm
  c} = -0.2$ (the median for the sample) gives a reasonable fit to the
relation for any value of core velocity $\beta_{\rm c} \ga
0.94$. Fig.~\ref{fig:tfc}(a) shows an example for the best fit,
$\beta_{\rm c} = 0.98$ ($\Gamma_{\rm c} = 4.8$).  The rms scatter in
$\log f_0$ is 0.26 for this speed.  For comparison, \citet{LPdRF}
derived $\beta_{\rm c} = 0.91 \pm 0.05$ from a similar analysis of the
relation between core fraction and the jet/counter-jet intensity ratio
at the brightness flaring point for the full B2 jet sample, but with a
larger scatter of 0.45 in $\log f_0$.  3C\,449 has a lower value of
$f$ than any of the modelled sources, consistent with the expected
large angle to the line of sight (we plot it with $\theta = 90^\circ$
in Fig.~\ref{fig:tfc}, but did not use it in the fit).

It is of interest to see how much the scatter in the relation between
core and extended luminosity is reduced by fitting out the dependence
on inclination in this way.  Fig.~\ref{fig:tfc}(b) shows a plot of
core luminosity, $P_{\rm c}$, against extended luminosity, $P_{\rm
  ext}$.  We have corrected $P_{\rm c}$ to luminosity in the rest
frame of the emitting material using the same assumptions as in
equation~(\ref{eq:corefraction}). The resulting quantity, $P^\prime_{\rm
  c}$ is given by
\begin{eqnarray}
P^\prime_{\rm c}  &=& \frac{2\Gamma_{\rm c}^{2+\alpha_{\rm c}}P_{\rm c}}{(1-\beta_{\rm c}\cos\theta)^{-(2+\alpha_{\rm c})} + (1+\beta_{\rm c}\cos\theta)^{-(2+\alpha_{\rm c})}}
\label{eq:debeam} \\ \nonumber
\end{eqnarray}
and is plotted against $P_{\rm ext}$ in
Fig.~\ref{fig:tfc}(c)\footnote{$P^\prime_{\rm c} > P_{\rm
    c}$ for $\theta > 29^\circ$ with this choice of parameters, so
  only 1553+24 has a smaller core luminosity in the rest frame
  compared with the observed frame.}.  The relations between core and
extended luminosity both before and after correction for Doppler
boosting are consistent with our
assumption of constant intrinsic ratio. The correction reduces the rms
dispersion about the best-fitting linear relation from 0.43 for $\log
P_{\rm c}$ to 0.20 for $\log P^\prime_{\rm c}$.  The best fit for the
core luminosity in the rest frame is $P^\prime_{\rm c} = 0.16 P_{\rm
  ext}$ for a frequency of 1.4\,GHz.  The implication for the type of
source we model is that the rest-frame emission produced on parsec
scales (which is known to vary on time-scales of years) is
surprisingly well correlated with emission extending in some cases to
enormous distances and which is presumably built up over the entire
source lifetime.

\subsection{Depolarization ratio distribution}
\label{DPdist}

The lobe containing the approaching jet will be seen through less
magnetoionic material associated with the host galaxy and will
therefore show lower fluctuations in foreground Faraday rotation than
the receding lobe \citep{L88}. The degree of polarization integrated
over the approaching lobe therefore decreases less rapidly with
increasing wavelength in the approaching lobe.  We define the average
depolarization between two frequencies DP = $<p(\nu_{\rm
  low})/p(\nu_{\rm high})>$.

\begin{figure}
  \epsfxsize=5cm
  \epsffile{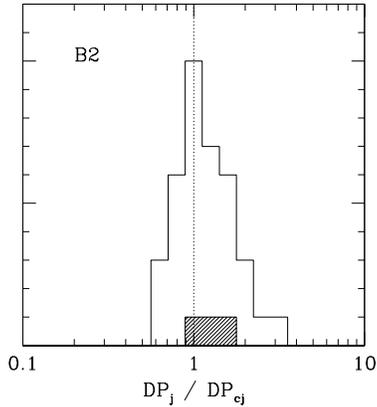}
  \caption{Histogram of depolarization ratio, DP$_{\rm j}$/DP$_{\rm
      cj}$ for the sample defined by \citet{Morganti97}.  DP is the
    ratio of scalar mean degrees of polarization at 1.4 and 4.9\,GHz,
    so a smaller value of DP corresponds to heavier Faraday
    depolarization. DP$_{\rm j}$ and DP$_{\rm cj}$ refer to the lobes
    containing the brighter and fainter jets, respectively.  The three
    sources in common with this study, 0206+35, 0755+37 and
    1553+24, are shown hatched. The vertical dotted line indicates DP$_{\rm j} =$ DP$_{\rm cj}$.\label{fig:dphist}}
\end{figure}

Measurements of the ratios of the mean scalar degrees of polarization at
frequencies of 4.9 and 1.4\,GHz for the lobes of 37
sources from the B2 sample were presented by \citet{Morganti97}.  They confirmed the
strong tendency for the lobe containing the brighter jet to be less
depolarized, and showed that this is due primarily to sources with
one-sided jet bases (or, almost equivalently, bright cores).
Fig.~\ref{fig:dphist} shows a histogram of depolarization ratio from
\citet{Morganti97}. The three sources in common with this study
(0206+35, 0755+37 and 1553+24) are indicated.  They have DP$_{\rm
  j}$/DP$_{\rm cj} \ga 1$, as expected.

\subsection{Faraday rotation asymmetry}
\label{RMrmsratio}

A more direct measure of Faraday rotation fluctuations is the rms
dispersion in RM across a lobe, $\sigma_{\rm RM}$,
determined at high spatial resolution. We have published high-quality
RM images for eight out of 11 of the sources discussed in this
paper. In addition, we made a two-frequency RM image for
NGC\,193 from observations at 4.9 and 1.365\,GHz \citep{lobes}.  For
nine sources, we could therefore derive $\sigma_{\rm RM}$ across the
main and counter-jet lobes with good sampling at high resolution.
$\sigma_{\rm RM}$ is a more sensitive measure of foreground Faraday
rotation than depolarization and allows us to probe much smaller
Faraday depths.  We evaluated it over all unblanked pixels, making a
first-order correction for fitting error to avoid positive bias. The
image resolutions and values of $\sigma_{\rm RM}$ are given in
Table~\ref{tab:RM}, along with references to the observations and data
reduction.

\begin{figure}
  \epsfxsize=6cm
  \epsffile{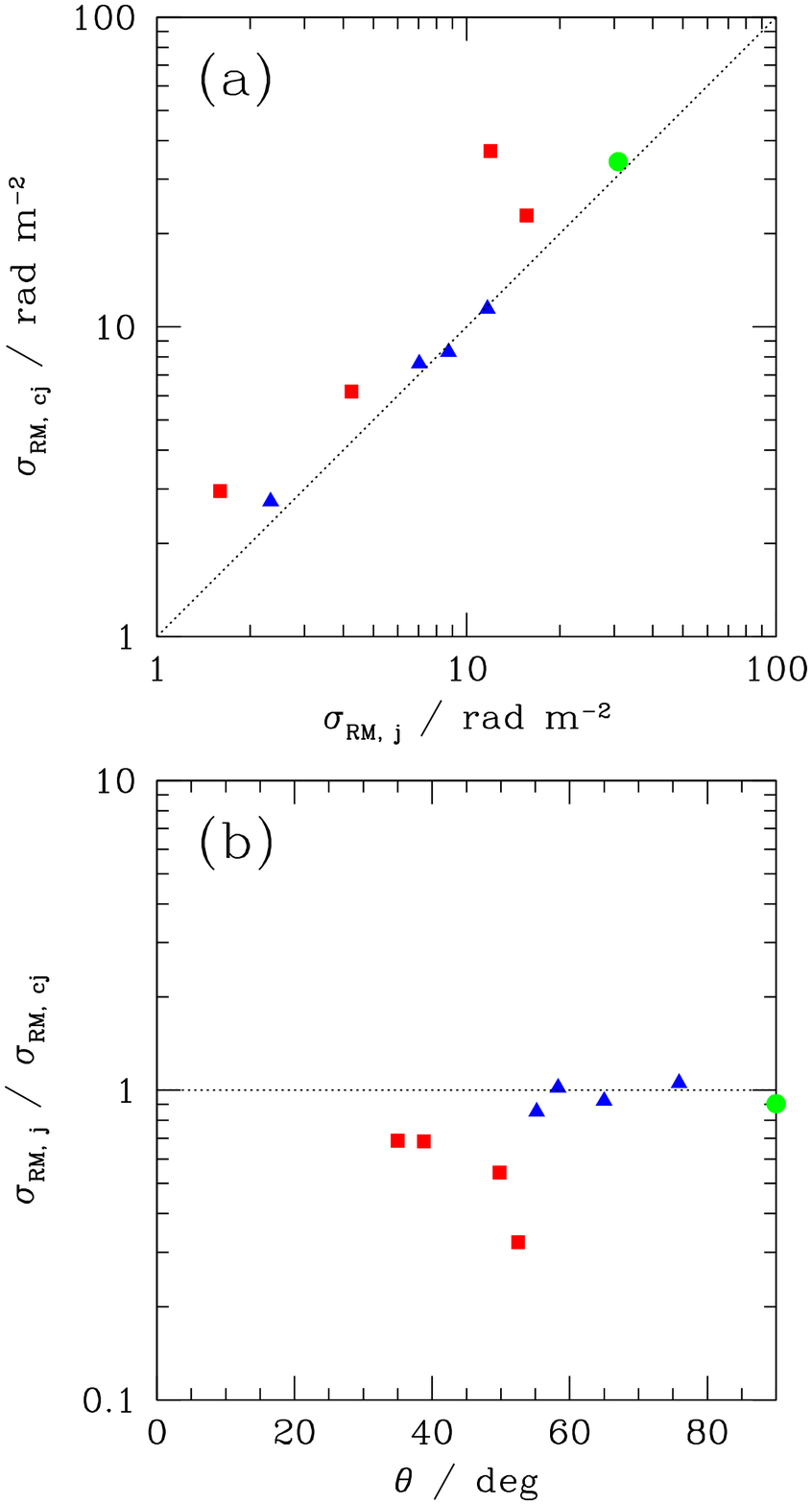}
  \caption{(a) A plot of the rms rotation measure for the 
    counter-jet lobe, $\sigma_{\rm RM, cj}$, against that for
    the main jet lobe, $\sigma_{\rm RM, j}$. (b) The ratio $\sigma_{\rm RM,
      j}/\sigma_{\rm RM, cj}$ plotted against $\theta$. In both panels,
    the points are coded by angle to the line of sight. Red squares: $\theta <
    55^\circ$; blue triangles: $55^\circ < \theta < 90^\circ$; green circles: $\theta
    \approx 90^\circ$ (3C\,449). The dotted lines represent $\sigma_{\rm
      RM, j} = \sigma_{\rm RM, cj}$.
    \label{fig:rmplot}}
\end{figure}

In Fig.~\ref{fig:rmplot}(a), we plot $\sigma_{\rm RM}$ for the main
and counter-jets against each other and in Fig.~\ref{fig:rmplot}(b),
we plot their ratio against inclination. There is a significant
asymmetry, in the sense that $\sigma_{\rm RM, j} < \sigma_{\rm RM,
  cj}$, for $\theta \la 55^\circ$ and the ratio is very close to unity
for larger angles to the line of sight. There are no examples where
$\sigma_{\rm RM, j}$ is significantly larger than $\sigma_{\rm RM,
  cj}$.  The significance of the correlation between $\sigma_{\rm RM,
  j}/\sigma_{\rm RM, cj}$ and $\theta$ is 97\% according to the
Spearman rank test.

This result, and the earlier measurements of depolarization asymmetry
for the B2 sample \citep{Morganti97}, are qualitatively consistent
with a simple picture in which the variations of Faraday depth across
the brightness distributions are produced by roughly spherical
distributions of ionized gas containing fluctuating magnetic fields.
Profiles of $\sigma_{\rm RM}$ for spherically-symmetric model gas
density profiles and power-law dependences of field strength on
density indeed show that significant asymmetries can be produced,
particularly for $\theta \la 50^\circ$
(e.g.\ \citealt{GC91,3c31RM}). We note a number of complications,
however.
\begin{enumerate}
\item The expansion of radio sources into the surrounding hot gas is
  expected to cause local increases in density and field strength,
  particularly if the expansion is supersonic \citep{HuarteE}; shells
  of denser gas are indeed observed around the lobes of M\,84
  \citep{Finoguenov}.
\item The present sample includes three examples of highly ordered RM
  distributions which must be affected by interactions between the
  sources and their local environments \citep[0206+35, M\,84 and
    3C\,270;][]{Guidetti11}.
\item Even for sources with chaotic RM distributions which might
  plausibly originate from undisturbed plasma, it is  necessary to take account
  of that fact that the relativistic particles evacuate cavities in the
  surrounding hot gas, causing deviations from spherical symmetry 
  (e.g.\ \citealt{3c31RM}).
\item There is a wide variation in measured external density profile 
  and in the size of the radio structure compared with the core
  radius of the surrounding hot gas.
\end{enumerate}
Nevertheless, our results are fully consistent with the idea that the
Faraday rotation is produced by distributed, local foreground
plasma\footnote{No asymmetry would be expected if the Faraday-rotating
  material is in a very thin shell or mixing layer around the radio
  lobes.}.  A difference between sources at $\theta < 55^\circ$ (which
show significant side-to-side differences) and those with $\theta >
55^\circ$ (which do not) is apparent from Fig.~\ref{fig:rmplot}. Such
a discontinuity could be produced by the type of cavity model developed
by \citet{3c31RM}, but observations of a larger sample would be needed
to establish the robustness of the result.

\begin{center}
  \begin{table}
    \caption{Rotation measure rms, $\sigma_{\rm RM}$, for the lobes
      associated with the main (approaching) and counter (receding)
      jets. (1) Source name; (2) angle to the line of sight, $\theta$,
      in deg; (3) resolution (FWHM, in arcsec); (4) rms rotation
      measure for the main jet lobe, in rad\,m$^{-2}$; (5) as (4), but
      for the counter-jet lobe; (6) reference.\label{tab:RM}}
    \begin{minipage}{80mm}
      \begin{tabular}{lllrrr}
        \hline
        &&&&&\\
        Source & $\theta$ & FWHM   & $\sigma_{\rm RM, j}$ & $\sigma_{\rm RM, cj}$ & Reference \\
        & (deg)      & (arcsec) & \multicolumn{2}{c}{(rad\,m$^{-2}$)}& \\
        &&&&&\\
        \hline
        &&&&&\\
        0755+37  &35.0 &1.3 &  4.3  &  6.2  & 3 \\
        0206+35  &38.8 &1.2 & 15.6  & 22.9  & 2 \\
        NGC\,315   &49.8 &5.5 &  1.6  &  3.0  & 5 \\
        3C\,31    &52.5 &1.5 & 12.0  & 37.0  & 6 \\
        NGC\,193   &55.2 &1.6 &  2.3  &  2.7  & 7 \\
        M\,84    &58.3 &1.65& 11.7  & 11.5  & 2 \\
        3C\,296   &65.0 &1.5 &  7.0  &  7.6  & 4 \\
        3C\,270   &75.9 &1.65&  8.8  &  8.3  & 8 \\
        3C\,449   &90.0 &1.25& 30.9  & 34.2  & 1 \\
        &&&&&\\
        \hline
      \end{tabular}

      References: 
      (1) \citet{Guidetti10};
      (2) \citet{Guidetti11};
      (3) \citet{Guidetti12};
      (4) \citet{LCBH06};
      (5) \citet{ngc315ls};
      (6) \citet{3c31RM};
      (7) \citet{lobes};
      (8) Laing et al. (in preparation).
    \end{minipage}
  \end{table}
\end{center}

\section{Intrinsic asymmetries}
\label{intrinsic}

We have shown that symmetrical, relativistic jet models can fit the
observed brightness and polarization distributions very well, and that
their use yields similar values for many of the physical parameters in all
cases that we have studied.  We are confident that relativistic aberration 
dominates close to the AGN and that our derived physical parameters are most
reliable there.  It is obvious, however, that the bending and
asymmetric morphologies of FR\,I jets on much larger scales are inconsistent with the
hypothesis of continuing symmetrical flow on those scales and that 
environmental effects eventually dominate.  We have
therefore restricted our modelling to the inner jet regions
(specifically, where bends in the jets are slight) and have ignored
very bent sources completely.  The jets are unlikely to be perfectly
symmetrical even where they form and our criteria for
selecting the regions to model are inevitably somewhat subjective.

For these reasons, we now attempt to quantify the effects of intrinsic
side-to-side differences on our results. It is difficult to be
definitive without a physical model for the deviations from intrinsic
symmetry (which could in principle affect any combination of geometry,
emissivity function, velocity or field structure in complicated ways). We have
therefore chosen to analyse three representative examples, in which the
deviations are only in one of velocity, rest-frame emissivity function or field
ordering.

We constructed a base model with representative
parameters, including: $\beta_1 = 0.8$, $v_1 = 1.0$, $\beta_0 = 0.32$
and $v_0 = 0.5$ (equivalent to an emissivity-weighted average of
$\beta = 0.24$ at $r \geq r_{v0}$), with $r_{v1}$ and $r_{v0}$ corresponding to
1.8 and 5.0\,kpc on a model grid of 10\,kpc (all in projection). We
took four representative angles to the line of sight, $\theta =
30^\circ$, $45^\circ$, $60^\circ$ and $75^\circ$.

\subsection{Velocity}

We modified the base model by multiplying all of the velocities on one
side of the nucleus by a constant factor, made model images and then
fit to them using our standard procedures. As expected, a velocity
asymmetry of this type is fit primarily by changes in a combination of
the on-axis velocities $\beta_1$ and $\beta_0$, the fractional edge
velocities $v_1$ and $v_0$, and the angle to the line of sight,
$\theta$. The fitted value of $\theta$ is biased in the obvious sense:
it is underestimated if the approaching jet is faster and
overestimated if it is slower.  For a 20\% difference in velocity, the
maximum error in $\theta$ ranges from 1\fdg6 at $\theta = 30^\circ$ to
2\fdg4 at $\theta = 70^\circ$.  The fitted velocities $\beta_1$ and
$\beta_0$ typically lie mid-way between the mean of the new main and
counter-jet velocities and the values for the base model, so the error
on the true mean velocity is about 5\% for a 20\% asymmetry, as is the
error on $v_0$ and $v_1$.

We conclude that the fits are robust to asymmetries in velocity, with
errors typically at the 5\% level for a 20\% asymmetry, comparable
with our estimated errors.

\subsection{Emissivity function}
\label{asymm-em}

This example is perhaps the most interesting, because we can constrain
the intrinsic emissivity function ratio between the two jets from the
statistics of reversals in sidedness in a sample of sources. Suppose
that all jets have a constant intrinsic sidedness ratio $R_{\rm int}$,
but that they have identical velocity fields. Provided that the
velocity is sufficiently high near the nucleus, relativistic effects
will dominate there except for jets which are very close to the plane
of the sky, so we will identify the near side correctly in almost all
cases. Farther from the nucleus, where the jets have decelerated, we
will observe reversals in the observed sidedness if the approaching
jet is the intrinsically fainter one and the angle to the line of
sight is sufficiently large.  Suppose that the jets decelerate from
$\beta = \beta_1$ to $\beta = \beta_0$. Then we will observe reversals for angles in the range
$\theta_1 > \theta > \theta_0$, where
\begin{eqnarray}
  R_{\rm int} & = & \left [ \frac{1 + \beta_1\cos\theta_1}{1 - \beta_1\cos\theta_1} \right ]^{2+\alpha}\nonumber \\
  & = &\left [ \frac{1 + \beta_0\cos\theta_0}{1 - \beta_0\cos\theta_0} \right ]^{2+\alpha}\label{eq:reverse}  \\ \nonumber
\end{eqnarray}
assuming isotropic emission in the rest frame
(equation~\ref{eq:iratio}).  The probability of observing reversals in
an isotropic sample is then
\begin{eqnarray}
  f_{\rm rev} & = & (\cos\theta_0 - \cos\theta_1)/2 \\ \nonumber
\end{eqnarray}
and the corresponding intrinsic sidedness ratio is 
\begin{eqnarray}
  R_{\rm int} & = & \left [\frac{1/\beta_0 - 1/\beta_1+2f_{\rm rev}}{1/\beta_0 - 1/\beta_1-2f_{\rm rev}} \right]^{2+\alpha}. \label{eq:emconst}\\\nonumber
\end{eqnarray}
If $\theta > \theta_1$ and the receding jet is intrinsically brighter
(probability $\cos\theta_1/2$), then it will appear brighter at all
distances from the nucleus and might be identified as the approaching
jet. The jet/counter-jet sidedness ratio will appear to increase with
distance from the nucleus in this case, however.

For an alternative model in which the jets are intrinsically
symmetrical at the base but develop an intrinsic asymmetry after
deceleration, the ratio is
\begin{eqnarray}
  R_{\rm int} & = & \left [\frac{1/\beta_0+2f_{\rm rev}}{1/\beta_0-2f_{\rm rev}} \right]^{2+\alpha}.\label{eq:emramp}\\\nonumber
\end{eqnarray}

For the B2 jet sample, we have measurements of sidedness ratio at a
projected distance of 14.3\,kpc\footnote{$H_0$ =
  70\,$\rm{km\,s^{-1}\,Mpc^{-1}}$; the original reference used a
  different Hubble constant.} from the nucleus for 25 sources, of
which 2 show reversals in sidedness compared to the brightness flaring
point \citep[Fig.~6c]{LPdRF}, so $f_{\rm rev} = 0.08$. The mean ratios
at 14.3\,kpc (averaged over all sources and also subdivided by
fractional core flux density) are consistent with $\beta_0 \approx
0.24$ for isotropic emission: this is an emissivity-weighted average
across the jets, and therefore corresponds to a somewhat higher
on-axis velocity, as in our base model.  For an initial velocity
$\beta_1 = 0.8$ and a model with a constant intrinsic asymmetry, we
find $R_{\rm int} \approx 1.3$ (equation~\ref{eq:emconst}); if the
jets are initially symmetrical, then $R_{\rm int} \approx 1.2$
(equation~\ref{eq:emramp}).  We use the latter model, since we have
found no cases of sidedness ratio increasing with distance in the B2
jet sample (although only $\approx 1/25$ would in any case be expected).
The sample size is small and the selection criteria for the B2 jet sample
include a wider variety of source types than we consider here, so our
estimate is very approximate. Nevertheless, it does indicate that
intrinsic emissivity function variations are fairly small on the typical scales
we model.

In order to test the effects of such an asymmetry on our derived
parameters, we started with the symmetrical base model and multiplied
the emissivities of one of the jets by a factor increasing linearly
from 1 at the nucleus to 1.2 at $r = r_0$ and thereafter remaining
constant.  The principal systematic errors are in the angle to the
line of sight, $\theta$ and the velocity variables after deceleration,
$\beta_0$ and $v_0$. These are in the obvious sense that sources with
the intrinsically brighter jet on the near side are fit as being
closer to the line of sight and/or faster, the deviations increasing
with $\theta$. If the intrinsically brighter jet is on the far side,
the effects are in the opposite sense. Errors in $\theta$ range from
$\approx$1\degr at $\theta = 30\degr$ to $\approx$4\degr at $\theta =
75\degr$; in $\beta_0$ from 0.04 to 0.09 and in $v_0$ from 0.04
to 0.08.  With the assumed form of variation in asymmetry, the errors
in quantities measured at the flaring point are negligible.  There are
also errors at the 10\% level in the edge emissivities and
radial/toroidal field ratios.

Emissivity function variations at the level we have simulated will limit our
ability to model jets with $\theta \ga 75\degr$, but are comparable with 
or less than other sources of error for lower inclinations.

\begin{figure*}
\epsfxsize=15cm \epsffile{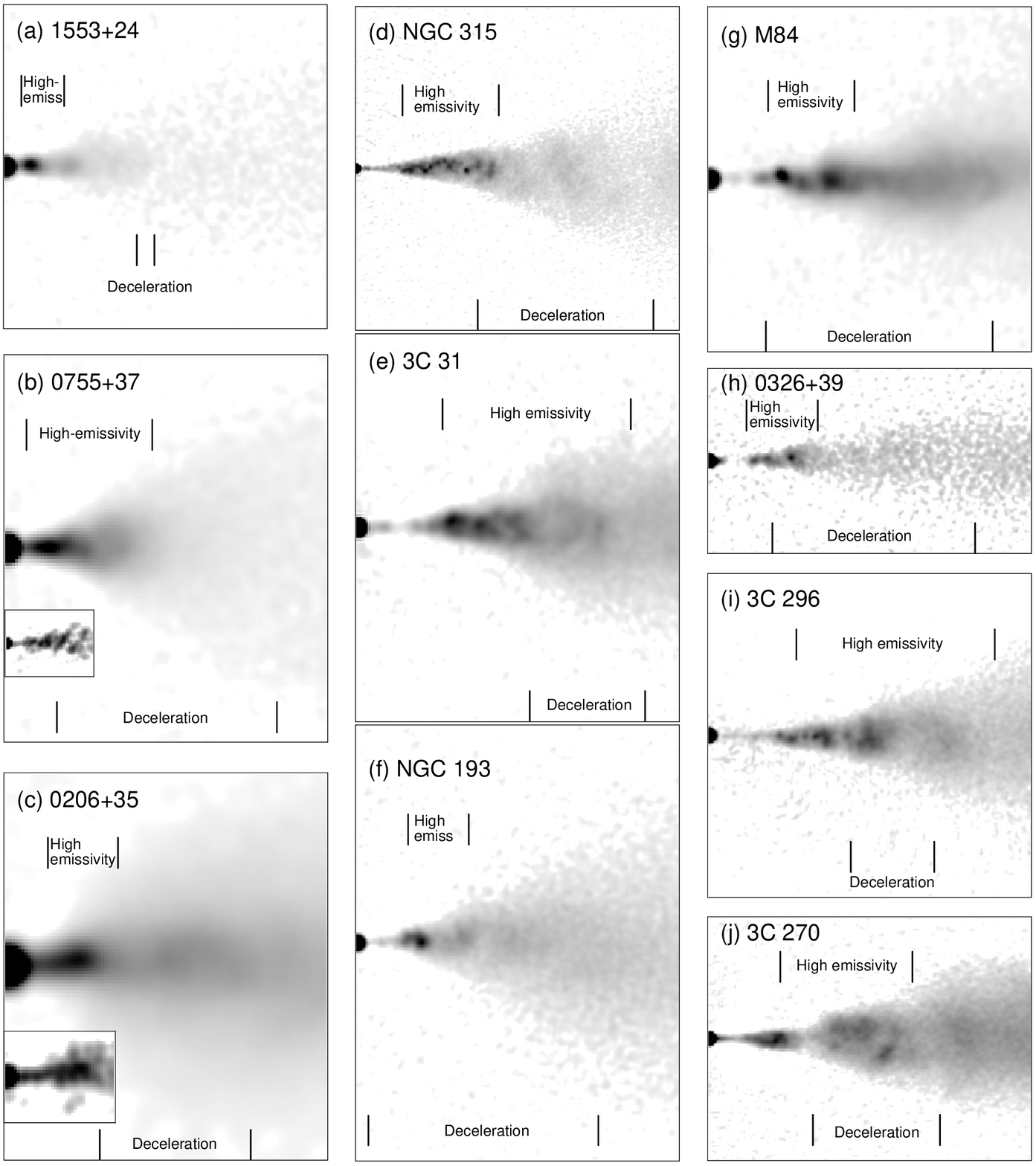}
\caption{Grey-scale plots of total intensity for the main jet bases,
  showing the locations of the high-emissivity and rapid deceleration
  zones. The main panels are VLA images at the highest available
  resolution (Table~\ref{tab:images}); the inserts in panels (b) and
  (c) are MERLIN images \citep{lobes}, plotted on the same scale.
\label{fig:ihires}}
\end{figure*}

\subsection{Field ordering}

Finally, we checked the effect of multiplying either the
radial/toroidal or longitudinal/toroidal field ratio on one side of
the source by a constant factor of 1.5.  There was no systematic
effect on either the derived inclination or the initial velocity
parameters $\beta_1$ and $v_1$.  The outer velocity parameters were
affected at the 5 -- 10\% level, but without much systematic dependence on
inclination.  The fitted values of the field ratios themselves were
close to the means of the values for the two jets.  The fits are
therefore fairly robust against asymmetries in field ratio.

\section{Discussion}
\label{discuss}

\subsection{The flaring region - a homologous structure}
\label{flarepoint}

Our models are constrained
mainly by properties we have measured in the well-resolved regions
downstream from the brightness flaring points, which evidently mark
an important transition in these FR\,I radio jets.  We now discuss
in more detail the evolution of these jets between their brightness flaring points
and recollimation.

Just downstream of the flaring point, there is an extended region over which
the emissivity remains high and the jet decelerates significantly
while undergoing geometrical flaring.  The relation between the
high-emissivity and deceleration zones is illustrated for the
individual sources in Fig.~\ref{fig:ihires}, where the fiducial
distances are marked on images of their main jets at the highest
available resolution(s).  The high-emissivity
regions identified by the model-fitting process are also obvious by
eye at high resolution. They all contain complex, non-axisymmetric,
high-brightness structures, the best resolved of these being the
quasi-helical filament in NGC\,315 \citep[Fig.~\ref{fig:ihires}d;][]{ngc315hires}.  The position
of the brightening in the main jet estimated by eye sometimes appears
slightly inconsistent with the brightness flaring point found by our
model. This is usually because the fit locates a very bright knot
slightly downstream of the true flaring point (e.g.\ 3C\,296 and
3C\,270; Figs~\ref{fig:ihires}i and j), but the differences are within
the errors quoted in Tables~\ref{tab:fullparams} and
\ref{tab:minparams}.

\begin{table*}
\caption{Correlations of fiducial distances in the geometrical flaring
  region.  (1) Fiducial distance; (2) symbol; (3) scaling with
  recollimation distance $r_0$; (4) significance level for correlation
  with $r_0$; (5) significance level for correlation with extended
  luminosity at 1.4\,GHz; (6) as (5), but for core luminosity, $P_{\rm
    c}$; (7) as (5), but for deboosted core luminosity $P^\prime_{\rm
    c}$.  The significance levels (in \%) are determined using the
  Spearman rank test and values in parentheses are with M\,84
  excluded.\label{tab:flare}}
\begin{tabular}{lllrrrr}
\hline
&&&&&&\\
\multicolumn{2}{c}{Quantity}& Scale &\multicolumn{4}{c}{Significance}\\
&&&versus $r_0$&versus $P_{\rm ext}$&versus $P_{\rm c}$&versus $P^\prime_{\rm c}$\\
&&&&&&\\
\hline
&&&&&&\\
Recollimation distance        &$r_0   $&      &         &80.0(45.4) &97.1(88.8)&98.1(92.3)\\
Flaring point distance        &$r_{e1}$& 0.095& 97.5    &88.3(64.4) &57.5( 3.4)&98.1(92.3)\\
End of high-emissivity region &$r_{e0}$& 0.32 & 96.7    &93.3(77.6) &82.6(51.2)&98.1(92.3)\\
Start of rapid deceleration   &$r_{v1}$& 0.23 & 95.7    &67.2(26.8) &91.8(77.6)&77.1(45.4)\\
End of rapid deceleration     &$r_{v0}$& 0.59 & $>$99.9 &72.4(30.0) &98.9(95.0)&95.2(83.0)\\
Start of magnetic evolution   &$r_{B1}$& 0.09 & 70.5    &29.2( 5.3) &60.7(37.8)&47.0(17.6)\\
End of magnetic evolution     &$r_{B0}$& 1.13 & $>$99.9 &80.0(45.4) &98.1(92.3)&97.1(88.8)\\
&&&&&&\\
\hline
\end{tabular}
\end{table*}

\begin{figure}
\epsfxsize=8.25cm
\epsffile{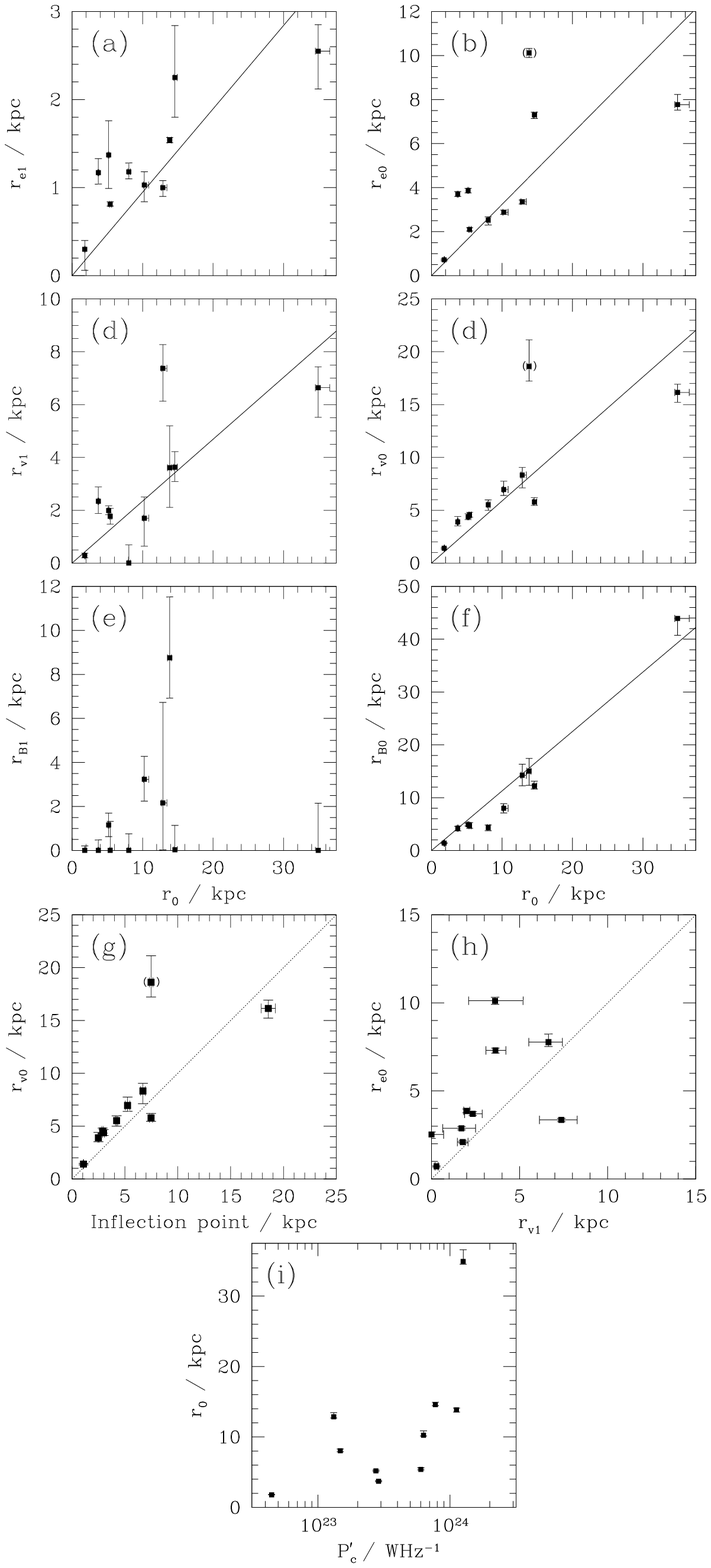}
\caption{(a) -- (f): plots of the fiducial distances against the
  recollimation distance, $r_0$.  (a) start of high emissivity region
  (brightness flaring point), $r_{e1}$; (b) end of high-emissivity
  region, $r_{e0}$; (c) start of deceleration, $r_{v1}$; (d) end of
  deceleration, $r_{v0}$; (e) start of magnetic evolution, $r_{B1}$;
  (f) end of magnetic evolution, $r_{B0}$.  The full lines plotted in all panels except (e)
  represent the linear fits from Table~\ref{tab:flare}.  (g) A plot of
  the position of the end of rapid deceleration, $r_{v0}$, against
  that of the point of inflection in the outer boundary of the flaring
  region (i.e.\ the distance at which the opening angle is a maximum).
  (h) a plot of the distance of the end of the high-emissivity region,
  $r_{e0}$, against that of the start of deceleration, $r_{v1}$. The
  line of equality is shown dotted in panels (g) and (h). The points
  for 0755+37 are bracketed on panels (b), (d) and (g), to emphasize
  that the fiducial distances $r_{v0}$ and $r_{e0}$ are anomalously
  high for this source. (i) a plot of recollimation distance, $r_0$,
  against deboosted core power, $P_{\rm c}^\prime$ (equation~\ref{eq:debeam}).
\label{fig:fidx0}}
\end{figure}
Figs~\ref{fig:fidx0}(a) -- (f) show plots of the positions of the
fiducial distances for velocity, emissivity function and field ordering against
the recollimation distance $r_0$. All of the fiducial distances except
$r_{B1}$ are correlated with $r_0$: the best-fitting linear relations
and the significance levels for the correlations, calculated using the
Spearman rank test, are given in Table~\ref{tab:flare}.  The main
points are as follows.
\begin{enumerate}
\item The distances of the start and end of the high-emissivity region,
  $r_{e1}$ and $r_{e0}$, and the start of deceleration, $r_{v1}$, are
  all well correlated with $r_0$ ($>$95\% significance;
  Figs~\ref{fig:fidx0}a -- c, Table~\ref{tab:flare}).
\item The end of rapid deceleration, $r_{v0}$, is even better correlated
  with $r_0$ ($>$99.9\% significance; Fig.~\ref{fig:fidx0}d): 9/10
  sources fall on a roughly linear relation (0755+37 is the
  conspicuous outlier).
\item The end of deceleration is also extremely close to the distance
  at which the jet has its maximum opening angle [the point of
  inflection at $r = -a_2(1)/3a_3(1)$ in the curve of
  equation~(\ref{eq:flare}) that defines the outer edge of the flaring
  region]. In other words, the jet starts to recollimate precisely
  where it stops decelerating. The relation between the inflection
  distance (close to $0.55r_0$) and $r_{v0}$ is plotted in Fig.~\ref{fig:fidx0}(g). 
  0755+37 is again the outlier from the linear relation.
\item The start of deceleration ($r_{v1} \approx 0.23r_0$) occurs in
  the middle of the high-emissivity region ($\approx 0.095r_0$ to $\approx 
  0.32r_0$). $r_{e0}$ is plotted against $r_{v1}$ in
  Fig.~\ref{fig:fidx0}(h): 9/10 sources have $r_{v1} \la r_{e0}$, the
  exception being 1553+24.
\item $r_{B1}$ is consistent with zero in the majority of cases and shows
  no correlation with $r_0$ (Fig.~\ref{fig:fidx0}e).  The region of
  rapid evolution in field structure therefore starts very close to
  the AGN, at the base of the geometrical flaring region.
\item In contrast, $r_{B0}$ is very well correlated with $r_0$, with a
  nearly linear relation $r_{B0} \approx 1.13 r_0$
  (Fig.~\ref{fig:fidx0}f and Table~\ref{tab:flare}).  Magnetic
  evolution therefore slows just after the transition between the
  flaring and outer regions, where  recollimation is complete.
\item The implicit correlation between $r_{e1}$ and $r_{v0}$ is
  equivalent to that found between jet-side gap length 
  and symmetrization distance for the B2 jet sample by
  \citet{LPdRF}, as these quantities are essentially the projections
  on the plane of the sky of $r_{e1}$ and $r_{v0}$, respectively.
\end{enumerate}

We have also checked for correlations between the fiducial distances
and the luminosities of the extended emission and the core (with and
without beaming corrections; all three luminosities are 
correlated, as demonstrated in Fig.~\ref{fig:tfc}).  The significance
levels given by the Spearman rank test are listed in
Table~\ref{tab:flare}.  M\,84 (by far the least luminous source) has a disproportionate influence on the
correlations, so we also tabulate the significance levels with it
excluded.  There is some evidence for positive correlations with luminosity
for all of the fiducial distances except $r_{B1}$.
The most significant correlations are with the deboosted
core luminosity $P^\prime_{\rm c}$ (equation~\ref{eq:debeam}). With
M\,84 included, the significance levels exceed 95\% for all of the
fiducial distances except $r_{B1}$ and $r_{v1}$.  Thus, although there
is considerable scatter, there is a general tendency
for the characteristic scales in the flaring region to increase with
luminosity.  As an example, we show a plot of the recollimation
distance $r_0$ against $P^\prime_{\rm c}$ (Fig.~\ref{fig:fidx0}i); the
remaining correlations are implicit.

To a first approximation, then, the flaring region is a {\em
  homologous} structure, in the sense that all of its characteristic
sizes -- its width and the fiducial distances for velocity and
emissivity function -- scale with its length, which in turn is weakly
correlated with luminosity.
 
\subsection{Jet deceleration}
\label{decel}

Where we first have good constraints on the jet velocity, just
downstream of the brightness flaring point, the transverse velocity
profiles are consistent with constant values and remarkably similar
between sources (Fig.~\ref{fig:beta3d}m). Whatever processes affect
the jet speed before that point cannot, therefore, lead to large {\it
  systematic} velocity gradients across the entire jet width.

The start of deceleration at $r = r_{v1}$ is accompanied by evolution
of the transverse velocity profile from flat to centrally peaked in at
least six and potentially all 10 of our sources.  This is prima facie
evidence that deceleration in the flaring region is dominated by
interaction with the environment and entrainment of surrounding
material.  Mass loss from stars within the jet volume
\citep*{Phi83,Kom94,BLK96} is an additional source of mass loading,
which may exceed boundary-layer entrainment for $r \la r_{v1}$ 
\citep{LB02b}, but which will not cause evolution of the transverse profile.

  Many authors
have discussed the development of surface instabilities and the
transition to fully-developed turbulence in astrophysical jets \citep[and references therein]{DeYoung10}, often
by analogy with non-relativistic, fluid flows observed in the
laboratory (e.g. \citealt{Dimotakis83, Dimotakis05}).  There are as
yet no models making testable predictions for the entrainment rate and
velocity evolution in the relativistic case, so we instead outline a
qualitative picture based on these general ideas, informed by the
results of our modelling, conservation-law analyses \citep{Bick94,LB02b}
and numerical simulations of light, relativistic jets
\citep{PM07,Rossi08}.

We first note that a flow that decelerates from relativistic to
sub-relativistic speeds by entraining external material must  be internally transonic
\citep{Bick94}.  The proper Mach number of the flow is
\begin{equation}
{\mathcal M} = \Gamma\beta/(\Gamma_{\rm s}\beta_{\rm s})
\end{equation}
where $\beta_{\rm s}c$ is the sound speed and $\Gamma_{\rm s} =
(1-\beta_{\rm s}^2)^{-1/2}$ \citep{Konigl80}.  $\beta_{\rm s}$ depends
on the jet composition, with an upper limit of $\beta_{\rm s} =
3^{-1/2} = 0.58$ for an ultrarelativistic fluid.  At the start of
deceleration, the mean velocity $\langle \beta_1 \rangle = 0.81$
corresponds to ${\mathcal M} = 2$ in this limit.  For any jet that
decelerates from such relativistic speeds by entrainment, conservation
of mass, momentum and energy alone imply that the Mach number drops to
${\mathcal M} \approx 1$ where $\beta \approx 0.3$, at the point where
the inertia of the jet becomes thermally dominated \citep[fig.\,2 of][]{Bick94}.
 
For 3C\,31, we argued from a similar conservation-law analysis that
the jet is mildly supersonic throughout the region that we model.  The
internal sound speed for the reference model of \citet{LB02b} is close
to the ultrarelativistic limit at the brightness flaring point, but
decreases with distance as the jet entrains external material. The
resulting Mach number on-axis at the end of the high-emissivity region
is ${\mathcal M} \approx 1.7$ for a velocity $\beta(r_{e0}) = 0.59$.
Other sources have similar velocities ($\langle \beta(r_{e0})\rangle =
0.61 \pm 0.07$ for the full sample) and are likely to have similar
Mach numbers.

Our results suggest the following conjecture for jet deceleration.
\begin{enumerate}
\item The jet has a mildly supersonic spine, with an initial speed $\beta
  \approx 0.8$ at the brightness flaring point. The equation of state
  is close to the ultrarelativistic limit, so ${\mathcal M} \approx
  2$.
\item An internally subsonic shear layer starts to penetrate the jet
  at or shortly downstream of the brightness flaring point. At a distance
  $r \approx r_{v1}$, two effects occur:
  \begin{enumerate}
  \item the shear layer reaches a significant fraction of the jet
    width, so we start to resolve a transverse velocity gradient and
  \item the spine flow also starts to decelerate.
  \end{enumerate}
\item The flow in the shear layer is turbulent, leading to
  isotropization of the magnetic field close to the edge of the jet
  (Section~\ref{field-results}).
\item The high-emissivity region corresponds to the portion of the
  flow in which the Mach number exceeds some critical value ${\mathcal
    M}_{\rm crit}$ in the range $1.5 \la {\mathcal M}_{\rm crit} \la 1.8$.
\item The end of rapid deceleration, at $r = r_{v0}$ occurs when the
  entrainment rate either:
  \begin{enumerate}
  \item drops to a negligible value, at which point
  the jet reaches an asymptotic velocity and the transverse profile
  `freezes out', or at least 
  \item decreases significantly, so the evolution
  of the velocity profile with distance is less rapid.
  \end{enumerate}
\item We can identify three distinct cases (see Fig.~\ref{fig:beta3d}):
  \begin{enumerate}
  \item the on-axis flow remains fast ($\beta \ga 0.5$) 
    throughout the modelled region, and the shear layer
    does not reach the axis before entrainment stops (0206+35, NGC\,315 and 3C\,296); 
  \item the on-axis flow is 
    still fast after deceleration, but continuing entrainment causes
    the shear layer to develop further after recollimation (3C\,31) and
  \item the shear layer expands to fill the entire volume of the jet
    before deceleration ceases, so the flow becomes slow, with $\beta
    \la 0.25$ (1553+24, 0755+37, NGC\,193, M\,84, 0326+39, 3C\,270).
  \end{enumerate}
\end{enumerate}
The obvious reason for the end of rapid deceleration is that the jet
is no longer propagating in a dense external environment, so the
entrainment rate drops abruptly.  Where this might happen depends on
the source morphology: seven of our sources (0326+39, M\,84, 3C\,296,
0206+35, 0755+37, NGC\,193 and 3C\,270) have well defined lobes which
appear to surround the jets: there is direct evidence for cavities in
the intergalactic medium (IGM) formed by the expansion of the lobes in
M\,84, 3C\,296 and 3C\,270
\citep{Cros08,Finoguenov,osullivan11}. Although projection could
mislead us about the spatial relationships between the jets and the
lobes in individual cases, this is unlikely in general.  It seems more
plausible that the jets are shielded from the surrounding IGM by
cocoons of relatively light (primarily relativistic) plasma in most,
if not all, of these sources. Within a few kpc of the nucleus, where
the IGM pressure is high, this is probably not the case: in all
sources with {\sl Chandra} observations, we see small, dense coronae
which do not appear to have been displaced by the radio jets or lobes
\citep{WBH01,H05,Finoguenov,osullivan11,Kharb12}.  

In 3C\,31, NGC\,315 and 1553+24, there are no lobes and the jets
instead appear to propagate in direct contact with the surrounding hot
plasma.  For the first two sources, external density profiles covering
our full modelled regions have been deduced from X-ray observations
\citep{H02,KB,Cros08}.  In 3C\,31, the external proton number
density decreases from 14000 to 3000\,m$^{-3}$ from the end of rapid
deceleration, $r_{v0}$ to the end of the grid. For NGC\,315, the
corresponding densities are 2500 and 500\,m$^{-3}$.  Thus the thermal
plasma around the jets in 3C\,31, which continue to decelerate for $r
> r_{v0}$, is roughly six times denser than the equivalent around
NGC\,315, whose jets maintain their speeds.  Note that both sources
also have dense coronae \citep{H02,WBH03,ngc315hires}.

In all sources except 3C\,31, it therefore seems likely that most of
the mass is ingested from these coronae, and that subsequent
entrainment into the jets (at least within the limited regions we
model) is slight.  As pointed out by \citet{DeYoung93}, this is
essential to produce the observed morphologies: deceleration to
transonic speeds should occur early in the flow, but further
deceleration must be minimal in order to prevent the outflow ceasing
completely.

Recollimation of the flow to form the conically spreading outer region
does not generate any structures which can unambiguously be identified with
recollimation shocks \citep{Sanders}: the brightness distributions are mostly quite
smooth at these distances (Fig.~\ref{fig:icomposite}). This is
consistent with the idea that the majority of the jets have become
subsonic across their full widths before recollimation.  The
most plausible candidate we have found for a recollimation shock in
any of the sources is the `on-axis enhancement' in the
brighter jet of NGC\,315 \citep[their Fig.4a]{ngc315ls}.  It may be
significant that the flow speed after recollimation remains high over
a substantial cross-section of the jets in this source\footnote{We also see
jet-crossing brightness steps or narrow features (`arcs') in 3C\,31
and 3C\,296, but most
of these are located well downstream of recollimation
\citep{LCBH06,3c31ls}.}.  

\subsection{Evolution of the magnetic field}
\label{Bevolution}

Our results show that the field evolution in FR\,I jets is, to
first order, from longitudinal to toroidal, occasionally with a
significant radial component.  The longitudinal component is indeed
expected to fall much more rapidly with distance than the two
transverse components in an expanding flow \citep{Burch79}. For a relativistic jet in the
quasi-one-dimensional approximation (neglecting variations across the
jet, such as velocity shear), the field evolution expected from
flux-freezing in a laminar flow is:
\begin{eqnarray}
B_l &\propto& x^{-2}              \label{eq:bl}\\
B_t,B_r &\propto& (x\beta\Gamma)^{-1} \label{eq:btr}\\ \nonumber
\end{eqnarray}
where $x$ is the jet radius \citep{Baum97}.

\begin{figure}
\epsfxsize=8.5cm
\epsffile{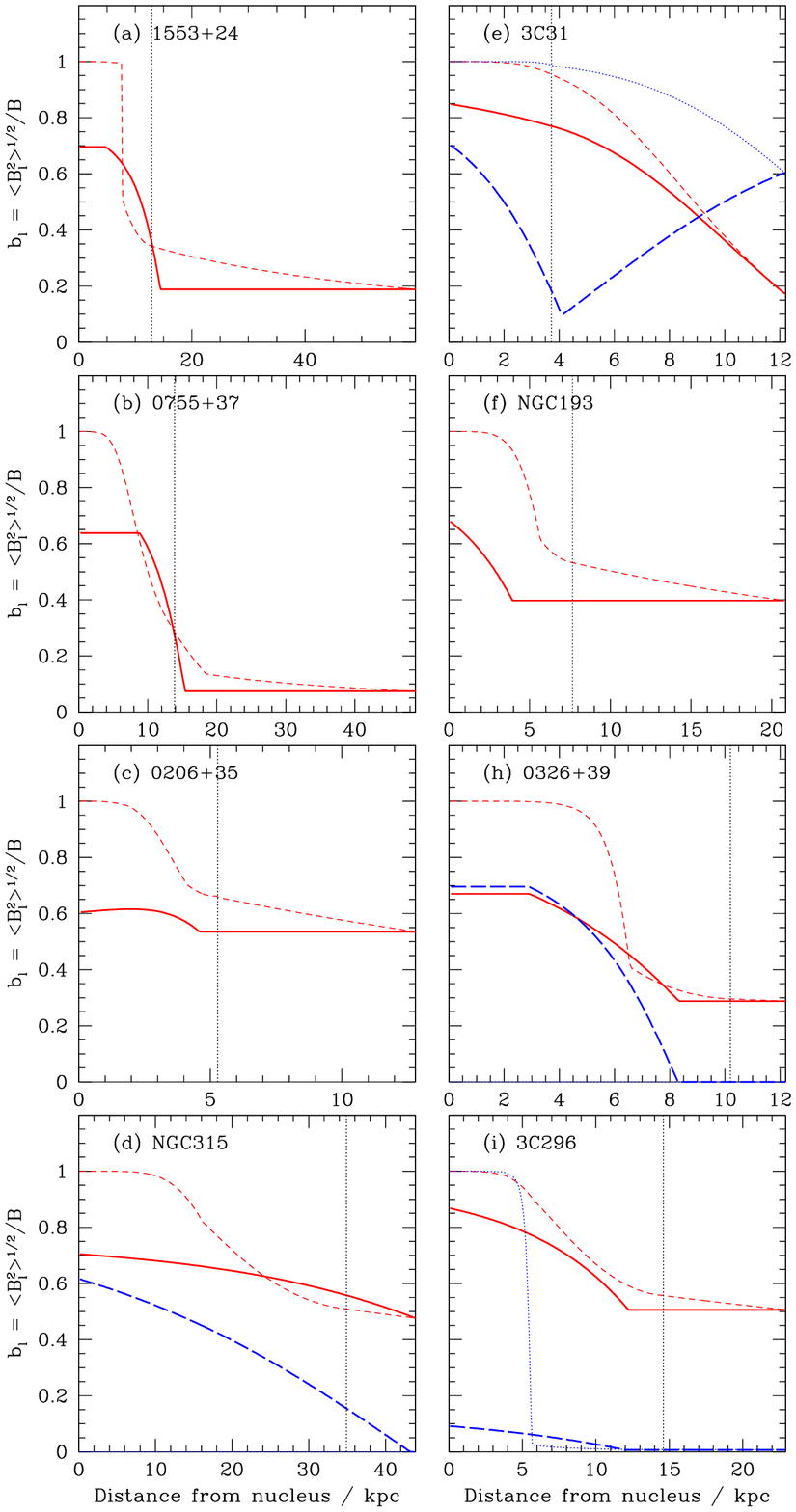}
\caption{Comparison of fitted profiles of fractional longitudinal
  field, $b_l = \langle B_{\rm l}^2 \rangle^{1/2}/B$, with the predictions
  of the simple flux-freezing model described in the text for on-axis
  and edge streamlines. Red, full, wide: fitted, on-axis; red,
  short-dashed, narrow: adiabatic on-axis; blue, long-dashed, wide:
  fitted, edge; blue, dotted,narrow: adiabatic, edge.  If the minimal
  models do not include transverse variations of field component
  ratio, then only one pair of profiles is plotted. The vertical
  dotted lines indicate the recollimation distance, $r_0$. Field
  parameters for the minimal models given in Table~\ref{tab:minparams}
  were used for all sources except 3C\,31.
\label{fig:blprof}}
\end{figure}

\begin{figure}
\epsfxsize=8.5cm
\epsffile{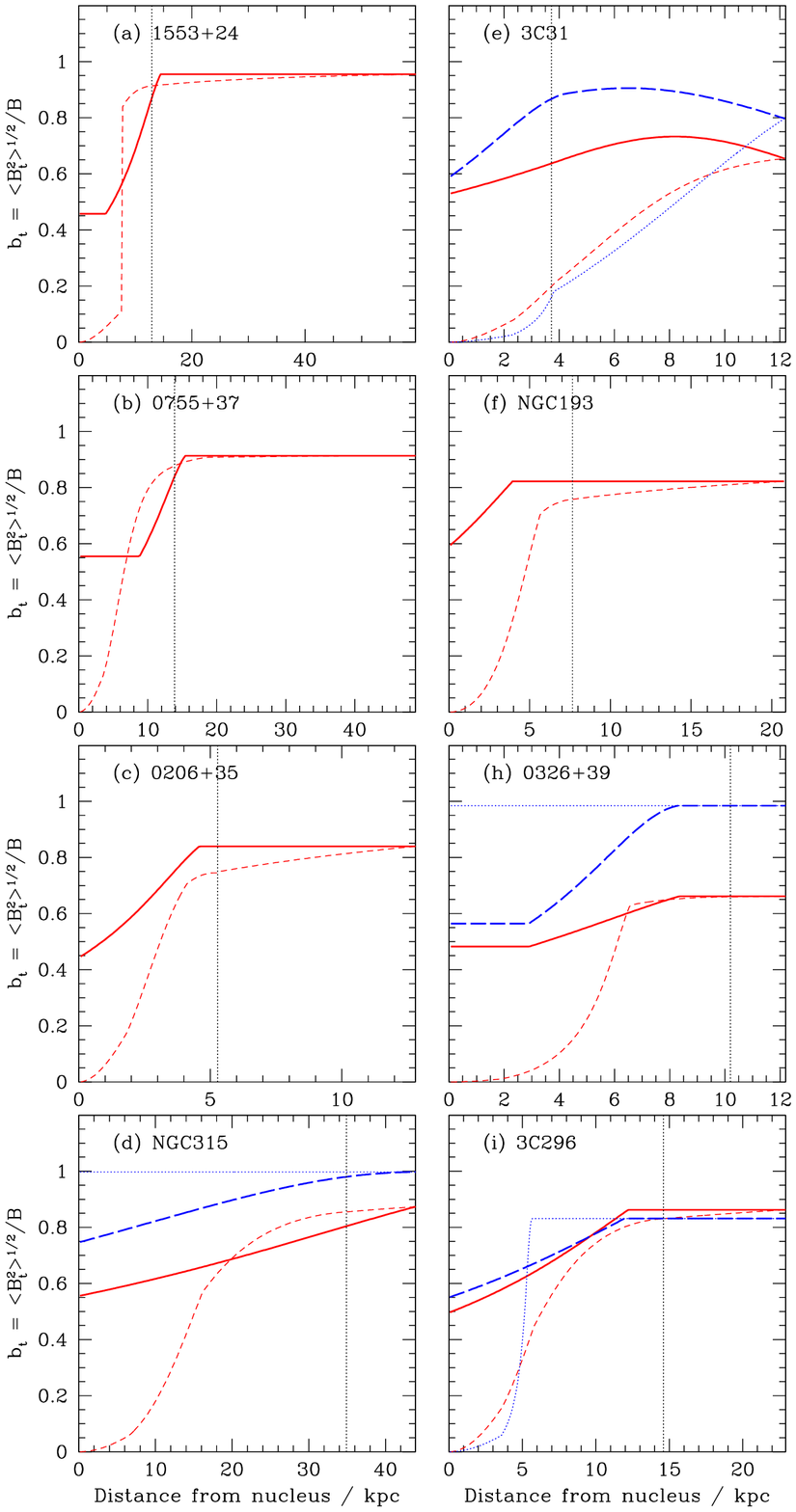}
\caption{As Fig.~\ref{fig:blprof}, but for the fractional toroidal
  field, $b_t = \langle B_{\rm t}^2 \rangle^{1/2}/B$.
\label{fig:btprof}}
\end{figure}

\begin{figure}
\epsfxsize=8.5cm
\epsffile{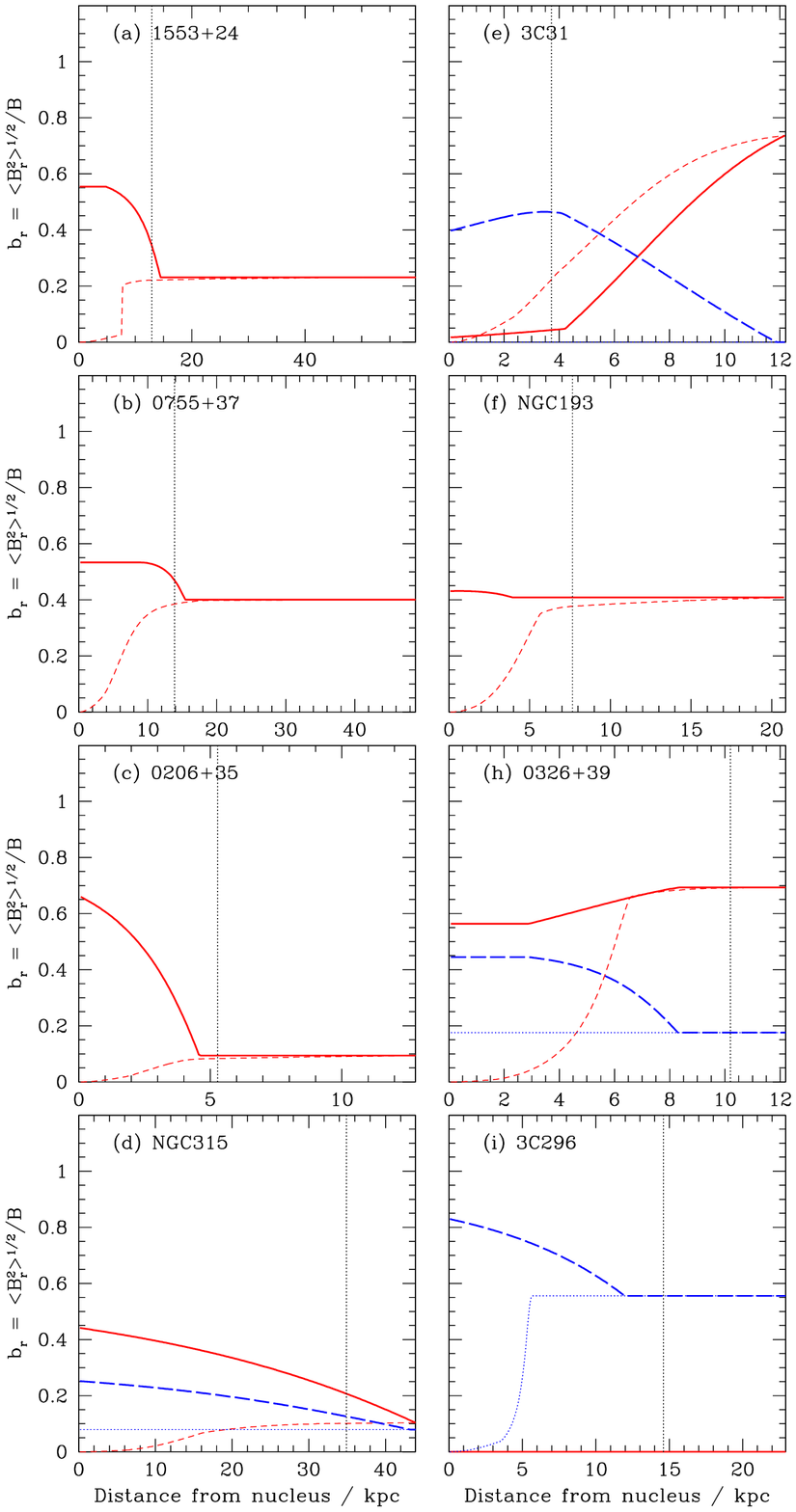}
\caption{As Fig.~\ref{fig:blprof}, but for the fractional radial
  field, $b_r = \langle B_{\rm r}^2 \rangle^{1/2}/B$.
\label{fig:brprof}}
\end{figure}

Figs~\ref{fig:blprof} --
\ref{fig:brprof} show comparisons of the predictions of
equations~(\ref{eq:bl}) -- (\ref{eq:btr}) with our model component ratios
for on-axis and edge streamlines.  The model and flux-freezing curves
are normalized at the edge of the model grid, where their slopes agree
quite well\footnote{This is not possible for M\,84 and 3C\,270, for
  which the fitted velocity becomes 0 before recollimation, so they
  are excluded.}.  We use minimal models (Table~\ref{tab:minparams})
in preference to full models except for 3C\,31.  The reason is that
the agreement between the slopes is significantly better for the
minimal models, which require the velocity to remain constant with
$\beta = \beta_0$ after deceleration (this is compatible with the data
except for 3C\,31).  The slopes for the flux-freezing model depend on
the velocity gradient (equations~\ref{eq:bl} -- \ref{eq:btr}) which in
some cases is not accurately determined from our fits.  This
problem occurs if the outer velocities, $\beta_0$, are low and the 
deceleration ends close to the edge of the model grid 
(Fig.~\ref{fig:beta3d}k and l). 
 A small random
or systematic error in $\beta_f - \beta_0$, for instance from an
intrinsic difference in emissivity function of the type described in
Section~\ref{asymm-em}, can then produce a significant, but spurious, change in
velocity gradient.  We suspect that the apparent accelerations in
1553+24, 0326+39 and 3C\,270 are examples of this effect \citep[Laing et al., in preparation]{CL}.

Close to the AGN, the field component ratios predicted by
flux-freezing vary much more rapidly with distance than our model
fits. In other words, the transition between longitudinal and
transverse field does not happen as abruptly as expected for jets
which are both expanding rapidly and decelerating. For the
normalization we have chosen, the flux-freezing approximation predicts
that the field becomes almost entirely longitudinal close to the AGN
(Fig.~\ref{fig:blprof}) and, conversely, that the toroidal and radial
components essentially vanish (Figs~\ref{fig:btprof} and
\ref{fig:brprof}).  The only exceptions occur where the longitudinal
component is close to zero at the edge of the model grid (the edge
streamlines for NGC\,315, 0326+39 and 3C\,296).  The simple reason for
the discrepancy is that the ratio of longitudinal to toroidal or
radial field is $\propto \Gamma\beta x^{-1}$ (equations~\ref{eq:bl} --
\ref{eq:btr}).  The jets expand rapidly with distance from the AGN in
the geometrically flaring region and also decelerate: both effects
cause $\Gamma\beta x^{-1}$ to decrease with distance (by a factor of $\approx$1300
over the model region for 3C\,31, for example).

At larger distances, particularly in the outer region, the model and
flux-freezing relations both tend to become flat.  The fitted
velocities for the minimal models are constant and the asymptotic
expansion rates in the outer regions are small, so
equations~(\ref{eq:bl}) -- (\ref{eq:btr}) also imply slow changes in the
component ratios.  To a reasonable approximation, the variation of
field component ratios after the end of rapid deceleration and
recollimation is consistent with flux-freezing (the one conspicuous
exception, 3C\,31, is the only source that continues to decelerate
significantly on scales $>r_{v0}$).

It is not surprising that the approximation of flux-freezing in a
quasi-one-dimensional flow fails for the high-emissivity and
rapid-deceleration regions, since this is where we infer shocks (Section~\ref{spectra}), strongly
evolving velocity shear and turbulence. We attempted to model the evolution of the field in 3C\,31
self-consistently assuming flux-freezing in a laminar velocity field
with a transverse gradient \citep{LB04}, but failed to get even
approximate agreement except in the outer region. One fundamental
problem is that the ratio of radial to toroidal field strength should
not change with distance even in the presence of shear in a simple
axisymmetric velocity field.  In 3C\,31 and some other sources, this
is not consistent with the model fits. Another issue is that the shear
vanishes on-axis in an axisymmetric velocity field of the type we
consider, so it is not possible to slow the decline in longitudinal
field there.  A likely explanation is that random (turbulent) motions
which we cannot model significantly affect the field component ratios
and maintain these ratios closer to equipartition between longitudinal
and transverse components than is expected from flux-freezing.

Finally, we note that the almost pure toroidal nature of the off-axis
magnetic field at large distances from the AGN (Fig.~\ref{fig:Btor}) could
have implications for jet collimation. If this field is
vector-ordered and sufficiently strong, then it could at least help to confine the on-axis
flow, which usually has higher emissivity (Fig.~\ref{fig:emiss}) and therefore
by implication higher particle pressure.  Vector-ordering of the
toroidal field component is consistent with, but not required by our
modelling.

\subsection{Adiabatic models}
\label{Adiabatic}

\begin{figure}
\epsfxsize=8.5cm
\epsffile{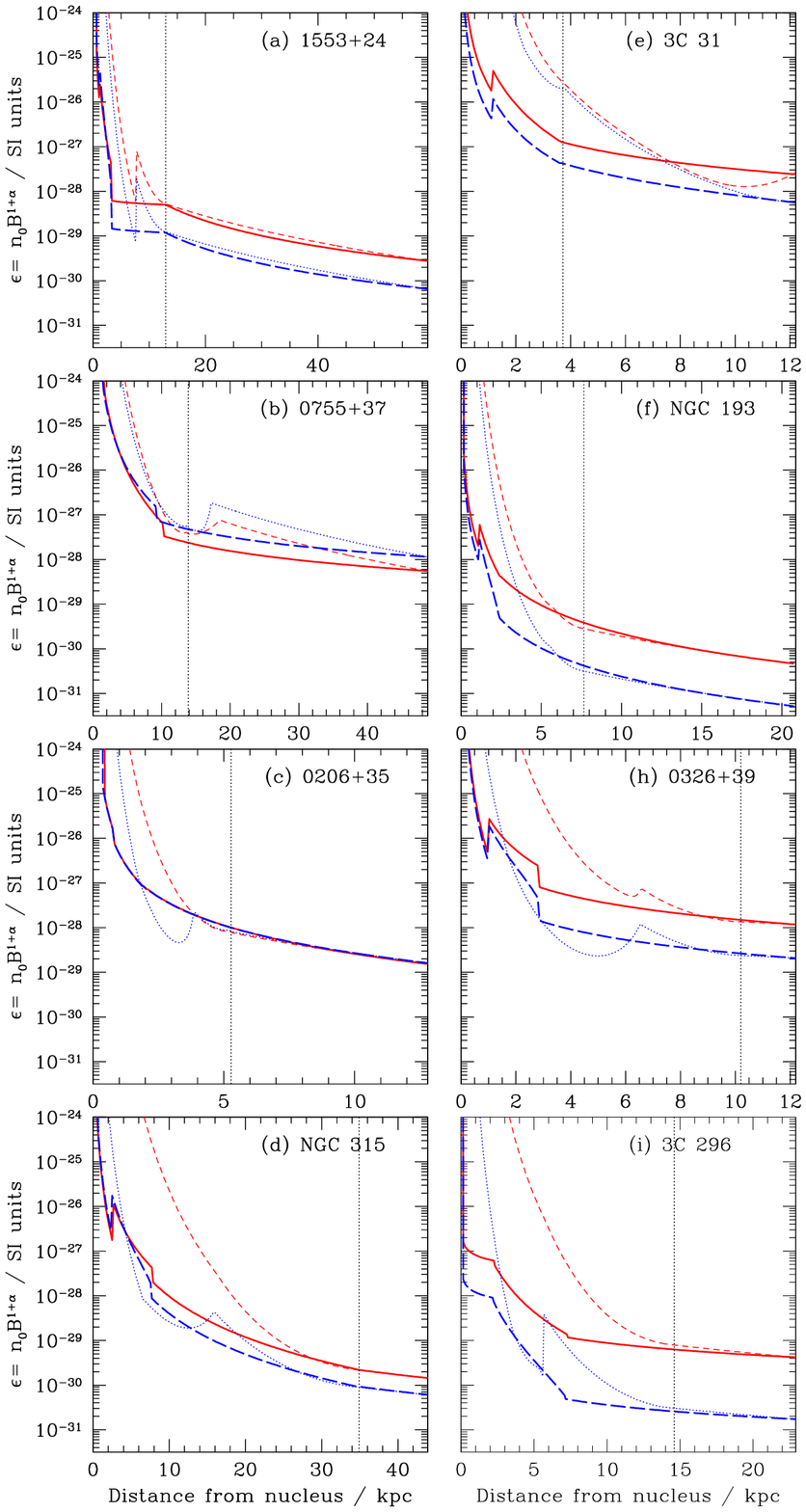}
\caption{Comparison of fitted emissivity function profiles with the predictions
  of adiabatic models for the emission on-axis and at the jet
  edges. Red full: fitted, on-axis; red short dashed: adiabatic
  on-axis; blue long dashed: fitted, edge; blue dotted: adiabatic,
  edge.  The vertical dotted lines indicate the recollimation distance, $r_0$.
\label{fig:emprof}}
\end{figure}

With the assumption that the energies of the radiating particles are altered only by
adiabatic losses (i.e.\ particle acceleration and radiative loss
processes can be ignored), the emissivity function $\epsilon$ can be
written in terms of the magnetic field $B$, as
\begin{eqnarray}
\epsilon \propto (x^2\beta\Gamma)^{-(1+2\alpha/3)}B^{1+\alpha}
\end{eqnarray}
\citep{Baum97,LB04}.  If the magnetic field follows the
quasi-one-dimensional flux-freezing relations of
Section~\ref{Bevolution}, then $B$ can be expressed in terms of the field
components $\langle \bar{B_l}^2 \rangle ^{1/2}$, $\langle \bar{B_t}^2
\rangle ^{1/2}$ and $\langle \bar{B_r}^2 \rangle ^{1/2}$, the radius
$\bar{x}$, velocity $\bar{\beta}$ and Lorentz factor $\bar{\Gamma}$ at
some fiducial location using equation (8) of \citet{LB04}:
\begin{eqnarray}
B &=& \left[ \langle \bar{B_l}^2\rangle\left(\frac{\bar{x}}{x}\right)^4
+ (\langle \bar{B_t}^2\rangle +\langle \bar{B_r}^2\rangle )\left(\frac{\bar{\Gamma}\bar{\beta}\bar{x}}{\Gamma\beta
x}\right)^2\right ]^{1/2} \label{eq:adiabat}
\end{eqnarray}
We can therefore predict the emissivity function using our fitted jet widths
and velocities together with the modelled field component ratios at
the fiducial location.  

We find that the {\it slopes} of the emissivity function variations fitted by
our jet models asymptotically approach those of equation~(\ref{eq:adiabat}) far 
from the AGN, in the region where the jets have decelerated and
recollimated.
In Fig.~\ref{fig:emprof}, we compare the emissivity function
variations for our  models with the predictions of
equation~(\ref{eq:adiabat}) for on-axis and edge streamlines\footnote{As in Section~\ref{Bevolution} and for the same reasons,
we used the minimal models for the sources in
Table~\ref{tab:minparams}.}.  The
magnitudes of the adiabatic and model-fitted emissivities have been
normalized to each other at the outer edge of the model grid for each
source (again leading to the exclusion of M\,84 and 3C\,270).  In most
cases, the two emissivity-function curves agree reasonably well after the jets
have both decelerated and recollimated.  
3C\,31 is again the exception (Fig.~\ref{fig:emprof}e); as for the field evolution, we 
suspect that this is related to continuing deceleration.  In every case the adiabatic approximation predicts a much
faster variation of emissivity function with distance from the AGN  than is actually observed within the
flaring region.  In other words, the
observed jets all fade much more slowly with increasing jet width in
their flaring regions than would be predicted from adiabatic losses alone, but
their brightness decrease becomes asymptotically adiabatic after they
have recollimated.  It is therefore unwise to attempt to determine the
variation of jet velocity with distance assuming adiabatic brightness
evolution in the flaring region, but such estimates may be valid after
recollimation.

The implication of our result is that there is little ongoing particle
acceleration or radiative energy loss for particles emitting at GHz
frequencies in the outer regions (radiative losses are not expected 
to be significant since there is no evidence for spectral steepening with distance from
the AGN; \citealt{spectra}).

In contrast, we would not expect the adiabatic relations to be
followed in the flaring regions, since these parts of the jets are
known to emit synchrotron radiation at all frequencies up to soft
X-rays. This requires ongoing particle acceleration, as we now
discuss.

\subsection{Particle acceleration and radio spectra}
\label{spectra}

Higher-frequency (mid- and near-IR, optical and especially X-ray)
emission has now been detected from many FR\,I main jet bases (and one
counter-jet) For the present sample, X-ray emission has been imaged
with {\em Chandra} in 0206+35, 0755+37, NGC\,315, 3C\,31, NGC\,193,
M\,84, 3C\,296 and 3C\,270
\citep*{WBH01,WBH03,ngc315hires,Worrall10,H02,H05,Harris02,Kharb12}. Optical
and/or mid-infrared detections have also been made for 1553+24,
0755+37, 3C\,31 and 3C\,296 \citep*{Croston03,H05,Lanz,Parma03}. The
high-frequency emission is thought to be synchrotron radiation from a
higher-energy extension of the relativistic electron population
responsible for the radio emission. High-frequency emission is
typically detected out to the end of the high-emissivity region in the
approaching jets. In two well-observed cases (NGC\,315 and 3C\,270),
it extends as far as the end of deceleration, albeit at a lower level
compared with the radio emission \citep[their table 3]{spectra}. The
synchrotron lifetimes for X-ray emitting electrons are tens to
hundreds of years -- considerably smaller than the light-travel times
across the jets.  This makes a cast-iron case for ongoing particle
acceleration in the regions we model.

We investigated the radio spectra of these jets over the frequency
range 1.4 -- 4.9 or 8.4\,GHz \citep{spectra}, with the
following conclusions.
\begin{enumerate}
\item The spectra always flatten slightly with increasing distance from the
  nucleus  between the brightness flaring point and
  the end of rapid deceleration.
\item The mean spectral indices are $\langle\alpha\rangle = 0.66 \pm
  0.01$ over the high-emissivity region and $\langle\alpha\rangle =
  0.59 \pm 0.01$ both immediately after deceleration and in the outer
  (recollimation) region.  The corresponding energy indices are $q =
  2.32$ and 2.18, respectively (equation~\ref{eq:energy}).
\item One source, NGC\,315, also shows transverse spectral gradients in the
  sense that $\alpha$ is higher on-axis \citep{ngc315ls}.
\item The steeper spectra close to the jet flaring points are 
  associated with typical bulk flow speeds $\beta \ga 0.5$.
\end{enumerate}

This radio spectral analysis is consistent with our results from
adiabatic models (Section~\ref{Adiabatic} and
\citealt{LB04}). Particle acceleration with $q = 2.32$ over the energy
range corresponding to GHz radio emission (electron Lorentz factors
$\gamma \sim 2000 - 30000$, assuming equipartition magnetic fields) is
required in the high-emissivity region.  The process must be capable
of accelerating electrons to $\gamma \sim 10^7 - 10^8$ for bulk flow
speeds $\beta \ga 0.5$ (or ${\mathcal M} \ga 1.5 - 2$, as we
conjecture in Section~\ref{decel}).  In the deceleration region, there
is a gradual transition to a characteristic energy index $q = 2.18$.
High-energy radiation is still produced, but is less prominent relative
to the radio emission.  The range of observed spectral indices and the
inferred dependence on velocity could result from first-order Fermi
acceleration by mildly relativistic shocks, in the limit that the
scattering mean free path is close to the electron gyro-radius
\citep{SB12}. If the bulk flow is at most mildly supersonic
(Section~\ref{decel}), this mechanism may not be efficient enough,
particularly in the slower, flatter-spectrum regions.  A second
mechanism would then be required, perhaps associated with increased
velocity shear, as suggested by the transverse spectral gradients in
NGC\,315.  After deceleration and recollimation, there is relatively
little ongoing particle acceleration\footnote{As noted earlier, 3C\,31
  may be an exception. In this regard, it is interesting that
  \citet{Lanz} find evidence for acceleration of electrons to $\gamma
  \approx 10^5 - 10^6$ (but not much higher) in its outer region.}.

\subsection{Brightness flaring and the high-emissivity region} 
\label{flare-discuss}

\begin{figure}
\epsfxsize=6.5cm
\epsffile{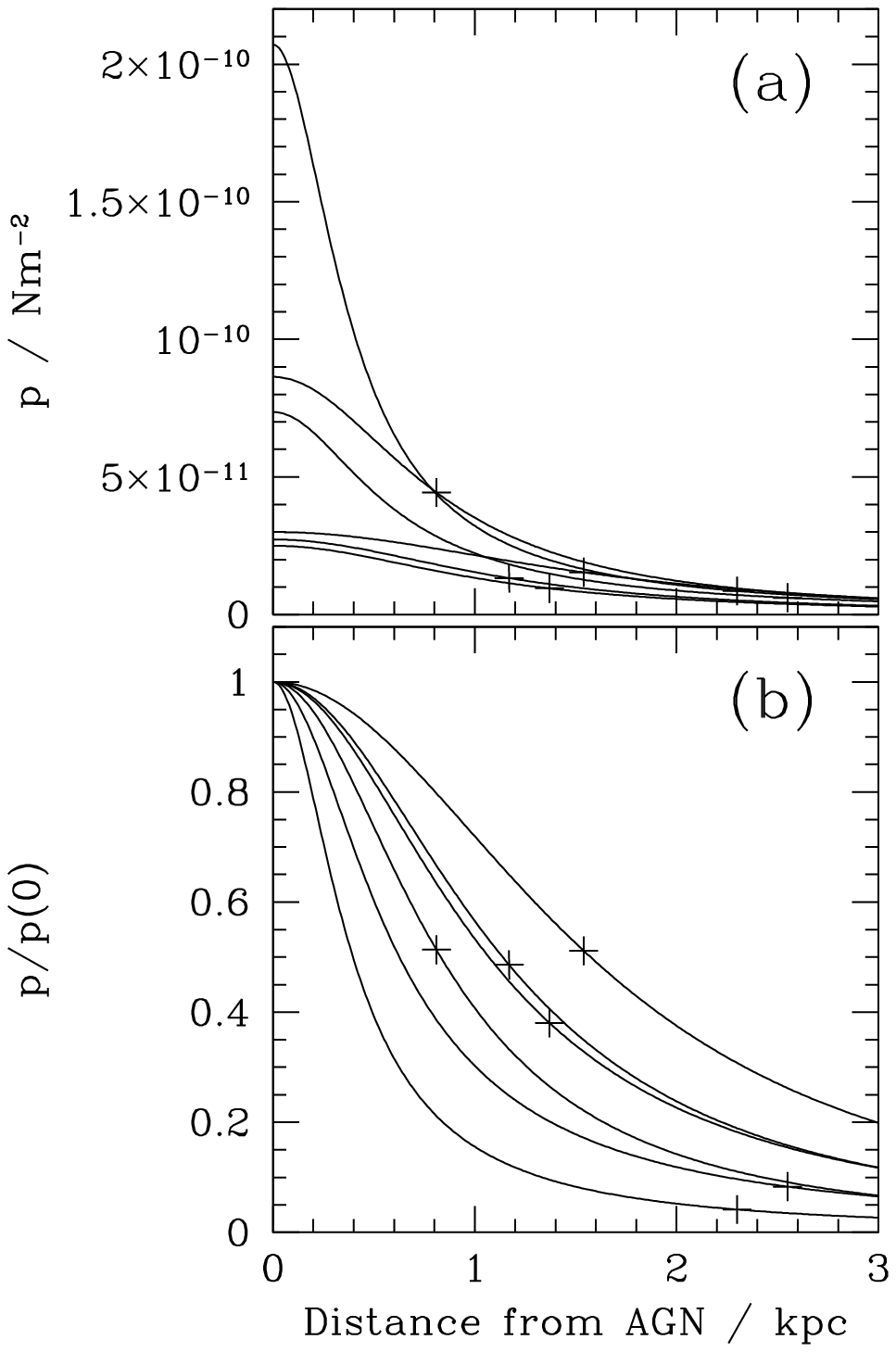}
\caption{Pressure profiles for the dense coronae of hot gas
  surrounding the jet bases in 0755+37, 0206+35, NGC\,315, 3C\,31,
  3C\,296 and 3C\,270. These were derived from isothermal beta-model
  fits to {\sl Chandra} observations
  \citep{WBH01,WBH03,H02,H05,Cros08}. (a) absolute pressures, $p(r)$;
  (b) relative pressures $p(r)/p(0)$. $+$ symbols mark
  the brightness flaring points.
\label{fig:pressures}}
\end{figure}

What might cause the jet brightening that marks the onset of the
extended flaring region?  We do not yet have an unambiguous explanation, but
can add some new constraints, as follows.
\begin{enumerate}
\item The outer isophotes of the jets expand at a {\it
    continually increasing} rate both upstream and immediately
  downstream of the brightness flaring points (Fig.~\ref{fig:ihires}).
\item The resolved radio structures in the jet bases downstream from
  the brightness flaring points, e.g.\ the quasi-helical chain of
  bright knots in the jet of NGC\,315 (Fig.~\ref{fig:ihires}d) and the
  knots in the high-emissivity regions of the jets of 3C\,31, 3C\,296
  and 3C\,270 (Figs~\ref{fig:ihires}e, i and j) are complex,
  non-axisymmetric and concentrated towards the jet axes (perhaps located at
  the spine/shear-layer boundary).  We find no convincing evidence for
  brightness enhancements that cross the entire width of the jets in
  their high-emissivity regions, although such structures could be present close to the flaring points 
  themselves, where the jets are too narrow for us to resolve transversely.
\item The flow immediately downstream of the flaring point must be at
  least transonic: $\beta \approx 0.8$, so ${\mathcal M} \ga 2$, the
  lower limit corresponding to a jet composition dominated by
  ultrarelativistic particles and field (Section~\ref{decel}).
\item The brightness flaring points are located in steeply falling
  external pressure gradients. All of the sources with published
  high-resolution X-ray images show dense, kpc-scale coronae of hot
  gas. Fig.~\ref{fig:pressures} shows the absolute and normalized
  pressure profiles derived from isothermal beta-model fits to {\sl
    Chandra} observations for 0755+37, 0206+35, NGC\,315, 3C\,31,
  3C\,296 and 3C\,270 \citep{WBH01,WBH03,H02,H05,Cros08}. For a
  pressure profile
\begin{equation}
p(r) = p(0) (1 + r^2/r_{\rm c}^2)^{-3\beta_{\rm atm}/2}
\end{equation}
the steepest gradient is at $r = r_{\rm c}(1+3\beta_{\rm
  atm})^{-1/2}$, which is in the range 0.2 -- 1\,kpc for the sources
in Fig.~\ref{fig:pressures} (core radii between 0.35 and
1.8\,kpc). The corresponding flaring point distances are 0.8 --
2.6\,kpc.  In 0755+37, 0206+35, 3C\,31 and 3C\,270, the brightness
flaring points are located close to the steepest pressure gradients;
in NGC\,315 and 3C\,296, they are significantly farther out 
(Fig.~\ref{fig:pressures}).
\end{enumerate}

Three mechanisms have been suggested for the sudden brightening of
FR\,I jets, as follows.
\begin{enumerate}
\item The jet is overpressured within the corona and expands rapidly until its
  pressure falls below that of the ambient medium, at
  which point a stationary recollimation shock forms
  \citep{Sanders}. The recollimation shock marks the flaring point and
  the high-emissivity region contains a series of shocks formed as the
  jet oscillates around pressure equilibrium. Shocks naturally lead to
  brightness enhancements not only from compression, but also as a
  result of first-order Fermi acceleration.
\item Alternatively, if the jet is initially in pressure equilibrium with
  the corona and the external pressure drops
  sufficiently steeply with distance, a standing shock (or series of
  shocks) can again be formed \citep{Bick84}. \citet{DM88} analysed the case of a relativistic
  jet which encounters an instantaneous drop in external pressure and showed
  that jet-crossing shocks form when the pressure drops by more than a factor of $\approx$2.  
\item Jets are likely to be unstable to the growth of Kelvin-Helmholtz
  instabilities \citep[and references
    therein]{Perucho12,Salvesen13}. The pressure maxima associated
  with these instabilities will cause brightness enhancements, as
  would any particle acceleration processes associated with growth of
  instabilities and transition to turbulence, or small-scale shocks.
  The flaring point might then be the location where the growth
  becomes non-linear.
\end{enumerate}

Case (i) was considered by \citet[simulation
  1]{PM07}, who studied the evolution of a light, overpressured,
relativistic jet propagating in the external density and pressure
distribution estimated for 3C\,31 \citep{H02} and demonstrated the
formation of strong recollimation shocks.  One argument against this
idea is that we see no evidence for recollimation in the outer
isophotes at the brightness flaring point\footnote{This also argues
  against \citet{Krause2012}'s identification of the flaring point as
  a stationary termination shock in a flow with an intrinsically large
  opening angle} (in contrast to the situation at the end of
the flaring region).  This is not conclusive, since entrainment and
mixing with the external medium will cause the jet to expand again,
but we have shown that evolution of the transverse velocity profile
appears to begin significantly downstream of the flaring point and we
might expect the spreading rate to decrease before mixing becomes
important. Finally, there is no explanation for the origin of the
initial overpressure, which is imposed as an initial condition.

The alternative that the jet is in pressure equilibrium with the
corona until $r \approx r_{\rm c}$, where it becomes overpressured [case(ii)], is more plausible for two
reasons.  First, pressure decreases are inferred from X-ray
observations at approximately the correct locations
(Fig.~\ref{fig:pressures}). Second, the outer envelope of the jet is
expected to expand (on average) as it adjusts to pressure balance, as
observed. The pressure decrease may be too slow for strong shocks to
develop, however, so it is not clear how large a brightness
enhancement would be observed.

The absence of obvious jet-crossing shocks is a potential difficulty
for cases (i) and (ii).

The drop in external pressure may still be the trigger for flaring in
brightness even if strong shocks are not formed.  In a simulation of
an initially pressure-matched jet with initial velocity $\beta = 0.5$,
again in the external density and pressure gradients inferred for
3C\,31, \citet[simulation 4]{PM07} confirmed that the cross-section of
the jet oscillated smoothly and that strong shocks were not formed. In
this simulation, however, the coupling to Kelvin-Helmholtz
instabilities was enhanced [case (iii)]\footnote{It is possible,
  however, that the increased coupling is an artefact of the
  axisymmetric simulation}.  The simulated jet disrupted
shortly after the growth of instabilities, whereas the jets analysed here flare without 
disruption, maintaining fairly smooth outer isophotes
(Fig.~\ref{fig:ihires}). Thus if Kelvin-Helmholtz instabilities are
important at the brightness flaring points of the jets in our sample, then they are unlikely
to be dominated by low-order modes.

The idea that brightness flaring is triggered by rapid expansion in a
steep external pressure gradient therefore seems plausible, but the
precise mechanism remains obscure.

\subsection{Velocities upstream of the brightness flaring point}
\label{inner}

\begin{figure}
\epsfxsize=6.5cm
\epsffile{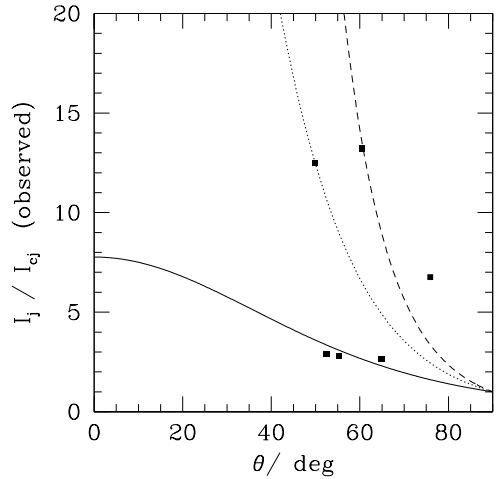}
\caption{Observed values of the jet/counter-jet sidedness ratio,
  $I_{\rm j}/I_{\rm cj}$ determined between 2 $\times$ FWHM and $0.6r_{e1}$ (in
  projection) from the core for NGC\,315, 3C\,31, NGC\,193, 0326+39,
  3C\,296 and 3C\,270, plotted against inclination, $\theta$ from our
  model fits.  The ratios for velocities of $\beta = 0.375$ (full),
  $\beta = 0.7$ (dotted) and $\beta = 0.94$ (dashed) predicted using
  equation~(\ref{eq:iratio}) are plotted for comparison.
\label{fig:side_theta}}
\end{figure}

We cannot address the question of the flow velocity upstream of the
flaring point satisfactorily using our models because the jets are
faint and poorly resolved transverse to their axes in these regions
(Section~\ref{narrow}).  We were, however, able to measure the jet and
counter-jet flux densities from where they are first separable from the
core (2 $\times$ FWHM) to $0.6r_{e1}$ in projection from the AGN for six sources
(we chose the outer limit to avoid any emission directly associated
with the flaring points at $r = r_{e1}$). The observed sidedness
ratios (plotted against $\theta$ in Fig.~\ref{fig:side_theta}) do not
show a consistent trend.  We can estimate the velocity roughly using
equation~(\ref{eq:iratio}). For three sources (3C\,31, NGC\,193 and
3C\,296), the ratios suggest a low value of $\beta \approx 0.4$; NGC\,315
and 0326+39 require $\beta \approx 0.7$ ($<\beta_1$) and $\beta
\approx 0.95$ ($>\beta_1$), respectively, and 3C\,270 has too high a
ratio to be consistent with any velocity.

The measurement of counter-jet flux densities close to the core is
difficult and spatial averages are poorly defined for irregular
brightness distributions.  Nevertheless, the slow speeds inferred
in three cases suggest two possible scenarios, which we cannot
presently distinguish.  The first is that the outflows as a whole are
accelerating with increasing distance from the AGN upstream of the
brightness flaring point, as might be the case for overpressured jets
propagating in steeply-decreasing pressure profiles (e.g.\ simulation
1 of \citealt{PM07}).  The alternative is that the emission from the
upstream region comes mainly from a slow-moving surface layer, and
that faster on-axis flow in the jet spine becomes visible only after
sudden deceleration at the brightness flaring point, as previously
suggested for 3C\,31 by \citet{LB02a}.  The emission from the surface
layer would have to dominate at least for $\theta \ga 50^\circ$.  We
might then expect to see emission from the spine components in the
approaching jets if they are orientated closer to the line of sight,
but projection would inevitably mean that they would be poorly
resolved, as is indeed the case for our sources with $\theta <
50^\circ$.  An argument against this alternative is that we see little
evidence for a slow boundary layer immediately downstream of the
brightness flaring point, where the transverse velocity profiles are
close to flat.

Independent evidence of slower speeds on parsec scales comes from
proper-motion measurements. The best-studied case is M\,87, which we
discuss below (Section~\ref{m87_cena}).  Proper motions have been
determined for the approaching jets in two of our modelled sources:
NGC\,315 \citep{Cotton99,Lister13} and 3C\,270 \citep*{Piner01}.  For
NGC\,315, the apparent speeds measured by \citet{Cotton99} are
$\beta_{\rm app} = 0.81$ at an angular separation of 4\,mas from the
core and $\beta_{\rm app} = 1.79$ at 10\,mas. For $\theta = 49\fdg8$,
the corresponding pattern speeds are $\beta_{\rm pattern} = 0.63$ at a
deprojected distance of 1.5\,pc and $\beta_{\rm pattern} = 0.93$ at
4.4\,pc. \citet{Lister13} find much slower speeds: $\beta_{\rm app} =
0.03 - 0.05$ from 2.5 to 6.4\,pc and $\beta_{\rm app} < 0.005$ from
0.4 to 1.7\,pc ($\beta_{\rm pattern} = 0.04 - 0.06$ and $\beta_{\rm
  pattern} < 0.007$); these are much slower than the flow speeds
inferred from the sidedness ratio \citep{Cotton99}.
We estimate a flow speed of $\beta = 0.85$ on kpc scales. In 3C\,270,
using the distance in Table~\ref{tab:sources}, the apparent speed is
$\beta_{\rm app} = 0.40$ at 7\,mas \citep{Piner01}, implying a pattern
speed of $\beta = 0.37$ at a deprojected distance of 1.1\,pc for
$\theta = 75\fdg9$, whereas we find a flow speed of $\beta = 0.92$ on
kpc scales.  In both cases, the pattern speed measured close to the
AGN can be significantly slower than the flow speed we estimate on
larger scales.  Again, it is not clear whether the inferred speeds are
consistent with an accelerating bulk flow with $\beta_{\rm pattern} =
\beta_{\rm flow}$: the discrepant speed estimates for NGC\,315 on pc
scales suggest a more complex situation.  Moving features in the jet
(e.g.\ shocks) may have $\beta_{\rm pattern} \la \beta_{\rm flow}$ or
could be preferentially located in a slower surface layer, as
suggested above.

\subsection{Comparison with other well-resolved sources}
\label{m87_cena}

As described in Section~\ref{sources}, the objects selected for this
study are the best-resolved and brightest of the nearby FR\,I radio
galaxies whose AGNs successfully formed large-scale radio structures
with bright, wide jets and counter-jets.  The jet bases of two other
nearby radio galaxies that do not have these defining characteristics
have also been studied in exquisite detail, and it is interesting to
examine how their properties relate to those we find for our sample.

The jet and counter-jet bases in Centaurus A \citep[and references
  therein]{Morganti10}, the closest radio galaxy to the Milky Way,
have been studied with high linear resolution.  Proper motions of
$\beta_{\rm app} = 0.3 - 0.8$ have been measured within its main jet
\citep{Hard03,Goodger10}, providing direct evidence for outflow at
near-relativistic velocities.  Two identifications have been proposed
for the brightness flaring point: the A1/AX1 knot complex
$\approx$0.2\,kpc from the AGN \citep{Hard03} and the region at
$\approx$3.5\,kpc where the jet enters the inner lobe
\citep*{Hard06}. While its inner jet structure may be analogous to the
flaring regions and fainter upstream precursors of the jet bases
studied here, Cen\,A is likely to be an example of an outflow that has
restarted in an environment disturbed by a merger of an
active galaxy with a gas-rich system.  It may not therefore be
a good case to compare in detail with the straight, relatively
undisturbed FR\,I jets studied here, even though it offers a close-up
view of entrainment and jet-gas interactions.  The faint, patchy
nature of its detected counter-jet precludes modelling by our methods
but if Cen\,A indeed contains a symmetrical decelerating outflow then
the overall faintness of its counter-jet features may require flow
velocities in its flaring region to exceed the pattern speeds measured
by \cite{Goodger10} and thus to be in the regime deduced here for
other sources. 

Studies of the spectrum of the main jet in Cen\,A between radio and
X-ray wavelengths have provided independent evidence for multiple
particle-acceleration mechanisms
\citep{Hard06,Hard07,Worrall08,Goodger10}. The X-ray emission out to
1.1\,kpc in projection from the AGN is dominated by knots and is
consistent with particle acceleration by small-scale shocks;
acceleration between 1.1 and 4.5\,kpc appears to require a distributed
mechanism and the spectral evolution at larger distances is consistent
with passive advection of particles. This picture is similar to the
one we propose in Section~\ref{spectra} and we tentatively associate
the two acceleration mechanisms with the characteristic radio spectral
indices of 0.66 (shocks) and 0.59 (distributed),
respectively\footnote{\citet{Goodger10} measure slightly steeper
 radio spectral indices, but the difference is marginally significant.}. We
cannot resolve structures on the scale of the knots in the Cen\,A jet
in our sources, and it could be that the large-scale spectral
gradients we observe result from gradual changes in the proportions of
the emission from these two mechanisms.

The jet in M\,87 has also been studied in detail at many wavelengths,
providing clear evidence for relativistic flow. Although the measured
proper motions are subrelativistic on pc scales
\citep{Reid89,Kellermann04,Kovalev07}, much larger speeds are found on
scales $\ga$60\,pc in projection with a maximum value of $\beta_{\rm
  app} \approx 6$ and a tendency to decrease with distance from the
AGN from 100\,pc outwards
\citep*{Biretta95,Biretta99,Cheung07,Meyer13}.  Whether the changes in
speed reflect true bulk acceleration and deceleration or merely
changes in pattern speed in an underlying fast flow remains
controversial (e.g.\ \citealt{Kovalev07,Nakamura13}): a
subrelativistic speed on pc scales would be inconsistent with the observed
sidedness ratio.  \citet{HE} argue from a conservation-law analysis
that there is bulk deceleration from $\Gamma \approx 4.4 - 7.5$ at a
projected distance of 80\,pc to $\Gamma \approx 1.8 - 2.7$ ($\beta
\approx 0.83 - 0.93$) at 1\,kpc (the spectacular `brightening point'
at Knot `A'). Although the latter velocity is similar to those we
infer just downstream of the brightness flaring points for the sources
in our sample, there are no collimated counter-jet features in M\,87
at distances from the AGN corresponding to the well-studied bright jet
(even well beyond the distance of knot A) and little evidence for the
geometrical flaring we observe.  This lack of counter-jet emission and
the small opening angle of its jet suggest that the structures in the
M\,87 jet are not those of a flaring region of the type described here
but may instead be an example of a faster, well-collimated
`strong-flavour' jet seen at a small angle to our line of sight.  If
so, then M\,87 may more closely resemble a `wide-angle-tail' source
seen at a small inclination angle, in which the jets disrupt rather
than flaring, decelerating and recollimating smoothly.

\section{Summary}
\label{Conclusions}

\begin{figure}
\epsfxsize=6.35cm 
\epsffile{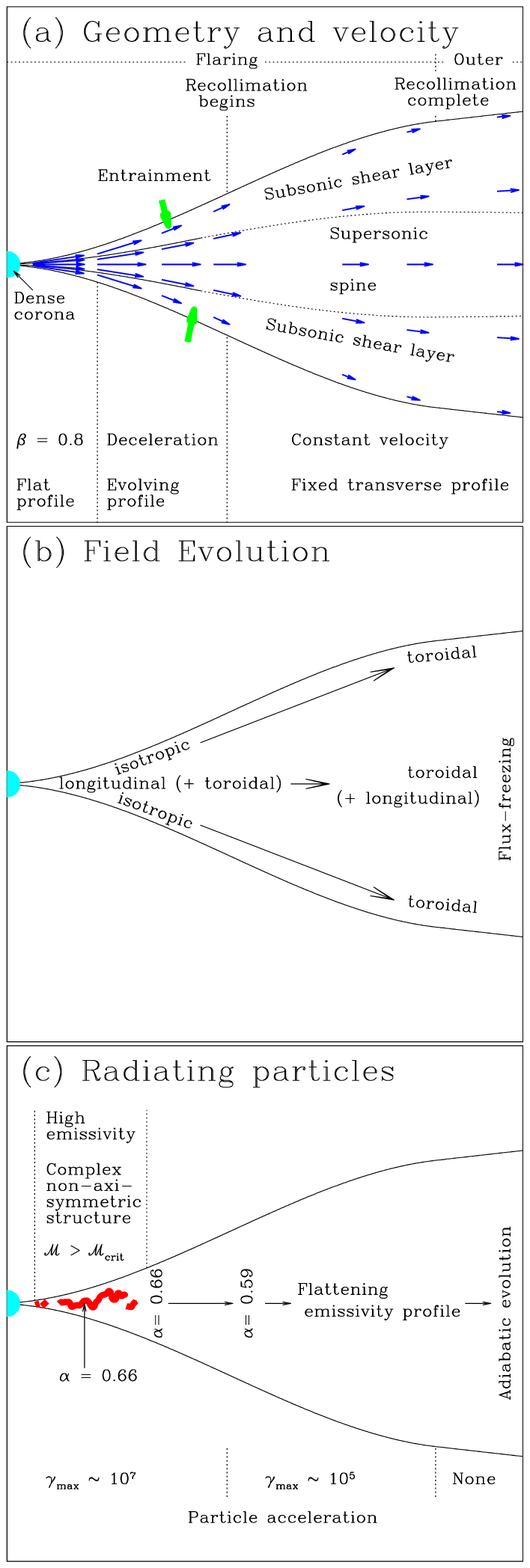}
\caption{Sketches showing the principal deductions from our model
  fits. (a) Geometry, velocity field and inferred boundary-layer
  entrainment.  The blue arrows represent the velocity field. The
  supersonic spine always extends into the deceleration region (where
  it is represented by full lines) but continues throughout the
  modelled region only in some cases (dotted lines).  (b) Evolution of
  field-ordering parameters. (c) Location of the high-emissivity and
  adiabatic regions; spectral variations. The red pattern
  schematically represents bright knots in the high-emissivity region.
\label{fig:jetsketch}}
\end{figure}

We have fit intrinsically symmetrical, axisymmetric relativistic flow
models to deep, high-resolution $I$, $Q$ and $U$ images of jets in ten
FR\,I radio galaxies, using the same parametrization in all cases. Our
conclusions are given below and key points are sketched in
Fig.~\ref{fig:jetsketch}.

\subsection{Direct inferences from the data}

\begin{enumerate}
\item The transverse-resolved sections of the jets start with geometrically flaring regions in
  which the spreading rates first increase rapidly and then
  decrease. The jets eventually recollimate to form conical outer
  regions.
\item The jet brightness distributions all show sudden brightness
  flaring following an initial dim, well-collimated region
  (Fig.~\ref{fig:icomposite}).  The brightness flaring point is not
  associated with a clear discontinuity in the spreading rate of the
  outer isophotes, but is within the regime of geometrical flaring in
  all sources (Fig.~\ref{fig:ihires}).
\item Immediately downstream of the brightness flaring points and
  within the region of geometrical flaring, we often see bright,
  non-axisymmetric, knotty substructures (e.g., NGC315, 3C\,31, 3C\,296 and 3C\,270 in
  Figs.~\ref{fig:icomposite} and \ref{fig:ihires}). These define the
  {\em high-emissivity region}.
\item The progression of collimation and brightness changes exhibited
  by the brighter jet is always followed on the same physical scales
  by the counter-jet in the same source, while the jet/counter-jet
  intensity ratio generally decreases with increasing distance from
  the AGN (Fig.~\ref{fig:icomposite}).
\item Near the brightness flaring point, transverse intensity profiles
  in the main jets tend to be centrally peaked whereas those in the
  counter-jets tend to be flat-topped or centrally darkened
  (Fig.~\ref{fig:icomposite}).
\item The jet/counter-jet asymmetries in linear polarization are well
  correlated with those in total intensity and follow a common pattern
  in all 10 sources: there is a progression along the jet axis from
  apparent magnetic field parallel to the axis to field perpendicular
  to the axis in the main jet, whereas the apparent field in the
  counter-jet is always perpendicular unless the jets are very
  symmetrical (Fig.~\ref{fig:qicomposite}).
\item The jets and counter-jets show systematic spectral variations in
  the flaring regions: there are small decreases in the
  radio spectral index with increasing distance from values near 0.66
  in the high-emissivity regions to 0.59 after recollimation.  The scale of this spectral variation appears
  to be tied to the recollimation distance.
\end{enumerate}

\subsection{Inferences from model fits}

\begin{enumerate}
\item Despite the wide range of linear scales, the geometrically
  flaring regions (after correcting for projection using the modelled
  inclination and scaling by the recollimation distance, $r_0$) have
  remarkably similar shapes (Fig.~\ref{fig:geomresults}b). The mean
  half-width/length ratio is 0.29 with a small dispersion.
\item Where it first becomes measurable near the AGN by our method,
  the outflow velocity has a mean value of $\langle\beta\rangle =
  0.81$ with an rms dispersion of 0.08 (Fig.~\ref{fig:beta3d}k).
\item At this point, the transverse velocity profiles are consistent
  with constant values in nine of the 10 jets
  (Fig.~\ref{fig:beta3d}m).
\item Farther downstream, all 10 jets decelerate with increasing
  distance from the AGN, although their deceleration rates vary widely
  (Fig.~\ref{fig:beta3d}). Rapid deceleration occurs across the entire
  widths of the jets.
\item After the end of rapid deceleration, the jet velocities on a
  given streamline are consistent with constant values except in
  3C\,31, which decelerates less rapidly.
\item In six cases, there is good evidence that a transverse velocity
  gradient develops during deceleration: the outflow at the jet edges
  is slower than on-axis (Fig.~\ref{fig:beta3d}n). Similar gradients
  could be present in any of the other jets.
\item In the four sources where jet speeds remain high after
  deceleration, the transverse velocity profiles are well
  determined. Their fractional edge velocities range from $\approx$0 to
  $\approx$0.5.
\item Jet magnetic fields are primarily longitudinal and toroidal, but not vector-ordered helices (see Section~\ref{assumptions} -- helical fields would
lead to unobserved asymmetries in the transverse total intensity and
polarization profiles). The
  toroidal component dominates at large distances from the AGN (Figs~\ref{fig:Blong} -- \ref{fig:Brad}). 
  The mean values of the rms
  fractional components (longitudinal:toroidal:radial) 
  evolve:
  \begin{enumerate}
  \item  on-axis from 0.78:0.55:0.29 close to the AGN to 0.55:0.80:0.23 after
  recollimation and 
  \item close to the edge of the jet from 0.62:0.61:0.50 (nearly isotropic) to 0.05:0.97:0.23 (almost purely
  toroidal).
  \end{enumerate}
\item Although the evolution from longitudinal to transverse field is
  expected in an expanding flow, the quasi-one-dimensional
  flux-freezing approximation predicts a much more rapid transition
  from longitudinal to transverse field than we infer
  (Figs~\ref{fig:blprof} -- \ref{fig:brprof}).  The slow evolution of
  field structure after recollimation is close to that expected from
  flux-freezing, however.
\item Downstream of the brightness flaring point, the emissivity
  function $\epsilon = n_0 B^{1+\alpha}$ declines with distance from the AGN. The
  slope of this decline tends to flatten with increasing distance
  (Figs~\ref{fig:emiss}k and l).
\item In the flaring region, and especially in the high-emissivity
  region, the slope of the emissivity function is flatter than that
  expected if the particle energies are affected only by adiabatic
  losses and the field is frozen into the flow (`adiabatic
  approximation').  After the jets have recollimated and decelerated,
  the two slopes are similar (Fig.~\ref{fig:emprof}). The
  implication is that particle acceleration is required throughout the flaring
  region, but not after recollimation.
\item The characteristic spectral index $\alpha \approx 0.66$ observed
  in the high-emissivity region is associated with jet speeds $\beta
  \ga 0.5$ (Section~\ref{spectra}; \citealt{spectra}).
\item The flaring regions are homologous structures, in the sense that
  the fiducial distances for velocity, emissivity function and magnetic field
  evolution scale linearly with the recollimation distance, $r_0$
  (Table~\ref{tab:flare}; Fig.~\ref{fig:fidx0}).  The brightness
  flaring point marks a discontinuity in some combination of speed and
  rest-frame emissivity function, located at $\approx$0.1$r_0$. The
  high-emissivity region runs from $\approx$0.1$r_0$ to
  $\approx$0.3$r_0$. Rapid deceleration starts midway along the
  high-emissivity region (at $\approx$0.2$r_0$; Fig.~\ref{fig:ihires})
  and lasts until $\approx$0.6$r_0$.  Magnetic evolution begins near
  the AGN and essentially stops just after the end of the flaring
  region, at $\approx$1.1$r_0$ (where flux-freezing becomes a
  reasonable approximation).
\item The end of rapid deceleration coincides accurately with the
  start of recollimation (i.e.\ where the spreading rate begins to
  decrease with distance; Fig.~\ref{fig:fidx0}g).
\item The inclination angles of the jets inferred from our modelling
  correlate well with other indicators of the jet orientation:
  fractional core flux density (Section~\ref{corefractioncorrelation})
  and Faraday rotation rms ratio (Section~\ref{RMrmsratio}).
\item We find a remarkably good correlation between the core luminosity
  (corrected for Doppler boosting) and the extended luminosity for our
  sample, despite the large difference in physical scale between the emitting regions (Fig.~\ref{fig:tfc}c).
\item We have analysed the effects of departures from intrinsic
  symmetry in emissivity function, using the statistics of reversals in jet
  sidedness to estimate the magnitude of the effect.  We find that the
  effects on our derived physical parameters are comparable with other
  uncertainties for $\theta \la 75^\circ$, but dominate at larger
  inclinations.
\end{enumerate}

\subsection{Further inferences about jet physics}

We conjecture the following about the internal physics of the jets from the
systematics given above.
\begin{enumerate}
\item The jet has a mildly supersonic spine, of which the
  high-emissivity region forms the base.  The
  composition is dominated by ultrarelativistic particles and magnetic
  field at the brightness flaring point, giving an internal Mach
  number ${\mathcal M} \approx 2$ for $\beta \approx 0.8$ and the
  sound speed decreases as matter is entrained. The high-emissivity
  region is the volume over which ${\mathcal M} \ga
  {\mathcal M}_{\rm crit} = 1.5 - 1.8$.
\item A subsonic shear layer forms at the edge of the jet at or
  slightly downstream of the brightness flaring point.  As it grows, a
  measurable transverse velocity gradient develops across the jet and the spine also
  decelerates.
\item Jet evolution is remarkably similar in twin-jet sources with and without
  lobes.  Jets in both classes of source propagate in
  direct contact with the external medium within the dense, hot,
  kpc-scale coronae that always surround their AGN and the
  majority of entrainment occurs in these regions.  As the external
  environment becomes more tenuous, owing to the jet entering a lobe
  or to a rapid decrease in the density of the external galactic
  atmosphere, the entrainment rate drops.  The flow velocity then
  usually reaches an asymptotic value, preserving its previously
  acquired transverse gradient, and the jet starts to recollimate,
  eventually spreading at a low and constant rate.  An exception is
  3C\,31, whose continuing deceleration on large scales can be
  accounted for by the availability of group-scale gas for
  entrainment.
\item Depending on the amount of entrainment, the shear
  layer may expand to fill the entire jet or the supersonic spine may
  persist after deceleration and recollimation.
\item Steeper radio spectra ($\alpha = 0.66$) and acceleration of
  particles up to Lorentz factors of $10^7 - 10^8$ occur in the
  supersonic flow before the jets decelerate significantly.  A
  possible mechanism is Fermi acceleration by mildly relativistic
  shocks \citep{SB12}. 
\item Flatter spectra ($\alpha = 0.59$) and lower maximum Lorentz
  factors ($10^5 - 10^6$) are associated with the flow after
  deceleration.  These could be produced by Fermi acceleration with a
  lower shock velocity or by a second acceleration mechanism, perhaps
  associated with velocity shear.
\item The trigger for jet deceleration remains unclear, but we note
  that the brightness flaring points are always located on the edges
  of the dense coronae, in steeply-falling external pressure
  gradients.  It is plausible that the jets become overpressured and
  that this results in the formation of internal shocks and/or in the
  non-linear growth of Kelvin-Helmholtz modes. Although the
  high-emissivity regions of several of the jets contain complex,
  non-axisymmetric brightness features (Fig.~\ref{fig:ihires}), these
  are not obviously consistent with either possibility, and
  observations at higher angular resolution may be needed to
  distinguish between different explanations for the flaring.
\end{enumerate}

\section*{Acknowledgements}

The National Radio Astronomy Observatory is a facility of the National
Science Foundation, operated under co-operative agreement by
Associated Universities, Inc. We thank the referee for
thought-provoking questions which caused us to improve the text in
several places. RAL thanks Mary and Alan Bridle for gracious
hospitality over the entire course of the project.

\appendix

\onecolumn

\section{Coordinate definitions and fitting functions}
\label{function-details}
\FloatBarrier

In this section, we tabulate the fitting functions for geometry,
velocity,  emissivity function and magnetic-field ordering. Expressions
for outflow and backflow components are given in
Tables~\ref{tab:functions} and \ref{tab:backflow_functions},
respectively.  

\begin{table*}
\caption{Coordinate definitions and functional forms for geometry,
  velocity,  emissivity function and magnetic-field ordering. (1)
  physical quantity; (2) symbol, as used in the text; (3) functional
  form; (4) range of distance coordinate, $r$, over which the expression is applicable. \label{tab:functions}}
\begin{minipage}[t][197mm][t]{180mm}
\begin{tabular}{llll}
\hline
&&&\\
Description & Quantity & Functional form & Distance range\\
&&&\\
\hline
&&&\\
Distance coordinate &
$r$ & $zr_0/[(r_0 + A)\cos\xi -A]$&$r \leq r_0$ \\
    && $(z+A)/\cos\xi - A$& $r \geq r_0$\\
    && $A = x_0/\sin\xi_0-r_0$ &\\
&&&\\
Streamline index & $s$ & by continuity  & $r \leq r_0$ \\ 
&& $\xi/\xi_0$ & $r \geq r_0$ \\
&&&\\
Radius & $x(z,s)$ & $a_2(s)z^2 + a_3(s)z^3$ & $r \leq r_0$ \\
         && $(z-r_0+x_0/\sin\xi_0)\tan\xi_0s$& $r \geq r_0$ \\
&&&\\
Velocity &$\beta(r,s)$ & $\beta_1 \exp(s^2\ln v_1)$ & $r \leq r_{v_1}$ \\
&&&\\
             && $\beta_1 \exp(s^2\ln v_1)\left (\frac{r_{v0} - r}{r_{v0} -r_{v1}}
             \right ) + \beta_0 \exp(s^2\ln v_0)\left (\frac{r - r_{v1}}{r_{v0} -r_{v1}}
             \right )$ & $r_{v1} \leq r \leq r_{v0}$ \\
&&&\\
             && $\beta_0 \exp(s^2\ln v_0)\left (\frac{r_{\rm grid} - r}{r_{\rm grid} -r_{v0}}
             \right ) + \beta_f \exp(s^2\ln v_f)\left (\frac{r - r_{v0}}{r_{\rm grid} -r_{v0}}
             \right )$ & $r_{v0} \leq r \leq r_{\rm grid}$ \\
&&&\\
Emissivity function &$\epsilon(r,s)$&$g_1 \left(\frac{r}{r_{e1}}\right)^{-E_{\rm in}}\exp( s^2\ln e_1)$&$r \leq r_{e1}$\\
&&&\\
               &&$~~~\left(\frac{r}{r_{e1}}\right)^{-E_{\rm mid}}\exp \left[\ln \left
                   ( s^2\frac{e_1(r_{e0}-r)+e_0(r-r_{e1})}{r_{e0}-r_{e1}} \right )
                 \right ]$&$r_{e1} < r \leq r_{e0}$\\ 
&&&\\
               &&$g_0 \left(\frac{r}{r_{e0}}\right)^{-E_{\rm out}}\left(\frac{r_{e0}}{r_{e1}}\right)^{-E_{\rm mid}}\exp \left[ s^2\ln \left
                   (\frac{e_0(r_{\rm grid}-r)+e_f(r-r_{e0})}{r_{\rm grid}-r_{e0}} \right )
                 \right ]$&$r_{e0} < r \leq r_{0}$\\
&&&\\
               &&$g_0 \left(\frac{r}{r_0}\right)^{-E_{\rm far}}\left(\frac{r_0}{r_{e0}}\right)^{-E_{\rm out}}\left(\frac{r_{e0}}{r_{e1}}\right)^{-E_{\rm mid}}\exp \left[ s^2\ln \left
                   (\frac{e_0(r_{\rm grid}-r)+e_f(r-r_{e0})}{r_{\rm grid}-r_{e0}} \right )
                \right ]$&$r_0 \leq r \leq r_{\rm grid}$\\
&&&\\
$\langle B_r^2/B_t^2\rangle^{1/2}$ &
$j(r,s)$&$j^{\rm cen} + (j^{\rm edge}-j^{\rm cen})s^{w_j} $&\\
      &&$j^{\rm edge} = j_1^{\rm edge}$&$r \leq r_{B1}$\\ 
      &&$j^{\rm edge} = \frac{j_1^{\rm edge}(r_{B0}-r)+j_0^{\rm edge}(r-r_{B1})}{r_{B0}-r_{B1}}$; $j^{\rm cen} = \frac{j_1^{\rm cen}(r_{B0}-r)+j_0^{\rm cen}(r-r_{B1})}{r_{B0}-r_{B1}}$&$r_{B1} \leq r \leq r_{B0}$\\
&&$j^{\rm edge} = \frac{j_0^{\rm edge}(r_{\rm grid}-r)+j_f^{\rm edge}(r-r_{B0})}{r_{\rm grid}-r_{B0}}$; $j^{\rm cen} = \frac{j_0^{\rm cen}(r_{\rm grid}-r)+j_f^{\rm cen}(r-r_{B0})}{r_{\rm grid}-r_{B0}}$&$r_{B0} \leq r \leq r_{\rm grid}$\\
$\langle B_l^2/B_t^2\rangle^{1/2}$ &
$k(r,s)$&$k^{\rm cen} + (k^{\rm edge}-k^{\rm cen})s^{w_k} $&\\
      &&$k^{\rm edge} = k_1^{\rm edge}$&$r \leq r_{B1}$\\ 
      &&$k^{\rm edge} = \frac{k_1^{\rm edge}(r_{B0}-r)+k_0^{\rm edge}(r-r_{B1})}{r_{B0}-r_{B1}}$; $k^{\rm cen} = \frac{k_1^{\rm cen}(r_{B0}-r)+k_0^{\rm cen}(r-r_{B1})}{r_{B0}-r_{B1}}$&$r_{B1} \leq r \leq r_{B0}$\\
&&$k^{\rm edge} = \frac{k_0^{\rm edge}(r_{\rm grid}-r)+k_f^{\rm edge}(r-r_{B0})}{r_{\rm grid}-r_{B0}}$; $k^{\rm cen} = \frac{k_0^{\rm cen}(r_{\rm grid}-r)+k_f^{\rm cen}(r-r_{B0})}{r_{\rm grid}-r_{B0}}$&$r_{B0} \leq r \leq r_{\rm grid}$\\
&&&\\
\hline
\end{tabular}
\end{minipage}
\end{table*}

\begin{table*}
\caption{Functional forms for geometry, velocity,  emissivity function
  and magnetic-field ordering in backflow components. These are
  exactly as in Table~4 of \citet{backflow}, and are given here for
  completeness.  (1)
  physical quantity; (2) symbol, as used in Table~\ref{tab:backflow}; (3) functional
  form; (4) range of distance coordinate, $r$, over which the expression is applicable.\label{tab:backflow_functions}}
\begin{tabular}{llll}
\hline
&&&\\
Description & Quantity & Functional form & Distance range\\
&&&\\
\hline
&&&\\
Backflow streamline index& $t$ & by continuity  & $r \leq r_0$ \\ 
&& $(\xi-\xi_0)/(\xi_{\rm b}-\xi_0)$& $r
\geq r_0$\\
&&&\\
Backflow radius & $x(z,s)$ & $a_2(t)z^2 + a_3(t)z^3$ & $r \leq r_0$ \\
         && $(z-r_0+x_0/\sin\xi_{\rm b})\tan\xi_{\rm b}s$& $r \geq r_0$ \\
&&&\\
Backflow velocity &
$\beta(t)$ & $\beta_{\rm b, in} + t(\beta_{\rm b, out}-\beta_{\rm b, in})$\\
&&&\\
Backflow emissivity function&
$\epsilon(r,t)$& 0              & $r < r_b$\\
             && $n_{\rm b}(r/r_0)^{-E_{\rm b}}\exp(\ln e_{\rm b}t^2)$& $r \geq r_b$ \\
&&&\\
Backflow $\langle B_r^2/B_t^2\rangle^{1/2}$&
$j$& $j_{\rm b}$ &\\
Backflow $\langle B_l^2/B_t^2\rangle^{1/2}$& $k$& $k_{\rm b}$ &\\
&&&\\
\hline
\end{tabular}
\end{table*}

\twocolumn

\section{$\chi^2$ values for the fits}
\label{chisq}
\FloatBarrier

In Table~\ref{tab:chisq}, we list the values of $\chi^2$ for the
best-fitting full and minimal models. These are summed over the Stokes
parameters $I$, $Q$ and $U$. The number of points is $\Sigma n_{\rm
  pixel} n_{\rm Stokes}$, where $n_{\rm pixel}$ is the number of {\em
  independent} sampling points and $n_{\rm Stokes}$ is the number of
Stokes parameters evaluated at that pixel.  The areas over which $I$,
$Q$ and $U$ were evaluated were always the same. Identical points were
normally used for all Stokes parameters (in which case the sum is just
$3\Sigma n_{\rm pixel}$), but we occasionally used the
higher-resolution image over a larger fraction of the area for $I$.

\begin{figure}
\epsfxsize=8.5cm
\epsffile{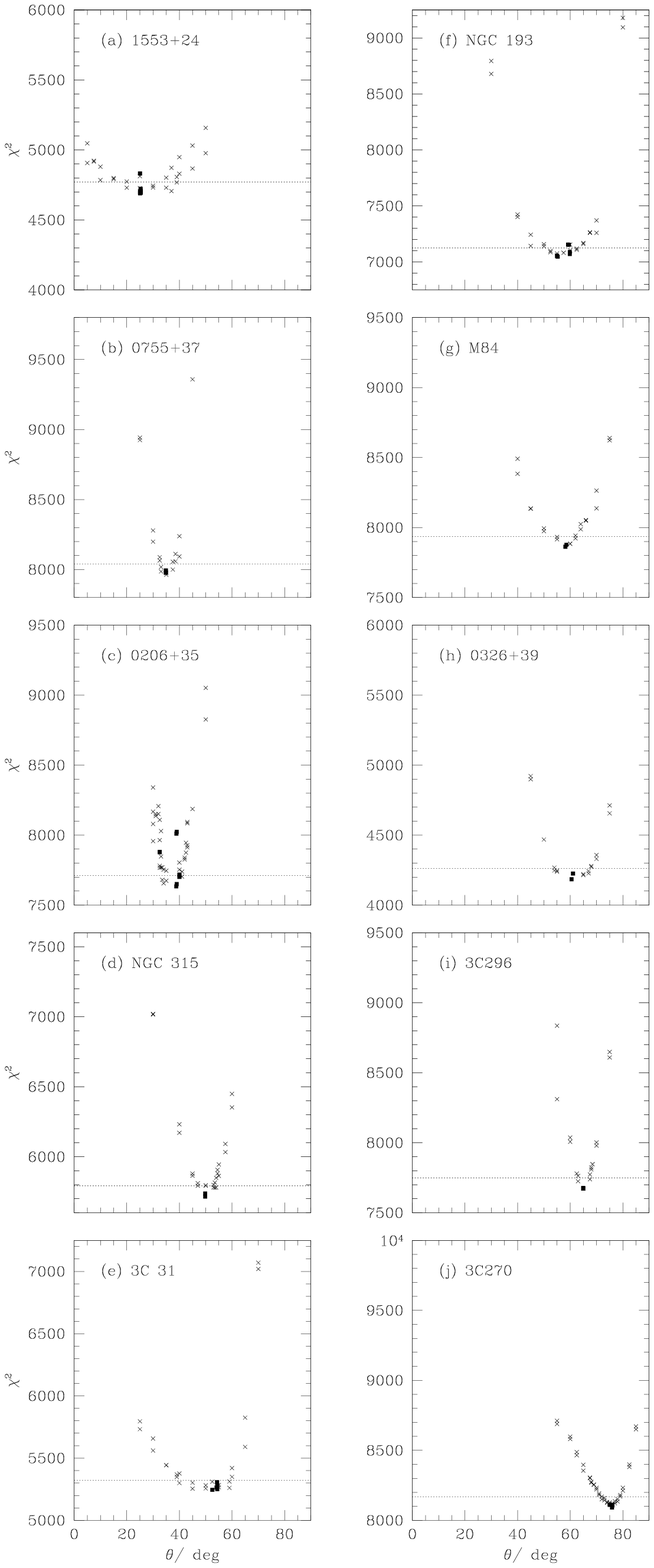}
\caption{Plots of $\chi^2$ against angle to the line of sight,
  $\theta$, for the full models. Crosses and filled squares are for optimizations with
  fixed and variable $\theta$, respectively. The numbers of
  independent points are given in Table~\ref{tab:chisq}. The horizontal, dashed lines represent the $\chi^2$ thresholds used to define error ranges (Section~\ref{errors}).
\label{fig:chisq}}
\end{figure}

Fig.~\ref{fig:chisq} shows plots of $\chi^2$ against inclination,
$\theta$, illustrating the lower bound described in Section~\ref{opt}.

\begin{table}
\caption{$\chi^2$ values for the fits. (1) Source name; (2) number of
  resolutions; (3) $\chi^2$ for the full models, summed over $I$, $Q$
  and $U$; (4) number of independent points. (5) and (6) as (3) and
  (4), but for the minimal models.\label{tab:chisq}}
\begin{tabular}{lcrrrr}
\hline
&&&&\\
Source&n$_{\rm res}$ &\multicolumn{2}{c}{Full}&\multicolumn{2}{c}{Minimal}\\
       &&$\chi^2$&N&$\chi^2$&N\\
&&&&\\
\hline
&&&&\\
1553+24&2& 4690.5 & 4566 & 4821.5 & 4566 \\
NGC\,315 &2& 5715.1 & 7368 & 5783.7 & 7368 \\
3C\,31   &2& 5244.8 & 3753 &        &      \\
NGC\,193 &2& 7047.0 & 5718 & 7236.9 & 5706 \\
M\,84    &1& 7857.1 & 5106 &        &      \\
0326+39&2& 4185.4 & 3594 & 4328.3 & 3594 \\
0755+37&2& 7962.6 & 5816 & 8022.4 & 5816 \\
0206+35&1& 7634.4 & 6708 & 8012.5 & 6696 \\
3C\,296  &2& 7671.5 & 8025 & 7677.5 & 8025 \\
3C\,270  &1& 8090.8 & 6384 & 8151.6 & 6384 \\
&&&&\\
\hline
\end{tabular}
\end{table}

\section{Notes on models for individual sources}
\label{notes}

\begin{description}
\item [{\bf 1553+24}] Our earlier model for this source gave a very
  small angle to the line of sight, $\theta \approx 8^\circ$
  \citep{CL}. A more systematic search of parameter space using
  optimizations with fixed values of $\theta$ (Section~\ref{opt})
  showed that this represented a local minimum in $\chi^2$.  The main
  reason for the difficulty in finding the global minimum is low
  signal-to-noise: 1553+24 is the weakest source in our sample.  The
  true minimum lies in a unusually broad range, $10^\circ \la \theta
  \la 40^\circ$ (Fig.~\ref{fig:chisq}a), but the derived velocity,
  emissivity function and field parameters are close to those given in
  Table~\ref{tab:fullparams} for any inclination in this range.  The
  new best-fitting inclination, $\theta = 25^\circ$ is more reasonable
  than the 8\degr\ found by \citet{CL}: 1553+24 has a bright core and a
  one-sided optical jet, consistent with a fairly low inclination, but
  does not show any of the nuclear properties of a BL Lac object, as
  would be expected if $\theta \la 10^\circ$, and the deprojected
  linear size of the extended structure is now less extreme.  As noted by
  \citet{CL}, the increase of jet/counter-jet sidedness ratio at large
  distances from the AGN is consistent with small intrinsic
  differences and may not require bulk acceleration.  The fitting
  functions used by \citet{CL}, in particular for emissivity function, differ
  significantly from those in this paper, making comparison
  difficult. The main substantive difference is that we find a 
  higher initial velocity, consistent with the increased inclination.
\item [{\bf 0755+37}] This is one of the two sources for which we
  include a mildly relativistic backflow in our models. The minimal
  model for this source is exactly as given by \citet{backflow}, who
  describe the backflow component in detail, and the full model is
  very similar.  The values of $r_{v0}$ and $r_{e0}$ in 0755+37 are
  anomalous (Fig.~\ref{fig:fidx0}b, d and g), perhaps because the
  counter-jet emission is dominated by the backflow component at large
  distances from the AGN, making it difficult to determine fiducial
  distances for the outflow.
\item [{\bf 0206+35}] This is the second source for which we include a
  backflow component in the model. The minimal model is as in
  \citet{backflow}; the full model differs only slightly.
\item [{\bf NGC\,315}] Our earlier model of NGC\,315 fit only the
  straight inner $\pm$70\,arcsec of the jets \citep{CLBC}. In order to
  constrain the flow parameters after recollimation, we have extended
  the model region to $\pm$100\,arcsec, after correcting for the slight bends
  observed in both jets.  The minimal model we tabulate for NGC\,315
  is described in detail by Laing \& Bridle (in preparation); the full model is very
  similar.  The technique of optimizing at fixed $\theta$
  showed that the inclination found by \citet{CLBC} corresponded to a
  local minimum in $\chi^2$ and that the true value is close to
  50\degr.  As a consequence, our fitted velocities are slightly
  faster than those given by \citet{CLBC}.  The fit places the outer
  magnetic fiducial distance close to the edge of the grid. We
  therefore fixed $r_{B0} = r_{\rm grid}$ to avoid convergence
  problems and no error is quoted for it.
\item [{\bf 3C\,31}] The fitting functions used for 3C\,31 by
  \citet{LB02a} differ significantly from those in this paper.
  \citet{LB02a} assumed a conical inner region close to the nucleus of
  3C\,31, but this is not required to fit the data and we found no
  evidence for such a structure in better-resolved cases.  Two forms
  of transverse structure were investigated by \citet{LB02a}:
  spine/shear-layer (in which parameter values in the two components
  are independent) and Gaussian (as in this paper).  The former
  gave only a slightly better fit, at the expense of a large number of
  additional parameters, so we do not consider it here.  Despite the
  different fitting functions, our new model is very similar to that
  described by \citet{LB02a}; for example, the smooth analytical form
  of the velocity profile assumed by \citet{LB02a} is close to the
  piecewise linear function of Fig.~\ref{fig:sketches}(a). $E_{\rm
      out}$ is not defined for 3C\,31 because the high-emissivity region ends
    very close to recollimation.
\item [{\bf NGC\,193}] The minimal model for this sources is described
  by Laing \& Bridle (in preparation); the full model is very similar.
\item [{\bf M\,84}] This small, intrinsically weak source is clearly
  interacting strongly with its environment, and the assumptions of
  intrinsic symmetry and axisymmetry fail closer to the nucleus than
  in any other case we have modelled. It is important to our study
  primarily because of its very low power.  The model is exactly as
  given by Laing \& Bridle (in preparation).  Although the total length of the model grid
  was $\pm20$\,arcsec, we only show a field of $\pm12.5$\,arcsec,
  since the fits are poor at larger distances from the AGN, where
  there is evidence for asymmetrical interactions with the local
  environment (Laing \& Bridle, in preparation).  The projections on the sky of the
  recollimation distance, $r_0$ and the fiducial distances for
  velocity and emissivity function, $r_{v1}$ and $r_{e1}$, are larger than
  12.5\,arcsec, so the parameters defined at $r_{\rm grid}$ (subscript
  $f$) do not influence the fits in the region we show.  They are
  therefore omitted from Table~\ref{tab:fullparams}.
\item [{\bf 0326+39}] The new model is very similar to that
  derived by \citet{CL}.
\item [{\bf 3C\,296}] Compared with the observed images fit by
  \citet{LCBH06}, those used here have been corrected for surrounding
  lobe emission and a slight bend in the outer counter-jet.  The new
  model includes a radial field component, and therefore provides a
  slightly better description of the polarization structure in the
  outer main jet.  Otherwise, the changes from \citet{LCBH06} are
  small. The prominent knot at the base of the counter-jet is not well
  fitted by any outflow model, as it is wider than the main jet at the
  same distance from the AGN.
\item [{\bf 3C\,270}] The model fit for 3C\,270 is described in detail
  by Laing et al. (in preparation).  It is a difficult source to model for several
  reasons. Firstly, systematic asymmetries in intensity and polarization are
  relatively small (the jet inclination is the largest in our sample)
  so intrinsic side-to-side differences have a significant effect even
  close to the AGN. The derived inclination of $\theta = 76\degr$ is
  at the practical upper limit for our technique
  (Section~\ref{asymm-em}).  Secondly, fits for fiducial distances are
  slightly different in the two jets. Finally, there are large areas
  of oblique apparent magnetic field which are inconsistent
  with any axisymmetric model (Fig.~\ref{fig:vectors}j), so the
  polarization fits are poor and the local degree of polarization is
  underestimated in the outer jets.  The fitted parameters describing
  the flow at projected distances $\ga$15\,arcsec (where the fit is
  poor) should be treated with caution.
\end{description}

\onecolumn
\section{Polarization vector plots}
\label{vectors}
\FloatBarrier

In Fig.~\ref{fig:vectors}, we show comparisons between the observed
and model linear polarizations, in the form of vector images.  The
format is similar to that of Figs~\ref{fig:icomposite} --
\ref{fig:qicomposite}, with panels (a) -- (j) showing model and
observed images arranged in order of increasing fitted angle to the
line of sight, but the individual panels are kept separate in the interests of legibility. The final
panel (k) shows the observations only for 3C\,449.

\begin{figure*}
\begin{minipage}{140mm}
\epsfxsize=13.5cm
\epsffile{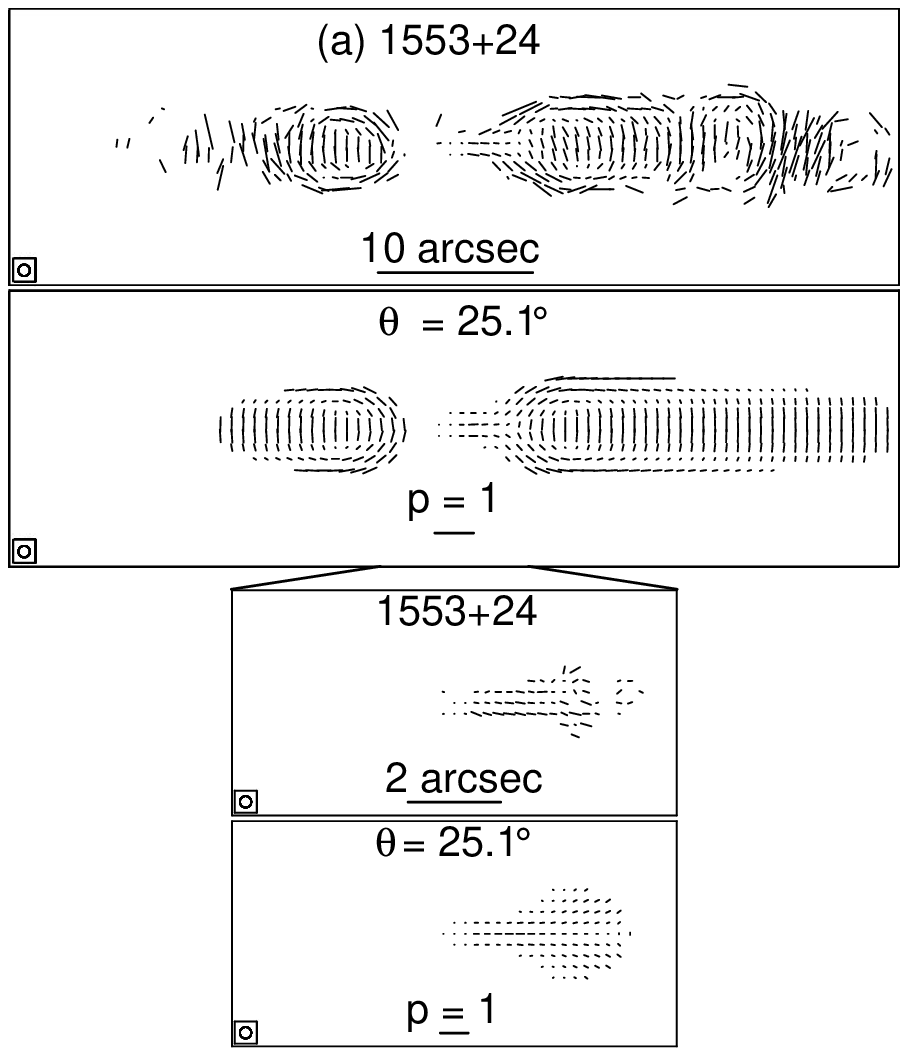}
\caption{(a) Comparison between observed and model polarization vector
  images for 1553+24. The lengths of the vectors are proportional to
  the degree of polarization, $p = P/I$, and their directions are
  those of the apparent magnetic field. The plots are arranged in
  pairs, as in Figs~\ref{fig:icomposite} -- \ref{fig:qicomposite}. The
  upper and lower panels of each pair show the observed image
  (labelled with the source name) and the model image (with the fitted
  value of $\theta$), respectively. The core is at the centre of each
  plot, and the brighter (approaching) jet is to the right. The angular scale is
  indicated by a labelled bar on the upper panel, the vector scale by
  a bar on the lower panel and the FWHM of the beam by a circle at the
  bottom of each plot.  The comparison at high resolution is shown
  below that at low resolution with the relative areas indicated.  The
  observed values of $P$ have been corrected for Ricean bias
  \citep{WK}.  The observed and model images are both blanked wherever
  $I < 5\sigma_I$ ($\sigma_I$ is the off-source noise level).  Field
  sizes and resolutions are given in
  Table~\ref{tab:images}.  \label{fig:vectors}}
\end{minipage}
\end{figure*}

\begin{figure*}
\begin{minipage}{140mm}
\epsfxsize=13.5cm
\epsffile{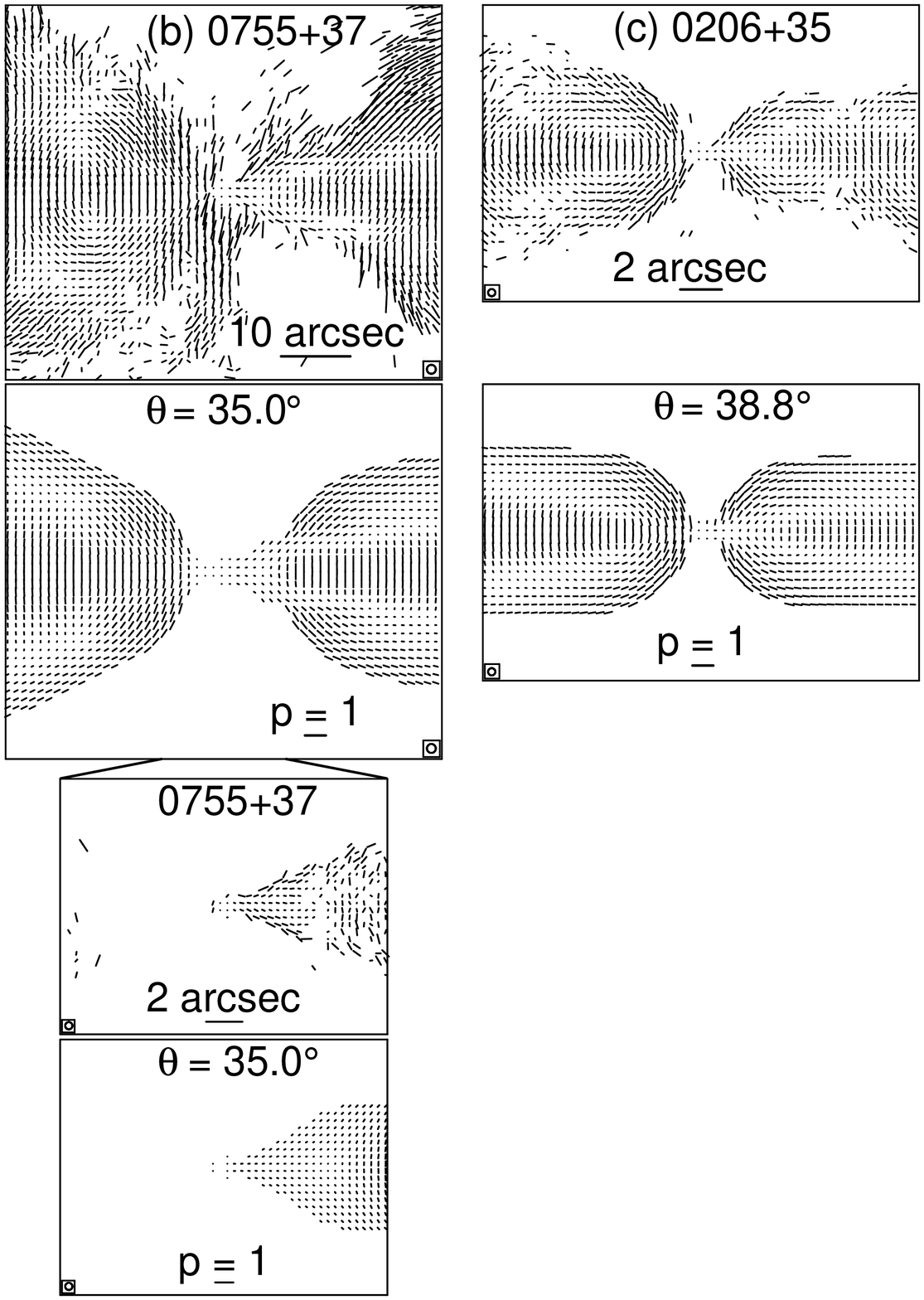}
\contcaption{(b) 0755+37; (c) 0206+35}
\end{minipage}
\end{figure*}

\begin{figure*}
\begin{minipage}{140mm}
\epsfxsize=13.5cm
\epsffile{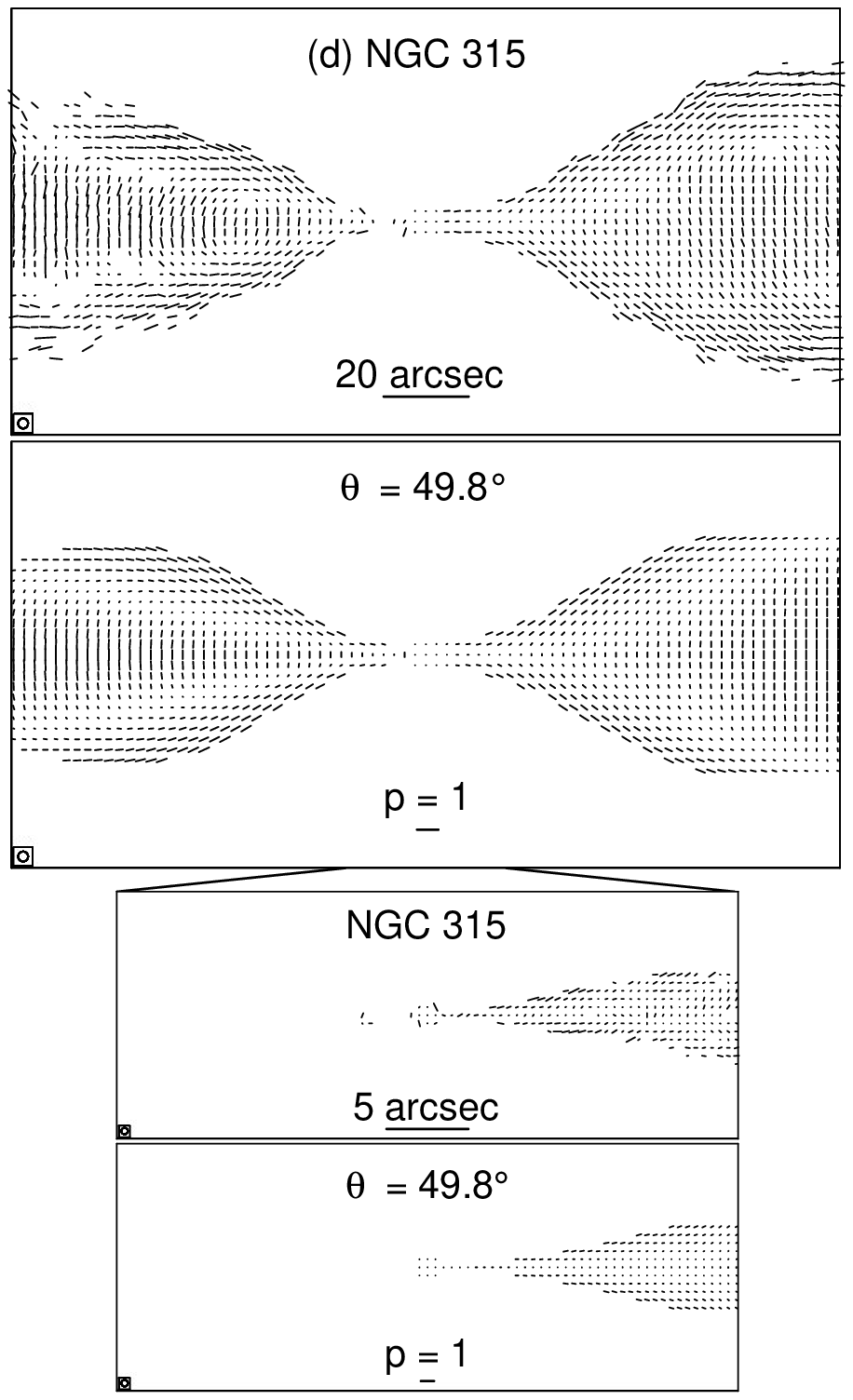}
\contcaption{(d) NGC\,315.}
\end{minipage}
\end{figure*}

\begin{figure*}
\begin{minipage}{140mm}
\epsfxsize=13.5cm
\epsffile{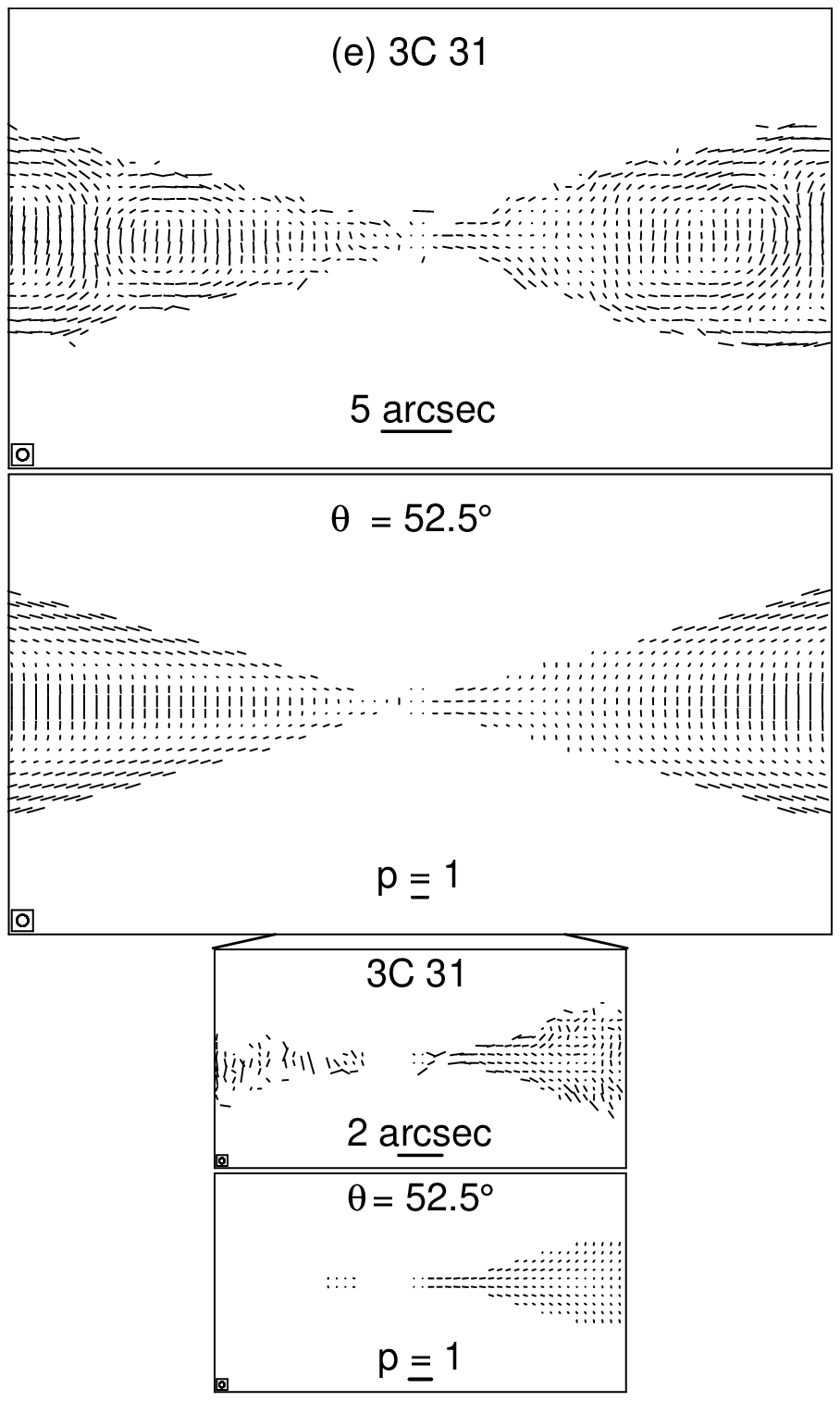}
\contcaption{(e) 3C\,31.}
\end{minipage}
\end{figure*}

\begin{figure*}
\begin{minipage}{140mm}
\epsfxsize=13.5cm
\epsffile{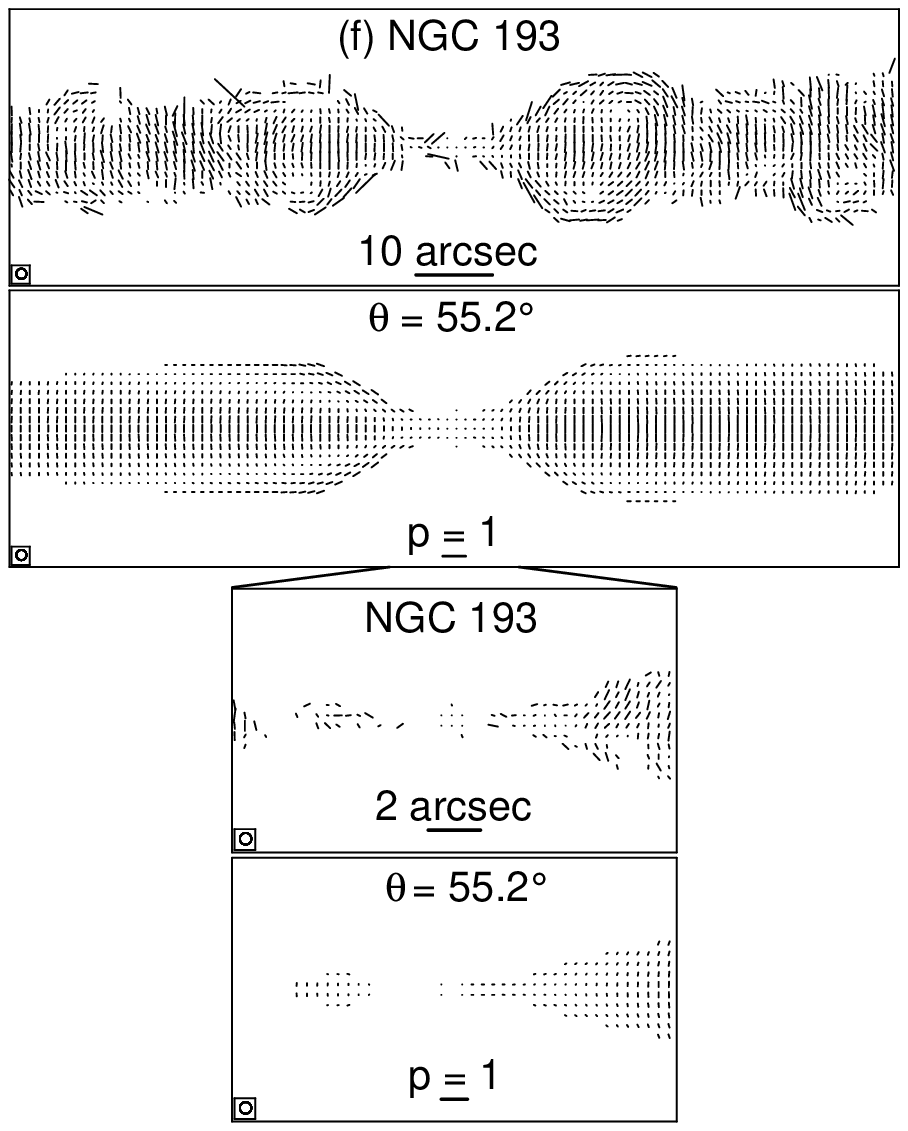}
\contcaption{(f) NGC\,193.}
\end{minipage}
\end{figure*}

\begin{figure*}
\begin{minipage}{140mm}
\epsfxsize=13.5cm
\epsffile{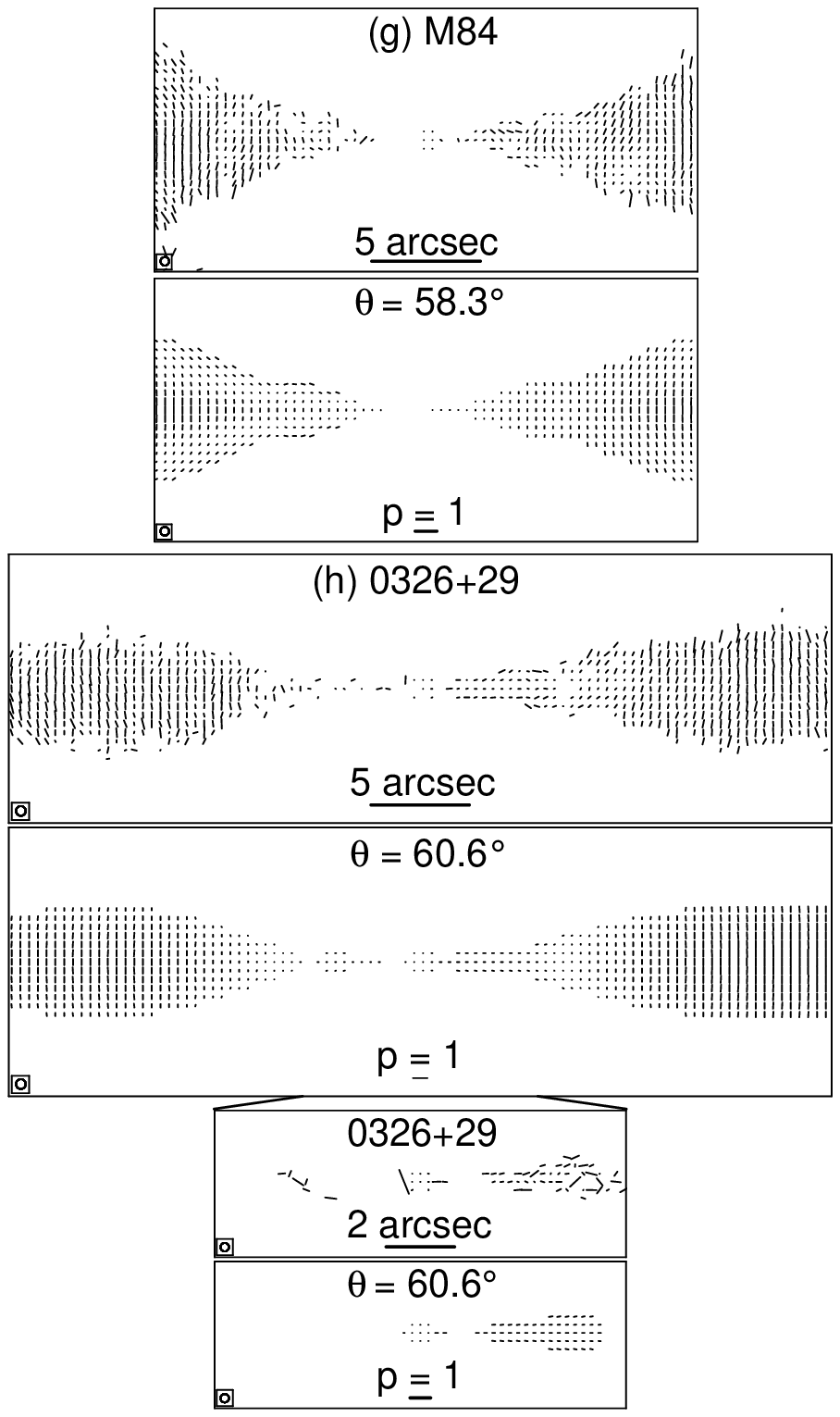}
\contcaption{(g) M\,84; (h) 0326+39.}
\end{minipage}
\end{figure*}

\begin{figure*}
\begin{minipage}{140mm}
\epsfxsize=13.5cm
\epsffile{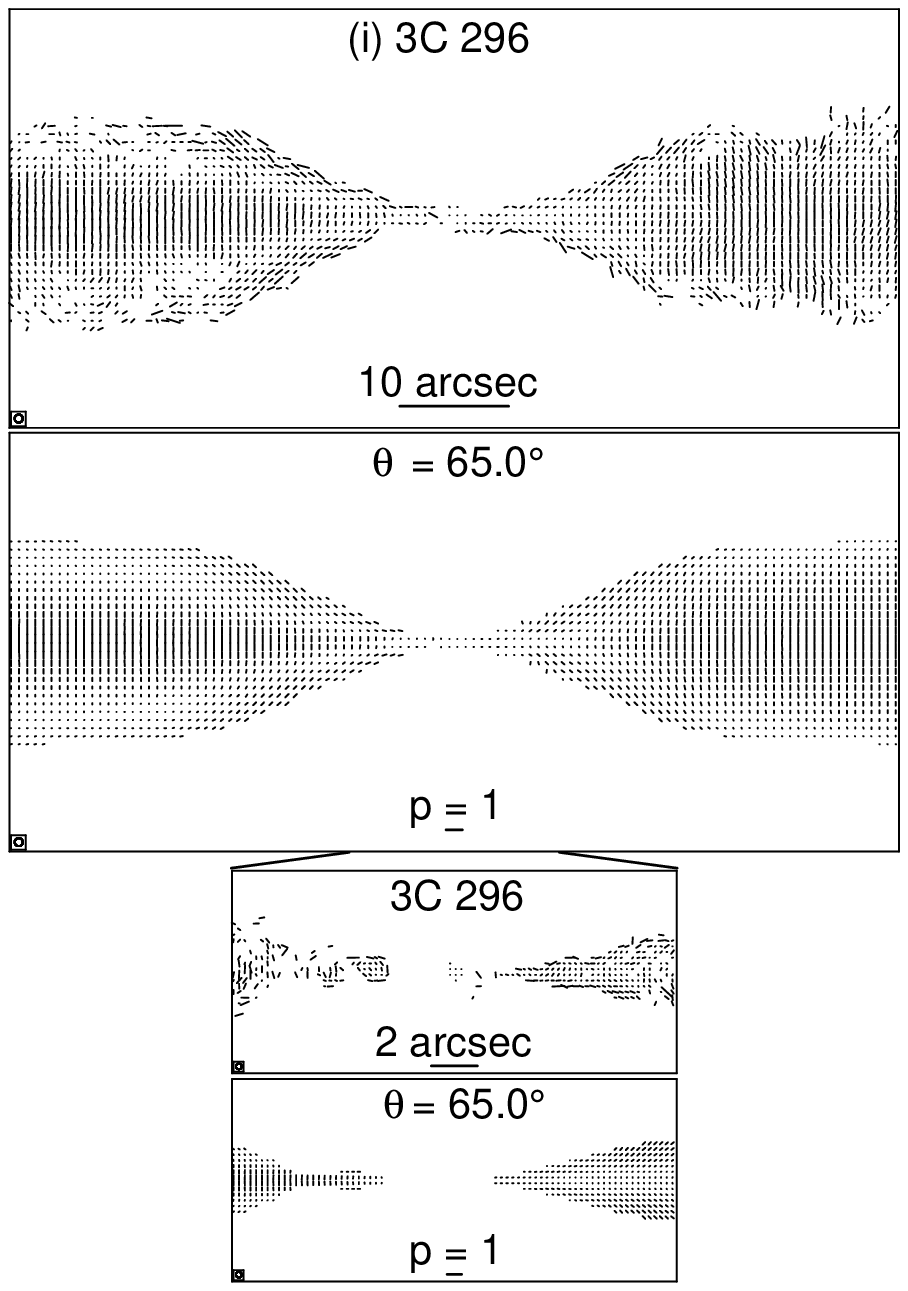}
\contcaption{(i) 3C\,296.}
\end{minipage}
\end{figure*}

\begin{figure*}
\begin{minipage}{140mm}
\epsfxsize=13.5cm
\epsffile{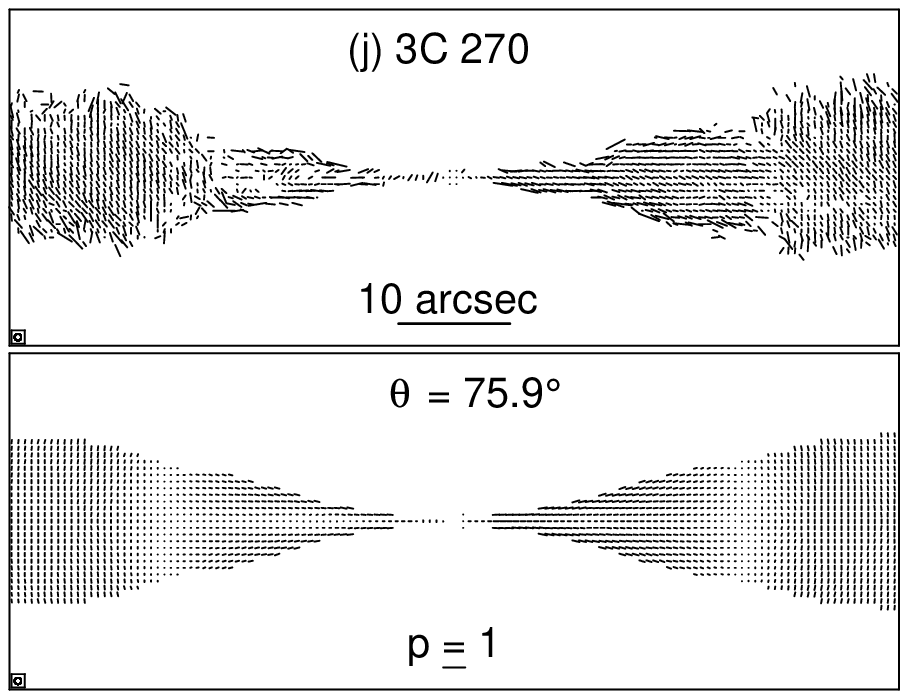}
\contcaption{(j) 3C\,270.}
\end{minipage}
\end{figure*}

\begin{figure*}
\begin{minipage}{140mm}
\epsfxsize=13.5cm
\epsffile{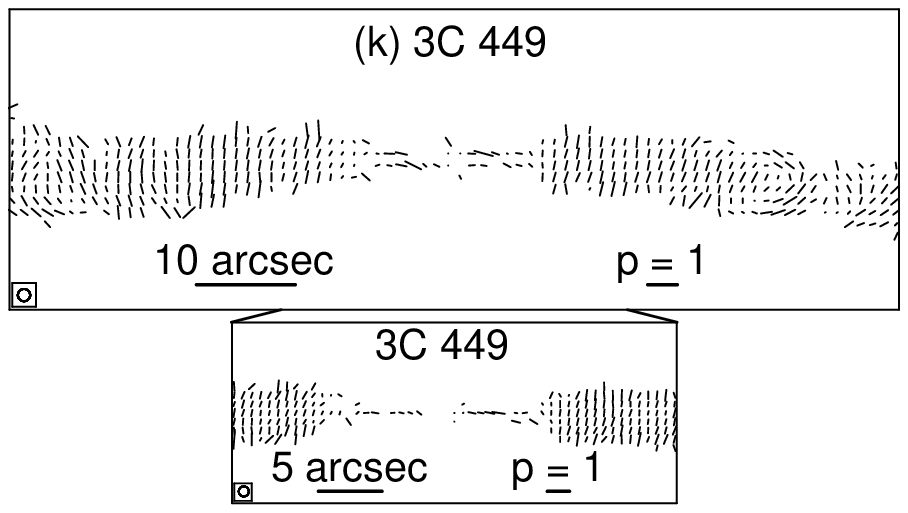}
\contcaption{(k) 3C\,449. Only observed images are shown for this source.}
\end{minipage}
\end{figure*}

\clearpage

\newpage

\section{Tabulated values of fiducial parameters}
\label{parameter-tables}
\FloatBarrier

Tables~\ref{tab:fullparams} and \ref{tab:minparams} list the values of
the fitted parameters and their errors for our full and minimal
models, respectively.  In Table~\ref{tab:backflow}, we give the
parameters of the backflow components for 0206+25 and 0755+37 as defined by \citet{backflow}. These
are not discussed in this paper, but are tabulated here for
completeness.

\begin{center}
\begin{table*}
\begin{minipage}[t][210mm][t]{180mm}
\caption{Fitted parameters for the full model outflows. The variables are defined in
  Section~\ref{functions} and Table~\ref{tab:functions}, except for
  the inclination range, $\Delta\theta$ (Section~\ref{errors}).
  Col.\,1 gives the variable name; the remaining 10 columns refer to
  the modelled sources, in order of increasing angle to the line of
  sight.  Angles are in degrees and fiducial distances are in
  kpc. Values in brackets were used to compute the brightness
  distributions in Figs~\ref{fig:icomposite} -- \ref{fig:qicomposite} and
  \ref{fig:vectors}, but have almost no effect on them and are
  effectively unconstrained.  The bracketed fractional edge velocities
  multiply central velocities which are close to zero and the
  bracketed values of $w_j$ and $w_k$ occur in cases where the centre
  and edge field ratios are nearly the same. $-$ means
  that the parameter is not defined (see
  Appendix~\ref{notes}).\label{tab:fullparams}}
\begin{tabular}{lrrrrrrrrrr}
\hline
&&&&&&&&&&\\
\multicolumn{1}{c}{Var-}  &\multicolumn{10}{c}{Source}\\
\multicolumn{1}{c}{iable} &1553+24&0755+37&0206+35&NGC\,315&3C\,31&NGC\,193&M\,84&0326+39&3C\,296&3C\,270\\
&&&&&&&&&&\\
\hline
&&&&&&&&&&\\
\multicolumn{11}{c}{Distances fixed in projection}\\
&&&&&&&&&&\\
$r_{\rm min}$&$0.48$&$0.73$&$0.39$&$0.07$&$0.12$&$0.16$&$0.06$&$0.12$&$0.14$&$0.18$\\
$r_{\rm grid}$&$59.5$&$48.6$&$13.1$&$43.9$&$12.2$&$21.9$&$2.12$&$12.2$&$22.9$&$6.26$\\
&&&&&&&&&&\\
\multicolumn{11}{c}{Geometry}\\
&&&&&&&&&&\\
$\theta$&$25.1_{-0.9}^{+0.8}$&$35.0_{-1.1}^{+0.8}$&$38.8_{-1.7}^{+1.0}$&$49.8_{-0.2}^{+0.5}$&$52.5_{-0.9}^{+0.5}$&$55.2_{-1.2}^{+3.5}$&$58.3_{-2.1}^{+2.8}$&$60.6_{-2.8}^{+3.0}$&$65.0_{-0.8}^{+1.4}$&$75.9_{-2.5}^{+2.6}$\\
$\Delta\theta$&$10 - 39$&$32.5 - 38.5$&$33 - 41$&$47 - 54$&$40 - 60$&$45 - 62.5$&$55 - 62$&$55 -
70$&$62.5 - 68$&$71 - 79$\\
$\xi_0$&$3.4_{-0.8}^{+0.4}$&$7.4_{-0.1}^{+0.2}$&$3.8_{-0.3}^{+0.9}$&$7.6_{-1.4}^{+3.5}$&$13.7_{-0.5}^{+0.3}$&$5.8_{-1.1}^{+0.7}$&$12.6_{-2.1}^{+3.4}$&$2.1_{-1.0}^{+4.1}$&$5.9_{-1.2}^{+1.1}$&$9.9_{-2.2}^{+1.9}$\\
$r_0$&$12.9_{-0.2}^{+0.6}$&$13.9_{-0.4}^{+0.2}$&$5.41_{-0.07}^{+0.21}$&$34.9_{-0.4}^{+1.7}$&$3.72_{-0.06}^{+0.07}$&$8.1_{-0.2}^{+0.2}$&$1.80_{-0.02}^{+0.05}$&$10.3_{-0.2}^{+0.6}$&$14.6_{-0.2}^{+0.3}$&$5.19_{-0.09}^{+0.10}$\\
$x_0$&$3.08_{-0.36}^{+0.06}$&$3.88_{-0.05}^{+0.09}$&$1.32_{-0.05}^{+0.02}$&$11.1_{-0.7}^{+0.2}$&$0.83_{-0.03}^{+0.02}$&$3.06_{-0.18}^{+0.07}$&$0.58_{-0.03}^{+0.01}$&$2.43_{-0.19}^{+0.04}$&$6.02_{-0.21}^{+0.06}$&$1.25_{-0.06}^{+0.04}$\\ 
&&&&&&&&&&\\
\multicolumn{11}{c}{Velocity}\\
&&&&&&&&&&\\
$r_{v1}$&$7.4_{-1.3}^{+0.9}$&$3.6_{-1.5}^{+1.6}$&$1.8_{-0.3}^{+0.3}$&$6.6_{-1.1}^{+0.8}$&$2.3_{-0.4}^{+0.6}$&$0.0_{-0.0}^{+0.7}$&$0.28_{-0.08}^{+0.10}$&$1.7_{-1.1}^{+0.8}$&$3.6_{-0.5}^{+0.6}$&$1.99_{-0.16}^{+0.18}$\\
$r_{v0}$&$8.3_{-1.2}^{+0.8}$&$18.6_{-1.4}^{+2.5}$&$4.6_{-0.3}^{+0.2}$&$16.2_{-1.0}^{+0.7}$&$3.9_{-0.4}^{+0.5}$&$5.5_{-0.5}^{+0.5}$&$1.41_{-0.08}^{+0.06}$&$7.0_{-0.6}^{+0.8}$&$5.8_{-0.3}^{+0.4}$&$4.4_{-0.3}^{+0.3}$\\
$\beta_1$&$0.95_{-0.09}^{+0.01}$&$0.88_{-0.04}^{+0.05}$&$0.83_{-0.05}^{+0.09}$&$0.85_{-0.03}^{+0.02}$&$0.71_{-0.04}^{+0.05}$&$0.72_{-0.07}^{+0.08}$&$0.76_{-0.06}^{+0.06}$&$0.82_{-0.11}^{+0.09}$&$0.70_{-0.07}^{+0.06}$&$0.92_{-0.02}^{+0.03}$\\
$\beta_0$&$0.19_{-0.03}^{+0.04}$&$0.25_{-0.10}^{+0.07}$&$0.72_{-0.10}^{+0.09}$&$0.55_{-0.05}^{+0.02}$&$0.57_{-0.04}^{+0.04}$&$0.18_{-0.02}^{+0.03}$&$0.00_{-0.00}^{+0.03}$&$0.02_{-0.02}^{+0.05}$&$0.58_{-0.05}^{+0.06}$&$0.00^{+0.06}$\\
$\beta_f$&$0.28_{-0.08}^{+0.10}$&$0.26_{-0.09}^{+0.15}$&$0.62_{-0.14}^{+0.17}$&$0.62_{-0.07}^{+0.07}$&$0.13_{-0.02}^{+0.04}$&$0.12_{-0.09}^{+0.10}$&$-$&$0.20_{-0.11}^{+0.08}$&$0.58_{-0.05}^{+0.07}$&$0.12_{-0.12}^{+0.20}$\\
$v_1$&$1.00_{-0.15}^{+0.00}$&$1.00_{-0.07}^{+0.00}$&$0.93_{-0.10}^{+0.07}$&$1.00_{-0.07}^{+0.00}$&$1.00_{-0.18}^{+0.00}$&$0.97_{-0.24}^{+0.03}$&$0.99_{-0.18}^{+0.01}$&$0.58_{-0.22}^{+0.24}$&$0.96_{-0.25}^{+0.04}$&$0.81_{-0.04}^{+0.05}$\\
$v_0$&$0.40_{-0.21}^{+0.36}$&$0.25_{-0.19}^{+0.21}$&$0.02_{-0.01}^{+0.01}$&$0.36_{-0.08}^{+0.04}$&$0.47_{-0.08}^{+0.12}$&$0.91_{-0.40}^{+0.09}$&$(0.07)$&$(0.64)$&$0.01_{-0.01}^{+0.00}$&$(0.31)$\\
$v_f$&$0.43_{-0.29}^{+0.57}$&$0.27_{-0.17}^{+0.48}$&$0.11_{-0.07}^{+0.09}$&$0.36_{-0.12}^{+0.11}$&$1.00_{-0.32}^{+0.00}$&$(0.86)$&$-$&$(0.70)$&$0.01_{-0.01}^{+0.00}$&$(0.43)$\\
&&&&&&&&&&\\
\multicolumn{11}{c}{Emissivity Function}\\
&&&&&&&&&&\\
$r_{e1}$&$1.00_{-0.10}^{+0.08}$&$1.54_{-0.03}^{+0.03}$&$0.81_{-0.02}^{+0.03}$&$2.6_{-0.5}^{+0.3}$&$1.17_{-0.13}^{+0.16}$&$1.18_{-0.08}^{+0.10}$&$0.30_{-0.24}^{+0.10}$&$1.03_{-0.19}^{+0.15}$&$2.3_{-0.5}^{+0.5}$&$1.4_{-0.4}^{+0.4}$\\
$r_{e0}$&$3.36_{-0.10}^{+0.08}$&$10.1_{-0.2}^{+0.2}$&$2.10_{-0.07}^{+0.08}$&$7.77_{-0.25}^{+0.46}$&$3.70_{-0.10}^{+0.12}$&$2.53_{-0.23}^{+0.14}$&$0.72_{-0.04}^{+0.03}$&$2.88_{-0.09}^{+0.10}$&$7.30_{-0.16}^{+0.13}$&$3.87_{-0.11}^{+0.10}$\\
$E_{\rm in}$&$4.2_{-0.8}^{+0.6}$&$2.4_{-1.1}^{+0.9}$&$(2.8)$&$4.15_{-0.15}^{+0.09}$&$(2.6)$&$(2.5)$&$(0.7)$&$4.5_{-0.3}^{+0.2}$&$(0.4)$&$(2.3)$\\
$E_{\rm mid}$&$4.37_{-0.14}^{+0.16}$&$3.76_{-0.04}^{+0.04}$&$2.64_{-0.09}^{+0.08}$&$3.07_{-0.16}^{+0.13}$&$3.23_{-0.09}^{+0.11}$&$3.60_{-0.30}^{+0.22}$&$0.60_{-0.55}^{+0.35}$&$2.44_{-0.31}^{+0.28}$&$3.10_{-0.15}^{+0.14}$&$3.28_{-0.19}^{+0.19}$\\
$E_{\rm
  out}$&$0.12_{-0.09}^{+0.08}$&$1.15_{-0.26}^{+0.25}$&$1.94_{-0.10}^{+0.08}$&$3.03_{-0.07}^{+0.08}$&$-$&$2.12_{-0.08}^{+0.07}$&$2.61_{-0.14}^{+0.10}$&$1.29_{-0.12}^{+0.07}$&$0.89_{-0.09}^{+0.06}$&$0.2_{-0.3}^{+0.3}$\\
$E_{\rm far}$&$1.99_{-0.10}^{+0.10}$&$1.17_{-0.10}^{+0.09}$&$2.23_{-0.16}^{+0.21}$&$2.08_{-0.92}^{+1.10}$&$1.39_{-0.06}^{+0.05}$&$2.16_{-0.11}^{+0.11}$&$-$&$1.82_{-1.22}^{+1.29}$&$0.89_{-0.11}^{+0.15}$&$2.1_{-0.5}^{+1.0}$\\
$e_1$&$1.6_{-0.4}^{+0.7}$&$1.1_{-0.3}^{+0.2}$&$1.2_{-0.4}^{+1.1}$&$1.4_{-0.4}^{+0.4}$&$0.26_{-0.07}^{+0.08}$&$0.6_{-0.2}^{+0.2}$&$0.4_{-0.1}^{+0.2}$&$0.8_{-0.3}^{+0.5}$&$0.14_{-0.04}^{+0.04}$&$0.72_{-0.17}^{+0.18}$\\
$e_0$&$0.21_{-0.07}^{+0.11}$&$2.30_{-0.50}^{+0.37}$&$1.19_{-0.22}^{+0.21}$&$0.44_{-0.07}^{+0.11}$&$0.35_{-0.06}^{+0.06}$&$0.11_{-0.03}^{+0.03}$&$0.00_{-0.00}^{+0.00}$&$0.19_{-0.09}^{+0.06}$&$0.04_{-0.01}^{+0.01}$&$0.12_{-0.03}^{+0.09}$\\
$e_f$&$0.33_{-0.13}^{+0.19}$&$2.30_{-0.43}^{+0.70}$&$0.97_{-0.28}^{+0.43}$&$0.44_{-0.10}^{+0.14}$&$0.25_{-0.05}^{+0.03}$&$0.13_{-0.04}^{+0.04}$&$-$&$0.17_{-0.05}^{+0.07}$&$0.04_{-0.01}^{+0.01}$&$0.14_{-0.05}^{+0.04}$\\
$g_1$&$0.17_{-0.07}^{+0.08}$&$1.73_{-0.49}^{+0.43}$&$1.61_{-1.41}^{+1.42}$&$0.13_{-0.03}^{+0.03}$&$0.31_{-0.18}^{+0.22}$&$0.29_{-0.18}^{+0.16}$&$(0.9)$&$0.15_{-0.07}^{+0.06}$&$1.2_{-0.6}^{+0.5}$&$1.2_{-0.3}^{+0.4}$\\
$g_0$&$0.16_{-0.02}^{+0.02}$&$0.52_{-0.03}^{+0.05}$&$1.14_{-0.10}^{+0.08}$&$0.49_{-0.05}^{+0.05}$&$1.03_{-0.05}^{+0.05}$&$1.17_{-0.11}^{+0.19}$&$0.81_{-0.07}^{+0.15}$&$0.35_{-0.04}^{+0.06}$&$0.87_{-0.05}^{+0.05}$&$1.12_{-0.10}^{+0.12}$\\
&&&&&&&&&&\\
\hline
\end{tabular}
\end{minipage}
\end{table*}
\end{center}

\begin{center}
\begin{table*}
\contcaption{}
\begin{tabular}{lrrrrrrrrrr}
\hline
&&&&&&&&&&\\
\multicolumn{1}{c}{Var-}  &\multicolumn{10}{c}{Source}\\
\multicolumn{1}{c}{iable} &1553+24&0755+37&0206+35&NGC\,315&3C\,31&NGC\,193&M\,84&0326+39&3C\,296&3C\,270\\
&&&&&&&&&&\\
\hline
&&&&&&&&&&\\
\multicolumn{11}{c}{Field component ratios}\\
&&&&&&&&&&\\
$r_{B1}$&$2.2_{-2.2}^{+4.5}$&$8.8_{-1.9}^{+2.7}$&$0.0_{-0.0}^{+1.3}$&$0.0_{-0.0}^{+2.2}$&$0.0_{-0.0}^{+0.5}$&$0.0_{-0.0}^{+0.8}$&$0.0_{-0.0}^{+0.2}$&$3.2_{-1.0}^{+1.1}$&$0.0_{-0.0}^{+1.1}$&$1.2_{-0.5}^{+0.5}$\\
$r_{B0}$&$14.2_{-1.9}^{+2.1}$&$15.0_{-2.7}^{+2.4}$&$4.7_{-0.5}^{+0.6}$&$43.9$&$4.2_{-0.5}^{+0.4}$&$4.3_{-0.5}^{+0.5}$&$1.38_{-0.06}^{+0.09}$&$8.0_{-0.9}^{+0.9}$&$12.2_{-0.6}^{+0.9}$&$4.9_{-0.2}^{+0.3}$\\
$j_1^{\rm cen}$&$0.6_{-0.5}^{+0.3}$&$1.0_{-0.4}^{+0.4}$&$1.5_{-1.3}^{+0.9}$&$0.8_{-0.3}^{+0.3}$&$0.0_{-0.0}^{+1.4}$&$0.9_{-0.8}^{+0.8}$&$1.3_{-0.5}^{+0.4}$&$1.3_{-0.5}^{+0.5}$&$0.0_{-0.0}^{+0.4}$&$0.6_{-0.6}^{+0.5}$\\
$j_1^{\rm edge}$&$2.1_{-0.3}^{+0.3}$&$1.0_{-0.2}^{+0.2}$&$1.5_{-0.4}^{+0.4}$&$0.3_{-0.1}^{+0.2}$&$0.7_{-0.4}^{+0.3}$&$0.9_{-0.5}^{+0.4}$&$0.7_{-0.5}^{+0.7}$&$0.8_{-0.4}^{+0.4}$&$1.5_{-1.5}^{+2.3}$&$0.6_{-0.3}^{+0.4}$\\
$j_0^{\rm cen}$&$0.4_{-0.4}^{+0.3}$&$0.6_{-0.6}^{+0.7}$&$0.2_{-0.2}^{+0.4}$&$0.1_{-0.1}^{+0.5}$&$0.1_{-0.1}^{+0.7}$&$0.7_{-0.2}^{+0.2}$&$0.4_{-0.1}^{+0.2}$&$1.1_{-0.4}^{+0.3}$&$0.0_{-0.0}^{+0.2}$&$1.0_{-0.2}^{+0.2}$\\
$j_0^{\rm edge}$&$0.0_{-0.0}^{+0.3}$&$0.3_{-0.3}^{+0.2}$&$0.1_{-0.1}^{+0.1}$&$0.0_{-0.0}^{+0.2}$&$0.5_{-0.1}^{+0.1}$&$0.3_{-0.2}^{+0.1}$&$0.3_{-0.2}^{+0.2}$&$0.2_{-0.2}^{+0.3}$&$0.7_{-0.7}^{+1.0}$&$1.1_{-0.3}^{+0.2}$\\
$j_f^{\rm cen}$&$0.9_{-0.9}^{+1.2}$&$0.5_{-0.5}^{+1.4}$&$0.2_{-0.2}^{+2.2}$&$-$&$1.1_{-1.1}^{+0.7}$&$1.0_{-0.6}^{+0.7}$&$-$&$1.0_{-0.5}^{+0.6}$&$0.0_{-0.0}^{+0.3}$&$0.9_{-0.3}^{+0.5}$\\
$j_f^{\rm edge}$&$0.1_{-0.1}^{+1.1}$&$0.5_{-0.3}^{+0.4}$&$0.1_{-0.1}^{+0.8}$&$-$&$0.0_{-0.0}^{+0.1}$&$0.5_{-0.5}^{+0.4}$&$-$&$0.1_{-0.1}^{+0.5}$&$0.7_{-0.7}^{+2.5}$&$1.0_{-0.4}^{+0.4}$\\
$k_1^{\rm cen}$&$1.8_{-0.8}^{+1.1}$&$1.2_{-0.2}^{+0.2}$&$1.5_{-0.4}^{+0.5}$&$1.3_{-0.1}^{+0.1}$&$1.6_{-0.3}^{+0.2}$&$1.4_{-0.3}^{+0.2}$&$1.5_{-0.2}^{+0.3}$&$1.3_{-0.4}^{+0.7}$&$1.8_{-0.1}^{+0.1}$&$1.7_{-0.4}^{+0.4}$\\
$k_1^{\rm edge}$&$1.7_{-0.1}^{+0.2}$&$1.16_{-0.09}^{+0.08}$&$1.4_{-0.2}^{+0.2}$&$0.8_{-0.1}^{+0.1}$&$1.2_{-0.3}^{+0.2}$&$1.0_{-0.3}^{+0.3}$&$0.9_{-0.2}^{+0.2}$&$1.2_{-0.1}^{+0.2}$&$0.2_{-0.2}^{+0.1}$&$1.6_{-0.2}^{+0.3}$\\
$k_0^{\rm cen}$&$0.3_{-0.3}^{+1.2}$&$0.1_{-0.1}^{+1.3}$&$0.9_{-0.2}^{+0.1}$&$0.6_{-0.1}^{+0.1}$&$1.2_{-0.1}^{+0.0}$&$0.5_{-0.1}^{+0.1}$&$0.8_{-0.1}^{+0.1}$&$0.5_{-0.5}^{+0.6}$&$0.59_{-0.06}^{+0.04}$&$0.78_{-0.10}^{+0.13}$\\
$k_0^{\rm edge}$&$0.2_{-0.2}^{+0.3}$&$0.1_{-0.1}^{+0.6}$&$0.5_{-0.2}^{+0.1}$&$0.0_{-0.0}^{+0.1}$&$0.1_{-0.1}^{+0.1}$&$0.3_{-0.2}^{+0.1}$&$0.0_{-0.0}^{+0.1}$&$0.0_{-0.0}^{+0.2}$&$0.01_{-0.01}^{+0.05}$&$0.75_{-0.11}^{+0.12}$\\
$k_f^{\rm cen}$&$0.1_{-0.1}^{+3.6}$&$0.1_{-0.1}^{+0.7}$&$0.8_{-0.3}^{+0.2}$&$-$&$0.3_{-0.2}^{+0.1}$&$0.6_{-0.4}^{+0.2}$&$-$&$0.3_{-0.3}^{+1.0}$&$0.58_{-0.07}^{+0.08}$&$0.8_{-0.2}^{+0.2}$\\
$k_f^{\rm edge}$&$0.3_{-0.3}^{+0.7}$&$0.1_{-0.1}^{+0.3}$&$0.6_{-0.4}^{+0.2}$&$-$&$0.7_{-0.1}^{+0.2}$&$0.5_{-0.5}^{+0.4}$&$-$&$0.2_{-0.2}^{+0.3}$&$0.00_{-0.00}^{+0.10}$&$0.75_{-0.20}^{+0.18}$\\
$w_j$&$1.5_{-0.5}^{+0.8}$&$(1.1)$&$(1.0)$&$0.7_{-0.5}^{+0.9}$&$0.3_{-0.3}^{+0.3}$&$0.8_{-0.4}^{+0.7}$&$(1.2)$&$1.1_{-0.4}^{+0.6}$&$(10)$&$(1.1)$\\
$w_k$&$(
0.3)$&$
(0.9)$&$1.0_{-0.4}^{+0.8}$&$1.3_{-0.2}^{+0.2}$&$1.4_{-0.2}^{+0.2}$&$1.5_{-0.9}^{+1.1}$&$0.6_{-0.1}^{+0.1}$&$0.4_{-0.4}^{+0.9}$&$1.2_{-0.1}^{+0.1}$&$(0.9)$\\
&&&&&&&&&&\\
\hline
\end{tabular}
\end{table*}
\end{center}

\begin{center}
\begin{table*}
\caption{Fitted parameters for minimal model outflows (see the
  description in the caption for
  Table~\ref{tab:fullparams}).\label{tab:minparams}}
\begin{tabular}{lrrrrrrrr}
\hline 
&&&&&&&&\\
\multicolumn{1}{c}{Var-}  &\multicolumn{8}{c}{Source}\\
\multicolumn{1}{c}{iable} &1553+24&0755+37&0206+35&NGC\,315&NGC\,193&0326+39&3C\,296&3C\,270\\ 
&&&&&&&&\\
\hline
&&&&&&&&\\
\multicolumn{9}{c}{Distances fixed in projection}\\
$r_{\rm min}$ &$0.48$&$0.73$&$0.38$&$0.07$&$0.16$&$0.12$&$0.14$&$0.19$\\
$r_{\rm grid}$&$62.0$&$48.8$&$12.8$&$43.9$&$20.9$&$12.2$&$22.9$&$6.42$\\
&&&&&&&&\\
\multicolumn{9}{c}{Geometry}\\
&&&&&&&&\\
$\theta$&$25.0_{-1.2}^{+0.8}$&$34.8_{-0.8}^{+0.7}$&$40.0_{-0.3}^{+0.3}$&$49.8_{-0.5}^{+0.3}$&$59.7_{-2.8}^{+1.6}$&$61.1_{-1.6}^{+3.2}$&$65.0_{-1.1}^{+1.3}$&$76.1_{-2.4}^{+2.9}$\\
$\Delta\theta$&$10 - 40$&$32.5 - 37.5$&$34 - 43$&$48 - 54.5$&$51 - 65$&$55 - 70$&$63 - 68.5$&$70 - 79$\\
$\xi_0$&$3.4_{-0.8}^{+0.4}$&$7.4_{-0.1}^{+0.2}$&$3.9_{-0.2}^{+0.2}$&$7.6_{-2.0}^{+2.6}$&$6.1_{-1.0}^{+0.9}$&$2.1_{-1.3}^{+3.0}$&$5.9_{-1.2}^{+1.1}$&$9.9_{-1.4}^{+3.0}$\\
$r_0$&$12.9_{-0.2}^{+0.6}$&$13.9_{-0.3}^{+0.3}$&$5.3_{-0.1}^{+0.1}$&$34.9_{-0.4}^{+1.3}$&$7.7_{-0.2}^{+0.1}$&$10.2_{-0.1}^{+0.5}$&$14.6_{-0.2}^{+0.3}$&$5.38_{-0.07}^{+0.14}$\\
$x_0$&$3.08_{-0.41}^{+0.06}$&$3.88_{-0.06}^{+0.08}$&$1.32_{-0.04}^{+0.02}$&$11.1_{-0.6}^{+0.2}$&$3.06_{-0.09}^{+0.10}$&$2.43_{-0.19}^{+0.03}$&$6.02_{-0.20}^{+0.07}$&$1.30_{-0.09}^{+0.03}$\\
&&&&&&&&\\
\multicolumn{9}{c}{Velocity}\\
&&&&&&&&\\
$r_{v1}$&$7.6_{-1.1}^{+0.6}$&$3.6_{-1.5}^{+1.6}$&$1.8_{-0.3}^{+0.3}$&$6.6_{-1.2}^{+0.7}$&$0.0_{-0.0}^{+0.7}$&$1.8_{-0.9}^{+1.0}$&$3.6_{-0.5}^{+0.6}$&$2.1_{-0.2}^{+0.2}$\\
$r_{v0}$&$7.7_{-0.7}^{+1.0}$&$18.5_{-1.5}^{+2.3}$&$4.1_{-0.2}^{+0.3}$&$16.2_{-1.1}^{+0.6}$&$5.7_{-0.5}^{+0.5}$&$6.6_{-0.7}^{+0.7}$&$5.8_{-0.3}^{+0.4}$&$4.5_{-0.2}^{+0.3}$\\
$\beta_1$&$0.95_{-0.07}^{+0.01}$&$0.88_{-0.04}^{+0.05}$&$0.86_{-0.07}^{+0.08}$&$0.85_{-0.03}^{+0.01}$&$0.75_{-0.06}^{+0.08}$&$0.81_{-0.10}^{+0.10}$&$0.70_{-0.07}^{+0.05}$&$0.92_{-0.02}^{+0.03}$\\
$\beta_0$&$0.21_{-0.03}^{+0.02}$&$0.25_{-0.05}^{+0.07}$&$0.68_{-0.05}^{+0.09}$&$0.55_{-0.02}^{+0.03}$&$0.18_{-0.02}^{+0.03}$&$0.07_{-0.03}^{+0.05}$&$0.58_{-0.03}^{+0.04}$&$0.00_{-0.00}^{+0.10}$\\
$v_1$&$0.98_{-0.10}^{+0.02}$&$1.00_{-0.06}^{+0.00}$&$0.95_{-0.13}^{+0.05}$&$1.00_{-0.08}^{+0.00}$&$0.98_{-0.18}^{+0.02}$&$0.55_{-0.20}^{+0.28}$&$0.96_{-0.26}$&$0.81_{-0.04}^{+0.06}$\\
$v_0$&$0.54_{-0.21}^{+0.25}$&$0.26_{-0.11}^{+0.19}$&$0.04_{-0.01}^{+0.02}$&$0.36_{-0.04}^{+0.05}$&$0.97_{-0.35}^{+0.03}$&$(0.82)              $&$0.01_{-0.01}^{+0.00}$&$(0.78)$\\
&&&&&&&&\\
\multicolumn{9}{c}{Emissivity Function}\\
&&&&&&&&\\
$r_{e1}$&$1.16_{-0.12}^{+0.11}$&$1.55_{-0.03}^{+0.04}$&$0.80_{-0.02}^{+0.02}$&$2.55_{-0.36}^{+0.35}$&$1.12_{-0.10}^{+0.08}$&$1.01_{-0.20}^{+0.13}$&$2.25_{-0.45}^{+0.58}$&$1.4_{-0.4}^{+0.5}$\\
$r_{e0}$&$3.34_{-0.06}^{+0.11}$&$10.2_{-0.3}^{+0.1}$&$2.04_{-0.06}^{+0.09}$&$7.77_{-0.23}^{+0.47}$&$2.46_{-0.19}^{+0.14}$&$2.85_{-0.10}^{+0.08}$&$7.30_{-0.15}^{+0.14}$&$4.00_{-0.14}^{+0.09}$\\
$E_{\rm
  in}$&$4.5_{-0.3}^{+0.3}$&$(2.4)$&$(3.1)$&$4.2_{-0.1}^{+0.1}$&$(2.7)$&$4.5_{-0.3}^{+0.2}$&&$(2.4)$\\
$E_{\rm mid}$&$4.4_{-0.1}^{+0.2}$&$3.76_{-0.04}^{+0.02}$&$2.59_{-0.08}^{+0.09}$&$3.07_{-0.14}^{+0.15}$&$3.6_{-0.2}^{+0.3}$&$2.4_{-0.4}^{+0.2}$&$3.1_{-0.1}^{+0.2}$&$3.3_{-0.2}^{+0.2}$\\
$E_{\rm out}$&$0.2_{-0.1}^{+0.1}$&$1.16_{-0.09}^{+0.05}$&$2.13_{-0.06}^{+0.08}$&$3.03_{-0.04}^{+0.12}$&$2.12_{-0.05}^{+0.05}$&$1.3_{-0.1}^{+0.1}$&$0.89_{-0.05}^{+0.05}$&$0.2_{-0.5}^{+0.2}$\\
$E_{\rm far}$&$1.9_{-0.1}^{+0.1}$&$-$&$-$&$2.1_{-0.9}^{+1.3}$&$-$&$-$&&$2.0_{-0.8}^{+0.7}$\\
$e_1$&$1.9_{-0.3}^{+0.9}$&$1.0_{-0.2}^{+0.3}$&$1.2_{-0.5}^{+0.6}$&$1.5_{-0.4}^{+0.5}$&$0.5_{-0.2}^{+0.2}$&$0.7_{-0.3}^{+0.4}$&$0.2_{-0.1}^{+0.1}$&$0.69_{-0.21}^{+0.14}$\\
$e_0$&$0.24_{-0.07}^{+0.07}$&$2.2_{-0.3}^{+0.5}$&$1.14_{-0.18}^{+0.16}$&$0.44_{-0.06}^{+0.09}$&$0.11_{-0.01}^{+0.05}$&$0.18_{-0.06}^{+0.06}$&$0.04_{-0.01}^{+0.01}$&$0.14_{-0.04}^{+0.05}$\\
$g_1$&$0.23_{-0.05}^{+0.07}$&$1.7_{-0.4}^{+0.5}$&$1.7_{-1.3}^{+0.8}$&$0.13_{-0.03}^{+0.03}$&$0.26_{-0.13}^{+0.22}$&$0.14_{-0.06}^{+0.05}$&$1.2_{-0.6}^{+0.5}$&$1.2_{-0.4}^{+0.3}$\\
$g_0$&$0.18_{-0.03}^{+0.02}$&$0.52_{-0.03}^{+0.06}$&$1.05_{-0.09}^{+0.08}$&$0.49_{-0.06}^{+0.05}$&$1.08_{-0.11}^{+0.16}$&$0.35_{-0.04}^{+0.06}$&$0.86_{-0.04}^{+0.06}$&$1.13_{-0.08}^{+0.16}$\\
&&&&&&&&\\
\multicolumn{9}{c}{Field component ratios}\\
&&&&&&&&\\
$r_{B1}$&$4.8_{-2.4}^{+4.0}$&$8.8_{-2.0}^{+2.8}$&$0.0_{-0.0}^{+1.4}$&$0.0_{-0.0}^{+2.1}$&$0.0_{-0.0}^{+0.8}$&$2.9_{-1.5}^{+0.7}$&$0.0_{-0.0}^{+1.2}$&$1.3_{-0.7}^{+0.4}$\\
$r_{B0}$&$14.5_{-2.4}^{+2.0}$&$15.4_{-3.2}^{+2.5}$&$4.6_{-0.5}^{+0.5}$&$43.9$&$3.9_{-0.5}^{+0.7}$&$8.3_{-0.9}^{+0.9}$&$12.2_{-0.6}^{+0.9}$&$5.1_{-0.4}^{+0.2}$\\
$j_1^{\rm cen}$&$1.21_{-0.18}^{+0.13}$&$0.96_{-0.09}^{+0.13}$&$1.50_{-0.22}^{+0.34}$&$0.8_{-0.3}^{+0.3}$&$0.7_{-0.2}^{+0.4}$&$1.2_{-0.4}^{+0.7}$&$0.0_{-0.0}^{+0.4}$&$0.62_{-0.19}^{+0.26}$\\
$j_1^{\rm edge}$&$-$&$-$&$-$&$0.3_{-0.1}^{+0.2}$&$-$&$0.8_{-0.3}^{+0.5}$&$1.5_{-1.5}^{+2.3}$&$-$\\
$j_0^{\rm cen}$&$0.24_{-0.24}^{+0.16}$&$0.44_{-0.15}^{+0.12}$&$0.11_{-0.11}^{+0.13}$&$0.1_{-0.1}^{+0.5}$&$0.50_{-0.05}^{+0.09}$&$1.1_{-0.3}^{+0.3}$&$0.0^{+0.2}$&$0.99_{-0.08}^{+0.11}$\\
$j_0^{\rm edge}$&$-$&$-$&$-$&$0.1_{-0.1}^{+0.1}$&$-$&$0.2_{-0.2}^{+0.2}$&$0.7_{-0.7}^{+0.9}$&$-$\\
$k_1^{\rm cen}$&$1.52_{-0.08}^{+0.13}$&$1.15_{-0.07}^{+0.08}$&$1.36_{-0.13}^{+0.13}$&$1.3_{-0.1}^{+0.1}$&$1.16_{-0.14}^{+0.15}$&$1.4_{-0.6}^{+0.4}$&$1.75_{-0.10}^{+0.14}$&$1.64_{-0.17}^{+0.14}$\\
$k_1^{\rm edge}$&$-$&$-$&$-$&$0.8_{-0.1}^{+0.1}$&$-$&$1.2_{-0.1}^{+0.3}$&$0.17_{-0.13}^{+0.17}$&$-$\\
$k_0^{\rm cen}$&$0.20_{-0.20}^{+0.20}$&$0.08_{-0.08}^{+0.22}$&$0.64_{-0.04}^{+0.05}$&$0.6_{-0.1}^{+0.1}$&$0.48_{-0.03}^{+0.04}$&$0.4_{-0.4}^{+0.4}$&$0.59_{-0.04}^{+0.03}$&$0.76_{-0.05}^{+0.04}$\\
$k_0^{\rm edge}$&$-$&$-$&$-$&$0.0_{-0.0}^{+0.1}$&$-$&$0.0_{-0.0}^{+0.1}$&$0.01_{-0.01}^{+0.04}$&$-$\\
$w_j$&$-$&$-$&$-$&$0.7_{-0.5}^{+1.0}$&$-$&$1.1_{-0.4}^{+0.6}$&$(9.6)$&$-$\\
$w_k$&$-$&$-$&$-$&$1.3_{-0.2}^{+0.3}$&$-$&$0.4_{-0.6}^{+0.6}$&$1.2_{-0.1}^{+0.1}$&$-$\\
&&&&&&&&\\
\hline
\end{tabular}
\end{table*}
\end{center}

\begin{center}
\begin{table*}
\caption{Fitted parameters for backflow components, as defined in
  \citet{backflow}.  (1) Variable name
  \citep[Table~\ref{tab:backflow_functions} of this
    paper]{backflow}; (2) unit; (3) values for minimal model of
  0755+37; (4) values for full model of 0755+37; (5) and (6) as (3)
  and (4), but for 0206+35. The parameters for the minimal models are
  exactly as given by \citet{backflow}. Those for the full models,
  which are very similar (columns 4 and 6) come from optimizations
  with the full set of outflow parameters listed in
  Table~\ref{tab:fullparams}. \label{tab:backflow}}
\begin{tabular}{lrrrrr}
\hline 
&&&&&\\
\multicolumn{2}{c}{Variable}&\multicolumn{4}{c}{Source}\\ 
&&\multicolumn{2}{c}{0755+37}&\multicolumn{2}{c}{0206+35}\\ 
&& min & full & min & full\\
&&&&&\\
\hline
&&&&&\\
\multicolumn{6}{c}{Geometry}\\
&&&&&\\
$\xi_{\rm back}$&deg&$15.6_{-0.1}^{+0.5}$&$15.7_{-0.1}^{+0.5}$&$10.9_{-0.5}^{+0.5}$&$10.6_{-0.8}^{+2.6}$\\
$r_{\rm back}$&kpc&$23.2_{-0.7}^{+0.8}$&$23.1_{-0.8}^{+0.8}$&$2.7_{-0.2}^{+0.1}$&$2.7_{-0.1}^{+0.2}$\\
&&&&&\\
\multicolumn{6}{c}{Velocity}\\
&&&&&\\
$\beta_{\rm back}^{\rm in}$&&$0.25_{-0.07}^{+0.04}$&$0.24_{-0.06}^{+0.05}$&$0.02_{-0.02}^{+0.03}$&$0.02_{-0.02}^{+0.04}$\\
$\beta_{\rm back}^{\rm out}$&&$0.35_{-0.05}^{+0.05}$&$0.36_{-0.05}^{+0.05}$&$0.20_{-0.07}^{+0.06}$&$0.18_{-0.07}^{+0.06}$\\
&&&&&\\
\multicolumn{6}{c}{Emissivity Function}\\
&&&&&\\
$n_{\rm back}$&$\times 100$&$0.094_{-0.010}^{+0.000}$&$0.094_{-0.005}^{+0.006}$&$2.3_{-0.2}^{+0.2}$&$2.3_{-0.1}^{+0.2}$\\
$E_{\rm back}$&&$1.81_{-0.05}^{+0.07}$&$1.82_{-0.06}^{+0.06}$&$1.66_{-0.07}^{+0.06}$&$1.64_{-0.08}^{+0.04}$\\
$e_{\rm back}$&&$0.79_{-0.14}^{+0.13}$&$0.80_{-0.11}^{+0.18}$&$0.05_{-0.01}^{+0.02}$&$0.06_{-0.02}^{+0.02}$\\
&&&&&\\
\multicolumn{6}{c}{Field component ratios}\\
&&&&&\\
$j_{\rm back}$&&$0.38_{-0.07}^{+0.07}$&$0.38_{-0.07}^{+0.07}$&$0.24_{-0.07}^{+0.08}$&$0.23_{-0.03}^{+0.10}$\\
$k_{\rm back}$&&$0.03_{-0.03}^{+0.15}$&$0.01_{-0.01}^{+0.17}$&$0.38_{-0.09}^{+0.08}$&$0.35_{-0.07}^{+0.12}$\\
&&&&&\\
\hline
\end{tabular}
\end{table*}
\end{center}

\end{document}